%% file: main.tex
\documentclass[11pt]{article}
\usepackage[a4paper]{geometry}
\usepackage{subfiles} % Remove
\usepackage{amsmath,amsfonts,amssymb}
\usepackage{authblk}%add affiliation
\usepackage{alltt}
\usepackage{bm,bbm,bbold}% bold math
\usepackage{booktabs}
\usepackage{comment}
\usepackage{dcolumn}% Align table columns on decimal point
\usepackage{empheq} 
\usepackage{enumitem}
\usepackage{float}  % image placement
\usepackage{graphicx}% Include figure files
\usepackage{ltablex}  
\usepackage[scr=boondox]{mathalpha}
\usepackage{mathtools}
\usepackage{physics}
\usepackage{relsize}
\usepackage{setspace}
\usepackage{txfonts}
\usepackage{xcolor}
\usepackage[most]{tcolorbox} % For shaded boxes, put after xcolor
\usepackage{upgreek}
\usepackage{slashed}
\numberwithin{equation}{section}
\usepackage[sort&compress,numbers,merge]{natbib}
\usepackage[colorlinks=true,breaklinks=true]{hyperref}
\hypersetup{
allcolors=[rgb]{0.0 0.0 0.6},
linkcolor=[rgb]{0.0 0.0 1}
}
\usepackage[capitalise]{cleveref}
\crefformat{pluralequation}{#2{\color{black} Eqs.~(}#1{\color{black} )}#3}
\Crefformat{pluralequation}
{#2{\color{black} Equations~(}#1{\color{black} )}#3}
\def\ann {{\rm ann}}
\def\arc{{\Lambda}}

\def\BS {{\cal B}}
\def\BSD {{\mathsmaller{\rm BSD}}}
\def\BSS {{\mathsmaller{\rm BSS}}}
\def\cuts {{\rm cuts}}
\def\Disc {{\rm Disc}}
\def\Div {{\rm div}}
\def\elas {{\rm elas}}
\def\ee{\upeta_\ell}
\def\eeA{\upeta_{A,\ell}}
\def\eeH{\upeta_{H,\ell}}
\def\etaA {\eta_{A}}

\def\etaAell {\eta_{A,\ell}}
\def\etaHell {\eta_{H,\ell}}
\def\etaIell {\eta_{I,\ell}}
\def\etaRell {\eta_{R,\ell}}
\def\etaren {\eta^{\ren}}
\def\etaAellren {\eta_{A,\ell}^{\ren}}
\def\etaHellren {\eta_{H,\ell}^{\ren}}
\def\hc{\rm h.c.}
\def\fin{{\rm fin}}
\def\fs{\mathscr{f}}
\def\full{{\rm full}}
\def\ID{\mathbbm{1}}
\def\inel {{\rm inel}}
\def\im	{\mathbbm{i}}
\def\ii	{\im}

\def\mT {m_{\mathsmaller{T}}}
\def\NR {{\mathsmaller{\rm NR}}}
\def\nn {\nonumber}
\def\opt{{\rm opt}}

\def\poles{{\rm poles}}
\def\PV{\mathtt{P.V.}}
\def\reg{{\rm reg}}
\def\ren{{\rm ren}}
\def\subsqrts{{\scalebox{0.8}{$\scriptscriptstyle\sqrt{s}$}}}
\def\sym{\mathscr{c}}
\def\tot {{\rm tot}}

\def\TRI {{\rm TRI}}
\def\twoPI{\mathsmaller{\rm 2PI}}
\def\unreg{{\rm unreg}}

\def\RR{\mathbbm{R}}
\def\CC{\mathbbm{C}}
\def\NN{\mathbbm{N}}

\def\WW{\mathbbm{W}}
\def\ZZ{\mathbbm{Z}}

\def\SS{\mathbb{S}}
\def\TT{\mathbbm{T}}
\newcommand{\myshadebox}[1]{%
\tcboxmath[
colback=gray!20, % background
colframe=gray!20, % same as back so the frame is invisible
boxrule=0pt,      % no visible rule
arc=0pt,          % sharp corners
left=1pt, right=1pt, top=1pt, bottom=1pt
]{#1}%
}
\colorlet{lightblue}{blue!30}
\colorlet{lightred}{red!30}

\input{TIKZ_prelim} %for Feynman diagrams
\title{\bf Unitarizing non-relativistic scattering}
\author[1,2]{\Large
Marcos M. Flores\footnote{marcos.flores@fys.uio.no}} 
\author{Kalliopi Petraki\footnote{kalliopi.petraki@phys.ens.fr}
}

\affil[1]{\large\it
Laboratoire de Physique de l'École Normale Supérieure, 
ENS, 
Université PSL, 
CNRS, 
Sorbonne Université, 
Université Paris Cité,
Paris, F-75005, France}

\affil[2]{\large\it Department of Physics, University of Oslo, Box 1048, N-0316 Oslo, Norway }

\begin{document}

\maketitle
\begin{abstract}
\noindent 
Unitarity imposes coupled constraints on elastic and inelastic amplitudes. Satisfying them requires resummation of the self-energy contributions from both elastic and inelastic channels. Inelastic channels generate anti-Hermitian contributions that can be consistently deduced from the unitarity relation underlying the optical theorem, leading to non-local separable potentials and a compact, unique and complete unitarization scheme in the non-relativistic regime. We present two alternative derivations of the anti-Hermitian kernel, from the continuity equation combined with LSZ reduction, and by integrating out inelastic channels. We further extend the unitarization framework to treat non-analytic and non-convergent behavior of inelastic amplitudes in the complex momentum plane and to incorporate bound states. For non-convergent amplitudes, we demonstrate two renormalization procedures in which anti-Hermitian separable potentials necessarily induce Hermitian separable counterterms, yielding finite cross-sections consistent with unitarity. These results provide a general tool for non-relativistic scattering, with clear applications to dark-matter phenomenology.

\end{abstract}

\clearpage
%\noindent\rule{\textwidth}{0.4pt}
\tableofcontents
%\noindent\rule{\textwidth}{0.4pt}

%%%%%%%%%%%%%%%%%%%%%%%%%%%%%%%%%%%%%%%%%%%%%%%%%%%%%%%
%%%%%%%%%%%%%%%%%%%%%%%%%%%%%%%%%%%%%%%%%%%%%%%%%%%%%%%

\clearpage
\section{Introduction \label{sec:Intro}}

%%%%%%%%%%%%%%%%%

Unitarity plays a central role in particle physics. A direct consequence of it, the optical theorem imposes \emph{coupled} constraints on elastic and inelastic partial-wave scattering amplitudes. 
Because the optical theorem and the resulting bounds are non-linear in the amplitudes, they cannot be exactly satisfied in calculations truncated at any finite order in perturbation theory. Preserving unitarity therefore requires resummation of the relevant interactions. In the presence of elastic interactions only, the self-energy kernel of a (multiparticle) state is Hermitian, and its resummation ensures consistency with the unitarity constraint on elastic amplitudes. Inelastic channels, however, generate anti-Hermitian contributions to self-energy kernels that encode the probability flux into other states. Consistency with the coupled unitarity constraints on elastic and inelastic amplitudes then requires a resummation of these contributions as well.

The anti-Hermitian part of the self-energy kernel can be deduced directly from the unitarity condition underpinning the optical theorem. On this basis, Ref.~\cite{Flores:2024sfy} formulated a unitarization scheme in the non-relativistic regime  for \emph{both elastic and inelastic} amplitudes. The anti-Hermitian potential obtained in this way has a distinctive structure: it is a sum of non-local but separable potentials, a consequence of the fact that it originates from contracting inelastic amplitudes at different momenta. Remarkably, this structure permits compact analytic solutions that can be readily applied in phenomenological investigations, accommodating all partial waves and multiple inelastic channels. The form of the anti-Hermitian potential constitutes an important difference between Ref.~\cite{Flores:2024sfy} and other unitarization approaches that include imaginary self-energy contributions in the computation of inelastic amplitudes, either with resummation~\cite{Blum:2016nrz,Braaten:2017dwq,Parikh:2024mwa,Watanabe:2025kgw} or without~\cite{Aydemir:2012nz,Kamada:2022zwb}.

The goal of this work is twofold. First, we generalize aspects of the formalism introduced in Ref.~\cite{Flores:2024sfy} to establish a solid foundation for phenomenological applications. In particular, we develop a framework that consistently encodes possible non-analytic and non-convergent behavior of inelastic amplitudes in the complex momentum plane, which, as already recognized in Ref.~\cite{Flores:2024sfy}, affects unitarization. When such amplitudes generate UV-divergent separable optical potentials, we demonstrate a simple renormalization procedure and show that the absorptive (anti-Hermitian) separable interactions necessarily require dispersive (Hermitian) separable counterterms. Including these contributions is essential for renormalizability, and extends the solution of Ref.~\cite{Flores:2024sfy} beyond the purely absorptive case; the converse need not hold, i.e.~Hermitian separable terms need not generate anti-Hermitian ones of the same form. Second, we expand on the theoretical underpinnings of the unitarization framework. We incorporate the possible presence of bound states, which introduce poles that contribute to the analytic structure, and present two alternative ways to derive the anti-Hermitian potential: one based on the continuity equation and LSZ reduction for two-particle states, and one based on the Feshbach formalism for integrating out states. Throughout the paper, we clarify various subtle points of the previous analysis.

The impetus for unitarizing inelastic cross-sections arose in large part from renewed interest in long-range interactions in the non-relativistic regime in the context of dark matter. Long-range interactions play an important role in several dark-matter frontiers, including multi-TeV thermal-relic dark matter~\cite{Hisano:2003ec}, self-interacting dark matter~\cite{Tulin:2017ara}, and primordial black holes~\cite{Flores:2020drq}. A strong connection has been drawn between long-range interactions and the saturation of non-relativistic unitarity bounds~\cite{vonHarling:2014kha,Baldes:2017gzw,Flores:2024sfy}. 

However, unitarity violations have been identified in several settings. Inelastic processes that are Sommerfeld-enhanced by light but massive mediators can exhibit parametric resonances that grow with decreasing velocity faster than unitarity allows~\cite{Blum:2016nrz}. Attractive Coulomb-like potentials can violate inelastic unitarity bounds at sufficiently large coupling~\cite{vonHarling:2014kha,Petraki:2016cnz}. Even more striking unitarity violations, \emph{for arbitrarily small couplings}, occur in radiative transitions between states governed by different Coulomb-like potentials, as in bound-state formation with emission of a charged scalar~\cite{Oncala:2019yvj,Oncala:2021swy,Oncala:2021tkz,Ko:2019wxq} or a non-Abelian gauge boson~\cite{Harz:2018csl,Harz:2019rro,Binder:2023ckj,Beneke:2024nxh}. These issues have highlighted theoretical deficiencies and impelled the development of a systematic and consistent unitarization procedure. The phenomenological significance of the unitarization prescription of  Ref.~\cite{Flores:2024sfy} has already been demonstrated in the context of dark-matter production via thermal freeze-out~\cite{Petraki:2025zvv}.

\bigskip

The paper is structured as follows. Starting with the unitarity relation, in \cref{sec:Unitarity}, we reproduce the bounds on elastic and inelastic processes and extract the anti-Hermitian part of the self-energy kernel. 
In \cref{sec:OpticalPotential}, we review the optical potential via the Feshbach projection, from which we derive, in a second way, the anti-Hermitian self-energy kernel in terms of inelastic amplitudes, as well as its Hermitian counterpart generated by inelastic vertices.  
In \cref{sec:Jost}, we review the analytic properties of wavefunctions of Hermitian potentials, introducing the Jost function formalism that allows us to incorporate bound states into the unitarization scheme of the next section. 
In \cref{sec:Resummation}, we deduce the anti-Hermitian kernel in a third way, using the continuity equation and LSZ reduction. We review the unitarization formalism of Ref.~\cite{Flores:2024sfy} and extend it to systematically incorporate possible non-analyticities and non-convergence of inelastic amplitudes. Two renormalization prescriptions for non-convergent amplitudes are presented and compared in \cref{sec:Renorm}. 
We conclude in \cref{sec:Discussion} by summarizing the methodology and discussing the implications of this work, with a focus on dark-matter-motivated examples. 
For convenience, \cref{app:Math} collects useful mathematical proofs and identities, while \cref{app:Notation} compiles the notation used in this document, with references to the defining equations.

%%%%%%%%%%%%%%%%%%%%%%%%%%%%%%%%%%%%%%%%%%%%%%%%%%%
%%%%%%%%%%%%%%%%%%%%%%%%%%%%%%%%%%%%%%%%%%%%%%%%%%%
%%%%%%%%%%%%%%%%%%%%%%%%%%%%%%%%%%%%%%%%%%%%%%%%%%%
\newpage
\section{Partial-wave bounds and self-energy kernel from unitarity \label{sec:Unitarity}}

Setting $\SS = \ID + \im \TT$, as is standard, the unitarity of the $\SS$ matrix, $\SS\SS^\dagger = \SS^\dagger\SS = \mathbbm{1}$, implies
\begin{align}
\label{eq:Unitarity_Tmatrix}
-\im (\TT - \TT^\dagger)  
= \TT^\dagger\TT
= \TT\TT^\dagger.   
\end{align}
We shall use this relation in two ways: first, to reproduce the known constraints on partial-wave amplitudes and cross-sections for elastic and inelastic processes; and second, to predict the contribution to the self-energy kernel generated by inelastic processes~\cite{Flores:2024sfy}. For the latter purpose, the $\TT$-matrix will be evaluated at general kinematics that do not necessarily satisfy the on-shell conditions. While asymptotic states obey on-shell conditions, the operator relation \eqref{eq:Unitarity_Tmatrix} implies constraints on the analytic continuation of amplitudes or kernels; in particular it determines their anti-Hermitian part through sums over on-shell intermediate states. Our analysis in this section largely follows Ref.~\cite{Flores:2024sfy}, but we provide additional details and clarify the conjugation properties of the interaction kernel.

\subsection{General analysis  \label{sec:Unitarity_General}}

In what follows, we denote by ${\cal M}^{ab}(\tau^a, \tau^b)$ the amplitude for the transition from state $a$ to state $b$, where $\tau$ collectively represents the particle momenta of each state in the center-of-momentum (CM) frame.  
For two-particle states, only one independent momentum exists, and we therefore write $\tau \to \vb{p}$.  
For simplicity, we neglect spin throughout.

\subsubsection*{Partial-wave expansion for 2-particle states}

For transitions between the 2-particle states $a$ and $b$, with momenta $\vb{p}^a$ and $\vb{p}^b$ in the CM frame, we analyze the amplitudes as follows
\begin{subequations}
\label{eq:Mcal2to2_PWanalysis}
\label[pluralequation]{eqs:McalElas_PWanalysis}
\begin{align}
\label{eq:Mcal2to2_PW}
{\cal M}^{ab} ({\bf p}^a,{\bf p}^b) 
&= 16\pi \sum_\ell (2\ell+1)
P_\ell (\hat{\bf p}^a\cdot\hat{\bf p}^b)     
{\cal M}_\ell^{ab} (p^a,p^b) ,
\\
{\cal M}^{ab}_{\ell}(p^a,p^b) 
&=
\dfrac{2}{(8\pi)^3} 
\int \dd\Omega_a \ \dd\Omega_b
P_{\ell}  (\hat{\vb{p}}^a\cdot\hat{\vb{p}}^b)
{\cal M}^{ab} (\vb{p}^a,\vb{p}^b) .   
\label{eq:Mcal2to2_PW_inv_Legendre}
\end{align}
\end{subequations}
where $p = |\vb{p}|$ denotes the magnitude of the corresponding 3-momentum.\footnote{
%%%%%%%%%%%%%%%%
We assume the orthonormality condition $\int d\Omega \, Y_{\ell m}(\Omega) \, Y_{\ell' m'}^*(\Omega) = \delta_{\ell\ell'}\delta_{mm'}$ for the spherical harmonics, and recall the useful decomposition 
$P_\ell(\hat{\vb{a}}\cdot\hat{\vb{b}}) = [(4\pi)/(2\ell+1)]
\sum_{m=-\ell}^\ell 
Y_{\ell m} (\Omega_{\vb{a}}) Y_{\ell m}^*(\Omega_{\vb{b}})$ 
for the Legendre polynomials.}
%%%%%%%%%%%%%%%% 
%
Note that if the incoming and/or outgoing particles are off-shell, $p^a$ and $p^b$ are in general different, and cannot be related to the total energy of the system, parametrized by the first Mandelstam variable, $s$. We leave implicit the dependence of the off-shell amplitudes on the zeroth components of the CM momenta.

\subsubsection*{Unitarity}
Projecting \cref{eq:Unitarity_Tmatrix} on the 2-particle state $a$, for different outgoing and incoming momenta, $\vb{p}$ and $\vb{p'}$, respectively, and inserting a complete set of on-shell states between $\TT$ and $\TT^\dagger$ on the right-hand side, we obtain 
\begin{align}
-\frac{\im}{2} \left[
{\cal M}^{aa} (\vb{p}',\vb{p}) - 
{\cal M}^{aa *} (\vb{p},\vb{p}') \right] 
= \sum_{c:~\text{on-shell}} 
{\cal X}^{ac} (\vb{p}',\vb{p})  
= \sum_{c:~\text{on-shell}} 
\tilde{\cal X}^{ac} (\vb{p}',\vb{p})  ,  
\label{eq:UnitarityRelation}
\end{align}
with 
\begin{subequations}
\begin{align}
{\cal X}^{ac}(\vb{p}', \vb{p}) 
\equiv 
\dfrac{1}{2} \int \dd_\subsqrts \tau^c  
\, {\cal M}^{ac*}(\vb{p},\tau^c)
\, {\cal M}^{ac}(\vb{p}',\tau^c),
\label{eq:Xcal_def}
\\
\tilde{\cal X}^{ac}(\vb{p}',\vb{p}) 
\equiv
\dfrac{1}{2} \int \dd_\subsqrts \tau^c   
\, {\cal M}^{ca}(\tau^c,\vb{p}) 
\, {\cal M}^{ca*}(\tau^c,\vb{p}'), 
\label{eq:XcalTilde_def}
\end{align}
\end{subequations}
where $\tau^c$ denotes collectively all the momentum variables that characterize the \emph{on-shell} state $c$, consisting of $N_c$ particles, and $\dd_\subsqrts\tau^c$ stands for the corresponding phase-space element with the energy-momentum conservation included, 
\begin{align}
\dd_\subsqrts \tau^c &\equiv
\dfrac{1}{\sym^c} 
\prod_{j=1}^{N_c} 
\dfrac{\dd^3 \vb{k}_j}{(2\pi)^3 \, 2E_j}   
(2\pi) \delta 
\left(\sqrt{s} -\sum_j (k_j)^0 \right)
(2\pi)^3 \delta^3 
\left(\sum_j 
\textbf{k}_j
\right) .
\label{eq:PhaseSpaceMeasure_def}
\end{align}
We have included the subscript $\subsqrts$ in the differential, to emphasize that the result of the integration depends on the energy imparted; this will become important in \cref{sec:OpticalPotential_H}.
Here, $E_j = \sqrt{\vb{k}_j^2 + m_j^2}$ is the on-shell energy for the $j$th particle, and $\sym^c$ is the symmetry factor of the state $c$ that ensures the phase space is not multiply counted ($\sym=j_1! j_2! \cdots$ if a state contains $j_1$, $j_2$, etc identical particles of species 1,~2, etc). 
For 1-particle and 2-particle states, the phase-space elements are
\begin{subequations}
\label{eq:PhaseSpaceMeasure}
\label[pluralequation]{eqs:PhaseSpaceMeasure}
\begin{align}
\text{1-particle states}: 
&\quad
\dd_\subsqrts\tau^c = 2\pi \, \delta (s - m_c^2) ,   
\label{eq:PhaseSpaceMeasure_1particle}
\\
\text{2-particle states}: 
&\quad
\dd_\subsqrts\tau^{c} = 
\dfrac{k^c(s)}{\sym^c \,16 \pi^2\sqrt{s}} \, \dd\Omega_c.   
\label{eq:PhaseSpaceMeasure_2particle}
\end{align}
\end{subequations}
For inverse decay processes, $m_c$ is the mass of the intermediate particle, while for 2-particle intermediate states, the magnitude of the momentum, $k^c=k^c(s)$, is fully specified by $s$ and the masses of the particles involved. We emphasize that in ${\cal X}^{ac}$, defined \cref{eq:Xcal_def}, the $c$ states are always on-shell, with their phase space fully integrated.

Acting on \cref{eq:UnitarityRelation} with 
$(8\pi)^{-2} 
\int \dd\Omega_{\vb{p}'} \, \dd\Omega_{\vb{p}} 
Y_{\ell m}^*  (\Omega_{\vb{p}'}) 
Y_{\ell m}^{} (\Omega_{\vb{p}})$, 
and using \cref{eq:Mcal2to2_PWanalysis,eq:Xcal_def}, 
the unitarity relation becomes
\begin{align}
-\dfrac{\im}{2} 
\left[
 {\cal M}^{aa}_{\ell} (p',p)
-{\cal M}^{aa*}_{\ell} (p,p')
\right] 
&= \sum_{c:~\text{on-shell}} 
{\cal X}_{\ell}^{ac} (p',p) ,
\label{eq:UnitarityRelation_PW}
\end{align} 
where 
\begin{subequations}
\label{eq:Xcal_PWanalysis}
\begin{align}
{\cal X}_{\ell}^{ac} (p',p) 
&\equiv
\dfrac{2}{(8\pi)^3}
\int \dd\Omega_{\vb{p}'} \dd\Omega_{\vb{p}}
\, P_\ell  (\hat{\vb{p}}' \cdot \hat{\vb{p}})
\, {\cal X}^{ac}  (\vb{p}',\vb{p}), 
\label{eq:Xcal_PW}
\\
{\cal X}^{ac} (\vb{p}',\vb{p}) &= 
16\pi \sum_\ell (2\ell+1)
\, P_\ell (\hat{\vb{p}}'\cdot\hat{\vb{p}})
\, {\cal X}^{ac}_\ell (p',p).
\label{eq:Xcal_PW_inv}
\end{align}
\end{subequations}
For 1-particle and 2-particle intermediate states, \cref{eq:Xcal_def,eq:PhaseSpaceMeasure,eq:Xcal_PWanalysis} yield
\begin{align}
\label{eq:Xcal_1Part_2Part}
{\cal X}_{\ell}^{ac} (p',p)
&= \left\{
\begin{array}{ll}
\dfrac{\delta_{\ell,0}}{16}
\, \delta(s-m_c^2)
\, {\cal M}^{ac \, *}(p,0)\, {\cal M}^{ac}(p',0),
& \text{1-particle}, 
\\[2ex]
\dfrac{2k^c}{\sym^c \sqrt{s}}
{\cal M}_\ell^{ac\,*} (p, k^c)
{\cal M}_\ell^{ac} (p', k^c) ,     
& \text{2-particle}.
\end{array}
\right.
\end{align}
Note that in the absence of spin or other internal degrees of freedom, the inverse-decay amplitudes cannot depend on the orientation of the three-momenta $\vb{p}$ and $\vb{p'}$ by rotational invariance, so only the $\ell = 0$ mode is present. Higher-$\ell$ modes can appear for resonances with spin or internal angular momentum (e.g. for composite states).

\subsubsection*{Rescaling of amplitudes}

For the 2-particle states $a$ and $b$, it is convenient to define the rescaled partial-wave amplitudes,
\begin{align}
M^{ab}_{\ell} (p^a,p^b) 
&\equiv 
\sqrt{\dfrac{4 p^a p^b}{\sym^a \sym^b s}}
\, {\cal M}^{ab}_{\ell} (p^a, p^b) .
\label{eq:Amplitudes_Rescaled_def}
\end{align}
For a 2-particle state $a$ and an on-shell state $c$ of any multiplicity, we also define
\begin{align}
X^{ac}_{\ell} (p',p)   
&\equiv
\sqrt{\frac{4 p' p}{(\sym^a)^2 s}}
\, {\cal X}_{\ell}^{ac} (p',p) 
\quad \overset{\text{c: 2-particle state}}{\longrightarrow} \quad
M^{ac*}_{\ell} (p, k^c) \, M^{ac}_{\ell} (p', k^c) ,
\label{eq:X_PW}
\end{align}
where the last expression was obtained using \cref{eq:Xcal_1Part_2Part,eq:Amplitudes_Rescaled_def}.

\subsubsection*{Cross-sections}

To compute cross-sections, we evaluate the amplitudes on-shell. 
The cross-section for a 2-to-$N$ process $a\to c$, is
\begin{align}
\sigma^{a\to c} (k) 
= \dfrac{1}{4k\sqrt{s}}
\int \dd_\subsqrts\tau^c \, |{\cal M}^{ac} ({\bf k},\tau^c)|^2 
= \dfrac{1}{4k\sqrt{s}} 
\, 2{\cal X}^{ac} (\vb{k},\vb{k})
=  \dfrac{8\pi}{k\sqrt{s}} 
\sum_\ell (2\ell+1)
{\cal X}^{ac}_\ell (k,k),
\label{eq:CrossSection_def}
\end{align}
where $\vb{k}$ is the on-shell momentum in the CM frame of the state $a$, determined by $s$, and we used \cref{eq:Xcal_def,eq:Xcal_PW_inv}. As in Ref.~\cite{Flores:2024sfy}, we define the \emph{unitarity cross-section}
\begin{empheq}[box=\myshadebox]{align}
\sigma_\ell^U (k) = \sym_\ell 
\, \dfrac{4\pi (2\ell+1)}{{\bf k}^2} ,   
\label{eq:sigmaU_def}
\end{empheq}
where $\sym_\ell$ is the symmetry factor of the incoming state.
Although the symmetry factor does not depend on the partial wave, we introduce the index $\ell$ to indicate that 
$\sym_\ell=1$ for distinguishable particles, 
$\sym_\ell=2$ for identical particles in allowed partial waves, while cross-sections for other partial waves vanish.  
With this, the partial-wave cross-sections, i.e. the individual terms in the sum of \cref{eq:CrossSection_def}, can be expressed as
\begin{align}
\sigma_\ell^{a\to c} (k) 
= \sigma_\ell^U (k)
\qty(\dfrac{2k}{\sym_\ell\sqrt{s}} {\cal X}_\ell^{ac} (k,k))
= \sigma_\ell^U (k) \, X_\ell^{ac} (k,k)
\ \overset{\text{c: 2-particle state}}{\longrightarrow} \
\sigma_\ell^U (k) \, |M_\ell^{ac} (k,k^c)|^2,
\label{eq:sigma=sigmaU*X}
\end{align}
where the rescaled quantities, $M_\ell^{ac}$ and $X_\ell^{ac}$, have been defined in \cref{eq:Amplitudes_Rescaled_def,eq:X_PW}.

\subsection{Partial-wave optical theorem and unitarity bounds  \label{sec:Unitarity_Bounds}}

We now return to the partial-wave unitarity relation \eqref{eq:UnitarityRelation_PW}, which we rescale using \cref{eq:Amplitudes_Rescaled_def,eq:X_PW}. Taking the incoming and outgoing particles to be on-shell, $p=p' \to k$, and setting $M_\ell^\elas (k) \equiv M_\ell^{aa}(k,k)$, we obtain
\begin{empheq}[box=\myshadebox]{align}
{\rm Im} [M_\ell^{\elas} (k)] 
= \sum_c X_\ell^{ac} (k,k) 
= \sigma_\ell^{\tot} (k) / \sigma_\ell^U (k),
\label{eq:OpticalTheorem_PW}
\end{empheq}
where the total partial-wave cross-section $\sigma_\ell^{\tot}$ includes all elastic and inelastic processes, and we used \cref{eq:sigma=sigmaU*X}. 
This is the partial-wave optical theorem. \Cref{eq:OpticalTheorem_PW} also implies that the rescaled elastic scattering amplitude obeys the inequality
\begin{align}
{\rm Im} [M_\ell^{\elas} (k)] \geqslant |M_\ell^{\elas} (k)|^2  ,
\end{align}
or, equivalently, it is bounded by the unitarity circle centered on $\im/2$ on the complex plane, 
\begin{empheq}[box=\myshadebox]{align}
|M_\ell^\elas (k) -\im/2| \leqslant 1/2 ,
\label{eq:UnitarityCircle_Elas}
\end{empheq}
with the equality holding in the absence of inelastic processes. For inelastic scatterings, the rescaled amplitudes  lie within a circle of radius $1/2$ centered on zero on the complex plane,
\begin{empheq}[box=\myshadebox]{align}
\sum_{c\neq a} X_\ell^{ac} 
= \Im(M_\ell^{aa}) - |M_\ell^{aa}|^2 
\leqslant  \Im(M_\ell^{aa})- [\Im(M_\ell^{aa})]^2 
\leqslant 1/4 .
\label{eq:UnitarityCircle_Inelas}
\end{empheq}
We can deduce a stronger, coupled constraint between elastic and inelastic processes, considering that
\begin{align}
\left(\sigma_\ell^\elas + \sigma_{\ell}^\inel\right) / \sigma_\ell^U 
=\sum_{c} X_\ell^{ac}
={\rm Im} M_{\ell}^{aa}
\leqslant |M_\ell^{aa}| 
=\left( \sigma_\ell^\elas / \sigma_\ell^U \right)^{1/2},
\end{align}
or equivalently
\begin{empheq}[box=\myshadebox]{align}
\sigma_\ell^\inel / \sigma_\ell^U \leqslant
\sqrt{\sigma_\ell^\elas / \sigma_\ell^U}
\left(1-\sqrt{\sigma_\ell^\elas / \sigma_\ell^U}\right) .
\label{eq:UnitarityLimit_Coupled}
\end{empheq}
The unitarity constraint \eqref{eq:UnitarityLimit_Coupled} implies the global upper bounds\footnote{
%%%%%%%%%%%%%%%%%%%% 
If the incoming particles carry spins, the orbital angular momentum in the unitarity constraints~\eqref{eq:UnitarityLimit_Coupled} and~\eqref{eq:UnitarityLimits_Global} should be replaced by the total angular momentum, $\ell \to j$. 
We may then attempt to derive constraints on the cross-sections for a given $\ell$ by averaging over all possible spin configurations. Considering that 
\begin{align*}
\dfrac{1}{(2s_1+1)(2s_2+1)}
\sum_{s=|s_1-s_2|}^{s_1+s_2} 
\sum_{j=|\ell-s|}^{\ell+s}(2j+1) = 2\ell+1,
\end{align*}
with $s_1$ and $s_2$ being the spins of the two interacting particles, the global constraints \eqref{eq:UnitarityLimits_Global} remain valid for the spin-averaged cross-sections for a given orbital angular momentum, $\ell$. This implies that the upper bounds on the mass of thermal-relic dark matter for a given $\ell$ remain as in Ref.~\cite[Section~8]{Flores:2024sfy}.
}
%%%%%%%%%%%%%%%%%%%% 
%
\begin{align}
\sigma_\ell^\elas / \sigma_\ell^U \leqslant 1
\quad \text{and} \quad
\sigma_\ell^\inel / \sigma_\ell^U \leqslant 1/4 .   
\label{eq:UnitarityLimits_Global}
\end{align}
Notably, the constraint \eqref{eq:UnitarityLimit_Coupled} implies also a \emph{lower} bound on $\sigma_\ell^{\elas}$ that depends on $\sigma_\ell^{\inel}$, attesting to the fact that inelastic interactions generate elastic scattering.

\subsection{Anti-Hermitian instantaneous self-energy kernel  \label{sec:Unitarity_Kernel}}

To properly compute the self energy of a state, the irreducible kernel must be first identified. For a 2-particle state, this consists of all 4-point diagrams that cannot be separated into two 4-point sub-diagrams (contributing to the same 4-point function) by cutting two internal propagators, and are thus 2-particle irreducible (2PI). 
Moreover, in the instantaneous approximation, any dependence of the kernel on the zeroth components of the incoming and outgoing CM momenta is neglected.

We may analyze the kernel in Hermitian and anti-Hermitian contributions, 
\begin{subequations}
\label{eq:K_decomp}
\label[pluralequation]{eqs:K_decomp}
\begin{align}
{\cal K}^{aa} (\vb{p}',\vb{p}) &=  
{\cal K}_H^{aa} (\vb{p}',\vb{p}) +  
{\cal K}_A^{aa} (\vb{p}',\vb{p}) ,
\\
{\cal K}_H^{aa} (\vb{p}',\vb{p}) &\equiv 
\dfrac{1}{2} \Big[ 
{\cal K}^{aa} (\vb{p}',\vb{p}) +
{\cal K}^{aa \, \dagger} (\vb{p}',\vb{p}) 
\Big],
\\
{\cal K}_A^{aa} (\vb{p}',\vb{p}) &\equiv 
\dfrac{1}{2} \Big[ 
{\cal K}^{aa} (\vb{p}',\vb{p}) -
{\cal K}^{aa \, \dagger} (\vb{p}',\vb{p}) 
\Big],
\end{align}
\end{subequations}
where 
\begin{align}
{\cal K}^{aa \, \dagger} (\vb{p}',\vb{p}) = 
{\cal K}^{aa \, *} (\vb{p},\vb{p}') .
\label{eq:Kdagger}
\end{align}
Evidently, 
${\cal K}_H^{aa \, \dagger} (\vb{p}',\vb{p})=
 {\cal K}_H^{aa} (\vb{p}',\vb{p})$
and
${\cal K}_A^{aa \, \dagger} (\vb{p}',\vb{p})=
-{\cal K}_A^{aa} (\vb{p}',\vb{p})$.
We emphasize that the Hermitian and anti-Hermitian components of the kernel are \emph{not} in general real and imaginary, respectively. They become so only under time-reversal invariance (TRI), which sets 
${\cal K} (\vb{p'},\vb{p}) 
\overset{\TRI}{=} 
{\cal K} (-\vb{p},-\vb{p'}) =
{\cal K} (\vb{p},\vb{p'})$, 
where the last equality holds for spinless particles due to rotational invariance.
We shall not assume TRI in this work.

The unitarity relations \eqref{eq:UnitarityRelation} and \eqref{eq:UnitarityRelation_PW} imply that inelastic processes participate in the self energy of a state. Their anti-Hermitian contribution can be deduced directly from the unitarity relation underpinning the optical theorem~\cite{Flores:2024sfy}. In particular,  \cref{eq:UnitarityRelation} implies that, under the instantaneous approximation, inelastic processes generate contributions to the self-energy kernel that satisfy the equation
\begin{subequations}
\label{eq:UnitarityRelation_Kernel}
\label[pluralequation]{eqs:UnitarityRelation_Kernel}
\begin{empheq}[box=\myshadebox]{align}
{\cal K}_A(\vb{p}',\vb{p})
&=
\dfrac{1}{2} \left[
{\cal K}^{aa} (\vb{p}',\vb{p}) - 
{\cal K}^{aa \, *} (\vb{p},\vb{p}') \right] 
\nn \\
&= 
\dfrac{\im}{2}
\sum_{c:~\text{on-shell}} 
\int \dd_\subsqrts\tau^c 
\, {\cal A}^{ac \,*} (\vb{p},\tau^c)  
\, {\cal A}^{ac} (\vb{p}',\tau^c) 
\label{eq:UnitarityRelation_Kernel_ac}
\\
&= 
\dfrac{\im}{2}
\sum_{c:~\text{on-shell}} 
\int \dd_\subsqrts\tau^c  
\, {\cal A}^{ca} (\tau^c,\vb{p})  
\, {\cal A}^{ca \,*} (\tau^c,\vb{p}'),  
\label{eq:UnitarityRelation_Kernel_ca}
\end{empheq}
\end{subequations}
where 
${\cal A}^{ac} (\vb{p},\tau^c)$ are the inelastic amplitudes, \emph{with all initial-state (elastic) 2PI factors amputated}.\footnote{ 
%%%%%%%%%%%%%%%%%
${\cal A}^{\inel}$ is often termed the `hard-scattering' amplitude, since it typically involves large momentum transfers, in contrast to the low-momentum (soft) interactions usually resummed in the wavefunction. However, some inelastic processes, such as bound-state formation with emission of an ultrasoft boson, involve only small momentum transfers. We thus avoid this terminology here, and  will instead refer to ${\cal A}^{\inel}$ as `inelastic vertex' or `irreducible inelastic amplitude'.} 
%%%%%%%%%%%%%%%%% 
The incoming and outgoing states on the left-hand side of \cref{eqs:UnitarityRelation_Kernel} are in general off-shell. On the other hand, the products of the inelastic interactions (i.e.~the intermediate states on the right-hand sides of \cref{eqs:UnitarityRelation_Kernel}) are on-shell, as mandated by the insertion of a complete set of physical states in the operator relation \eqref{eq:Unitarity_Tmatrix}. The momenta of the intermediate states, collectively denoted by $\tau^c$, are thus constrained by the energy imparted in the system and the on-shell dispersion relations. In terms of partial waves, \cref{eq:UnitarityRelation_Kernel_ac} becomes
\begin{subequations}
\label{eq:ImK=AA_PW}
\label[pluralequation]{eqs:ImK=AA_PW}
\begin{empheq}[box=\myshadebox]{align}
{\cal K}_{A, \ell}^{aa} (p',p) 
&=
\dfrac{\im}{(8\pi)^3}
\sum_{c:~\text{on-shell}} \int \dd_\subsqrts\tau^c
\int \dd\Omega_{\vb{p}'} \, \dd\Omega_{\vb{p}} 
\, P_{\ell} (\hat{\vb{p}}'\cdot\hat{\vb{p}}) 
\, {\cal A}^{ac\,*} (\vb{p},\tau^c)
\, {\cal A}^{ac} (\vb{p}',\tau^c)
\tag{\theequation}
\label{eq:ImK=AA_PW_MultiPart}
\\[1ex]
&\supset 
\sum_{r:~\text{1-particle}} 
\dfrac{\im \delta_{\ell,0}}{16} 
\ \delta(s-m_r^2) 
\ {\cal A}^{ar\,*} (p,0) 
\ {\cal A}^{ar} (p',0)
\label{eq:ImK=AA_PW_1part}
\\
&+
\sum_{b:~\text{2-particle}} 
\dfrac{\im 2k^b(s)}{\sym^b \sqrt{s}}
\ {\cal A}_\ell^{ab\,*} (p, k^b)
\ {\cal A}_\ell^{ab} (p', k^b) ,
\label{eq:ImK=AA_PW_2part}
\end{empheq}
\end{subequations}
where in the last two lines we included the phase-space integrated forms due to 1-particle and 2-particle intermediate states, following directly from \cref{eq:Xcal_1Part_2Part}.

%%%%%%%%%%%%%%%%%%%%%%%%%%%%%%%%%%%%%%%%%%%%%%%%%%%
%%%%%%%%%%%%%%%%%%%%%%%%%%%%%%%%%%%%%%%%%%%%%%%%%%%
%%%%%%%%%%%%%%%%%%%%%%%%%%%%%%%%%%%%%%%%%%%%%%%%%%%
\newpage
\section{Optical potential  \label{sec:OpticalPotential}}

In a Hermitian theory, the contribution of inelastic processes to the self-energy kernel can be obtained by isolating the state of interest and integrating out all other degrees of freedom, namely the products of the inelastic interactions, via the Feshbach projection~\cite{Feshbach:1958nx,Feshbach:1962nra}. We review the standard derivation of the optical potential in this framework, and then extend it to express the optical potential in terms of irreducible inelastic amplitudes. To our knowledge, this explicit representation has not previously appeared in the literature.

\subsection{Feshbach projection}
 
We consider the full Hermitian Hamiltonian of a system, $H_{\full}^{} = H_{\full}^\dagger$, acting on a Hilbert space ${\cal H}$ decomposed into orthogonal subspaces, with projection operators $P$ and $Q$,
\begin{align}
{\cal H} = P {\cal H} \oplus Q {\cal H},    
\qquad
P+Q=1, \quad P^2=P,\quad Q^2=Q,\quad PQ=0,
\label{eq:HilbertSpace_projections}
\end{align}
i.e.~a general state can be decomposed as
\begin{align}
|\Psi \rangle = |\Psi_P\rangle +  |\Psi_Q\rangle, 
\qquad \text{where} \quad
|\Psi_P\rangle \equiv P |\Psi\rangle, \quad  
|\Psi_Q\rangle \equiv Q |\Psi\rangle.    
\label{eq:Psi_projections}
\end{align}
Then, applying the two projection operators, $P$ and $Q$, on the Schrödinger equation,
\begin{subequations}
\begin{align}
H_{\full} \,|\Psi\rangle = E \,|\Psi\rangle ,   
\label{eq:SE}
\end{align}
we obtain the projected equations
\begin{align}
P H_{\full} P \,|\Psi_P\rangle + 
P H_{\full} Q \,|\Psi_Q\rangle 
&= E \,|\Psi_P\rangle, 
\label{eq:SE_P}
\\
Q H_{\full} P \,|\Psi_P\rangle + 
Q H_{\full} Q \,|\Psi_Q\rangle 
&= E \,|\Psi_Q\rangle. 
\label{eq:SE_Q}
\end{align}
\end{subequations}
We may solve \cref{eq:SE_Q} for $|\Psi_Q\rangle$,
\begin{align}
|\Psi_Q\rangle = 
\dfrac{1}{E - QH_{\full} Q + \im\epsilon} \,
Q H_{\full} P \,|\Psi_P\rangle ,
\label{eq:PsiQ_sol}
\end{align}
with the prescription $\epsilon \to 0^+$ ensuring an outgoing spherical wave. Substituting into \cref{eq:SE_P},
\begin{align}
\left(PH_{\full}P + V_\opt \right)
\,|\Psi_P\rangle = E \,|\Psi_P\rangle ,   
\label{eq:SE_P_Optical}    
\end{align}
where $V^\opt$ is the optical potential,
\begin{align}
V^\opt (E) \equiv 
PH_{\full}Q    
\dfrac{1}{E - QH_{\full}Q +\im \epsilon}
QH_{\full}P .
\label{eq:Vopt}
\end{align}

\subsubsection*{Hermitian and anti-Hermitian contributions}

As any operator, the optical potential can be decomposed into Hermitian and anti-Hermitian parts,
\begin{align}
V^\opt = V_H^\opt + V_A^\opt,
\quad \text{where} \quad
V_H^\opt \equiv 
\dfrac{1}{2} (V^\opt+V^{\opt \, \dagger}), 
\quad
V_A^\opt \equiv 
\dfrac{1}{2} (V^\opt-V^{\opt \, \dagger}).
\label{eq:Vopt_decomp}
\end{align}
Considering \cref{eq:Vopt}, and using
\begin{align}
\dfrac{1}{x+\im 0^+}=
\PV \left(\dfrac{1}{x}\right)
-\im\pi\delta(x),    
\label{eq:PrincipalValue+iDelta}
\end{align}
where $\PV$ denotes the Principal Value, we find
\begin{subequations}
\begin{align}
V_H^\opt (E) &= 
PH_{\full}Q 
\ \PV 
\left( \dfrac{1}{E-QH_{\full}Q} \right) 
QH_{\full}P ,
\label{eq:Vopt_H}
\\
V_A^\opt (E) &= -\im \pi
\, PH_{\full}Q 
\,\delta \Big(E - QH_{\full}Q \Big) 
\, QH_{\full}P .
\label{eq:Vopt_A}
\end{align}
\end{subequations}
$V_H^\opt$ and $V_A^\opt$ are the dispersive and absorptive parts of the optical potential, respectively. For any state in $P$ space,  $\im V_A^{\opt}$ is positive semi-definite, $\langle P\Psi |\im V_A^{\opt} |P\Psi\rangle \geqslant 0$. Moreover, considering that
\begin{align} 
\dfrac{1}{E-E_0} =  
\int_{-\infty}^{\infty} d\hat{E}
\ \dfrac{\delta(\hat{E}-E_0)}{E-\hat{E}} ,
\end{align}
we find that $V_H^\opt$ is the Hilbert transform of $V_A^\opt$, 
\begin{align}
V_H^\opt (E)
&=
\dfrac{1}{\im \pi} 
\ \PV 
\int_{-\infty}^\infty d\hat{E}
\ \dfrac{V_A^\opt (\hat{E})}{\hat{E}-E} .
\label{eq:VoptH_HilbertTransform}
\end{align}
Note that the converse is not necessarily true: $V_A^\opt$ is not the negative Hilbert transform of $V_H^\opt$ unless $V^\opt(E)$ is holomorphic in $E$ and decreases as $E\to \infty$.

\subsection{Absorptive potential \label{sec:OpticalPotential_A}}

Moving beyond the standard Feshbach formalism, we shall now project $V_A^\opt$ of \cref{eq:Vopt_A} onto $P$-space states, and aim to express it in terms of $P$-state interaction amplitudes. 

Let $\chi^j$ be a complete set of $Q H_\full Q$ eigenstates spanning the $Q$ subspace
\begin{align}
(Q H_\full Q) \,\chi^j = E^j \, \chi^j, 
\qquad 
Q\chi^j = \chi^j, 
\label{eq:FeshbachResonances}
\end{align}
where $j$ collects the discrete indices of the spectrum, but the eigenfunctions and eigenvalues may also depend on continuous variables, such as the independent momenta characterizing scattering states. The $Q$ operator can be expanded as
\begin{align}
Q = \sum_j \dfrac{1}{\sym^j} 
\int \dfrac{\dd^3 \vb{k}}{(2\pi)^3} \,
| \chi^j (\vb{k}) \rangle_{\NR} \ 
{}_{\NR}\langle \chi^j (\vb{k})|,
\label{eq:Qspace_Expansion}
\end{align}
where we assumed for simplicity that $|\chi^j (\vb{k}) \rangle_{\NR}$ are 2-particle states, with $\pm{\vb k}$ being the momenta of the constituent particles in the CM frame; for $Q$ eigenstates of different particle multiplicity, the phase-space integration can be adapted accordingly. As previously, $\sym^j$ is the symmetry factor of the state $j$.
The index $\NR$ denotes the normalization of states in  non-relativistic quantum mechanics, 
${}_{\NR}\langle \chi^i (\vb{k}) | \chi^j(\vb{\tilde k}) \rangle_{\NR}
= \sym^j (2\pi)^3 \delta^3 (\vb{k}-\vb{\tilde k}) \, \delta^{ij}$ (without implying that $| \chi^j(\vb{\tilde k}) \rangle_{\NR}$ is non-relativistic).
Inserting the above into \cref{eq:Vopt_A}, and setting ${\cal T} \equiv QH_\full P$, we find 
\begin{align}
&2\im V_A^\opt (E) =
\\
&= 
\sum_{i,j}
\dfrac{1}{\sym^i \sym^j}
\int
\dfrac{\dd^3 \vb{k}}{(2\pi)^3}
\dfrac{\dd^3 \vb{\tilde k}}{(2\pi)^3} 
\ {\cal T}^\dagger 
\, | \chi^i (\vb{k})\rangle_{\NR} 
\ {}_{\NR}\langle \chi^i(\vb{k})|    
\, (2\pi)\delta (E - QH_{\full}Q) 
\, | \chi^j (\vb{\tilde k}) \rangle_{\NR} 
\ {}_{\NR}\langle \chi^j (\vb{\tilde k})|    
\, {\cal T}
\nn \\
&= 
\sum_{j} 
\dfrac{1}{\sym^j}
\int \dfrac{\dd^3 \vb{k}}{(2\pi)^3} \, 
\ (2\pi)\delta (E - E^j(\vb{k}))
\ {\cal T}^\dagger \,  
| \chi^j (\vb{k})\rangle_{\NR} \
{}_{\NR}\langle \chi^j (\vb{k})|    
\ {\cal T} 
\nn \\
&= 
\sum_{j} 
\dfrac{1}{\sym^j}
\!\!\int\! \dfrac{\dd^3 \vb{k}_1}{(2\pi)^3}
\!\!\int\! \dfrac{\dd^3 \vb{k}_2}{(2\pi)^3}  
(2\pi) \delta [E - E^j(\vb{k}_1,\vb{k}_2)]
(2\pi)^3 \delta^3 (\vb{k}_1 + \vb{k}_2)
{\cal T}^\dagger 
| \chi^j (\vb{k}_1,\vb{k}_2)\rangle_{\NR} \ 
{}_{\NR}\langle \chi^j (\vb{k}_1,\vb{k}_2)|    
{\cal T} ,
\nn
\end{align}
where in the third step, setting $\vb{k} \to \vb{k}_1$, we re-expressed the phase-space integration over intermediate states in the more standard form   
$\int \dd^3\vb{k}/(2\pi)^3 =
\int \dd^3\vb{k}_1/(2\pi)^3
\int \dd^3\vb{k}_2/(2\pi)^3 
\ (2\pi)^3 \delta^3 (\vb{k}_1+\vb{k}_2)$.

Next, we project on incoming and outgoing $P$ states, with CM momenta $\vb{p'}$ and $\vb{p}$ respectively. Taking into account that, 
\begin{subequations}
\label{eq:Amplitudes_NonRelToRel}
\label[pluralequation]{eqs:Amplitudes_NonRelToRel}
\begin{align}
{}_{\NR}\langle \chi^j (\vb{k})| 
\, {\cal T} \, 
| \vb{p} \rangle_{\NR} 
&= 
\dfrac{{\cal A}^{\inel,j} (\vb{p},\vb{k})}
{\sqrt{ 
2E_1(\vb{p}) \, 2E_2(-\vb{p})\, 
2E_1^j(\vb{k}) \, 2E_2^j(-\vb{k}) 
}} ,
\\
{}_{\NR}\langle \vb{p}| 
\, {\cal T}^\dagger \, 
|  \chi^j (\vb{k}) \rangle_{\NR} 
&= 
{}_{\NR}\langle \chi^j (\vb{k})| \, {\cal T} \, | \vb{p} \rangle_{\NR}^* ,
\end{align}   
\end{subequations}
where $E_{1,2}$ and $E_{1,2}^j$ are the energies of the particles in the $P$ and $Q$ states, respectively, 
we obtain
\begin{align}
&{}_{\NR}\langle \vb{p}| V_A^\opt (E) | \vb{p'} \rangle_{\NR}
= -\dfrac{\im}{2} 
\sum_j 
\dfrac{1}{\sym^j}
\int \dfrac{\dd^3 \vb{k}_1}{(2\pi)^3 2E_1^j(\vb{k}_1)}
\int \dfrac{\dd^3 \vb{k}_2}{(2\pi)^3 2E_2^j(\vb{k}_2)} 
\nn
\\
&\times 
(2\pi) \delta [E - E^j(\vb{k}_1,\vb{k}_2)]
\ (2\pi)^3 \delta^3 (\vb{k}_1 + \vb{k}_2)
\ \dfrac{
{\cal A}^{\inel,j\,*} (\vb{p},\{\vb{k}_1,\vb{k}_2\})
{\cal A}^{\inel,j}   (\vb{p'},\{\vb{k}_1,\vb{k}_2\})}
{\sqrt{2E_1(\vb{p'})2E_2(-\vb{p'}) \, 2E_1(\vb{p})2E_2(-\vb{p})}}.
\end{align}
For non-relativistic $P$ states, 
$2E_1 (\vb{p}) 2E_2 (-\vb{p}) = 
2E_1 (\vb{p'}) 2E_2 (-\vb{p'}) \simeq 4\mT\mu$. 
Recalling the definition of the phase-space measure \eqref{eq:PhaseSpaceMeasure_def}, we arrive at
\begin{align}
{}_{\NR}\langle \vb{p}| V_A^\opt (E) | \vb{p'} \rangle_{\NR}
&\simeq -
\dfrac{1}{4\mT \mu}
\left[\dfrac{\im}{2} 
\sum_j
\int \dd_{_E} \tau^j \
{\cal A}^{\inel,j \, *} (\vb{p},\tau^j) \,
{\cal A}^{\inel,j} (\vb{p'},\tau^j) 
\right].
\label{eq:VoptA_Projected}
\end{align}
The factor in the square brackets reproduces the anti-Hermitian kernel as obtained from the unitarity relation, \cref{eq:UnitarityRelation_Kernel_ac}. Together with the non-relativistic pre-factor $-(4\mT\mu)^{-1}$  [cf.~\cref{eq:SchrodingerEq_momentum}], it gives the momentum-space anti-Hermitian potential.

\subsection{Dispersive potential \label{sec:OpticalPotential_H}}

The Hermitian part of the optical potential, \cref{eq:Vopt_H}, projected onto $P$ states, can be expressed in terms of the irreducible inelastic amplitudes using \cref{eq:VoptA_Projected,eq:VoptH_HilbertTransform}.

If ${\cal A}^{\inel,j}$ can be approximated by an expansion of the following form, 
\begin{align}
{\cal A}^{\inel,j} (\vb{p},\tau^j) = 
\sum_{\ell,m} 
Y_{\ell m} (\Omega_{\vb{p}})
\, a_{\ell m}^j (p/\mu) 
\, b_{\ell m}^j (\tau^j) ,
\label{eq:Ainel_ExpandAndFactor}
\end{align}
i.e.~for every angular mode, the dependence on $p$ and on the final-state phase space, $\tau^j$, is encapsulated in two separate factors, then it follows from \cref{eq:VoptA_Projected,eq:VoptH_HilbertTransform} that the dispersive and absorptive parts of the optical potential have the same dependence on the incoming and outgoing momenta of the $P$ states, albeit with different energy-dependent strengths. Indeed, using \cref{eq:Ainel_ExpandAndFactor} in \cref{eq:VoptA_Projected,eq:VoptH_HilbertTransform}, we find
\begin{align}
{}_{\NR}\langle \vb{p}| V^\opt (E) | \vb{p'} \rangle_{\NR} 
&\simeq 
-\dfrac{1}{4\mT\mu} \ (8\pi)^2 
\sum_j
\sum_{\ell,m}
\sum_{\ell',m'}
Y_{\ell m}^* (\Omega_{\vb{p}})
Y_{\ell' m'} (\Omega_{\vb{p'}})
\nn \\
&\times
[\Gamma_{H,\ell m \ell' m'}^j (E) + 
\im \Gamma_{A,\ell m \ell' m'}^j (E)]
\, a_{\ell m}^{j*} (p/\mu)
\, a_{\ell' m'}^j (p'/\mu), 
\label{eq:OpticalPot_Expansion}
\end{align}
where 
\begin{subequations}
\begin{align}
\Gamma_{A,\ell m \ell' m'}^j (E)
&\equiv  
\dfrac{1}{2}
\dfrac{1}{(8\pi)^2}
\int \dd_{_E}\tau^j
\, b_{\ell m}^{j\,*} (\tau^j)
\, b_{\ell' m'}^j (\tau^j)
= \delta_{\ell\ell'} \delta_{mm'}
\Gamma_{A,\ell}^j (E),
\label{eq:OpticalPot_AntiHermitianCoupling}
\\
\Gamma_{H,\ell m \ell' m'}^j (E)
&\equiv  
\dfrac{1}{\pi}
\ \PV
\int_{-\infty}^{+\infty} \!\!
\dfrac{\dd \hat{E}}{\hat{E}-E}
\Gamma_{A,\ell m \ell' m'}^j (\hat{E})
=
\dfrac{1}{\pi}
\ \PV
\int_{-\infty}^{+\infty} \!\!
\dfrac{\dd \hat{E}}{\hat{E}-E}
\Gamma_{A,\ell}^j (\hat{E}) 
\ \delta_{\ell\ell'} \delta_{mm'},
\label{eq:OpticalPot_HermitianCoupling}
\end{align}
\end{subequations}
where the last equality in \cref{eq:OpticalPot_AntiHermitianCoupling} is mandated by the rotational invariance of \cref{eq:OpticalPot_Expansion}, which implies not only the orthogonality of $b_{\ell m} (\tau^j)$ with respect to the phase-space measure $d_E\tau^j$, but also that both $\Gamma_{A,\ell}^j (E)$  and 
$a_{\ell m}^j (p/\mu) \to a_{\ell}^j (p/\mu)$ 
are independent of $m=m'$. 
Note that \cref{eq:OpticalPot_AntiHermitianCoupling} implies  $\Gamma_{A,\ell}^j (E) \geqslant 0$, while no such conclusion follows for $\Gamma_{H,\ell}^j (E)$.
The partial-wave contributions to the optical potential are then (cf.~\cref{eq:Mcal2to2_PW_inv_Legendre} for the convention for partial-wave analysis used throughout the paper)
\begin{multline}
\dfrac{1}{(8\pi)^2}
\int 
\dd\Omega_{\vb{p}} \,
\dd\Omega_{\vb{p'}}
\sum_{m=-\ell}^\ell
\dfrac{1}{2\ell+1}
\ Y_{\ell m}^{} (\Omega_{\vb{p}})
\ Y_{\ell m}^* (\Omega_{\vb{p'}})
\ {}_{\NR}\langle \vb{p}| V^\opt (E) | \vb{p'} \rangle_{\NR}
= \\ =
-\dfrac{1}{4\mT\mu} 
\sum_j
[\Gamma_{H,\ell}^j (E) + \im \Gamma_{A,\ell}^j (E)]
\ a_{\ell}^{j*} (p/\mu)
\ a_{\ell}^j (p'/\mu) .
\label{eq:OpticalPot_PW_separable}
\end{multline}

The factorized form \eqref{eq:Ainel_ExpandAndFactor} is naturally realized for contact-type interactions, where the irreducible inelastic vertex ${\cal A}^{\inel,j}$ is dominated by momentum transfers of order a hard scale $\Lambda$  (e.g. the mass scale of the interacting particles), and contains no light or massless particle exchange, nor other infrared singularities. For each partial wave, rotational invariance fixes the angular dependence to be $Y_{\ell m}(\Omega_{\vb p})$, with the corresponding radial dependence starting at order $p^\ell$ and higher-derivative corrections suppressed by powers of $p/\Lambda$. Equation~\eqref{eq:Ainel_ExpandAndFactor} further assumes that, for each angular mode, the dependence on the final-state kinematics factorizes into an overall function $b_{\ell m}^j(\tau^j)$; this is ensured if the expansion is truncated to the leading $p$ dependence.

%%%%%%%%%%%%%%%%%%%%%%%%%%%%%%%%%%%%%%%%%%%%%%%%%%%
%%%%%%%%%%%%%%%%%%%%%%%%%%%%%%%%%%%%%%%%%%%%%%%%%%%
%%%%%%%%%%%%%%%%%%%%%%%%%%%%%%%%%%%%%%%%%%%%%%%%%%%
\clearpage
\section{Analytic properties of radial wave functions for real potentials \label{sec:Jost}}

In this section we examine the analytic properties of solutions to the radial Schr\"odinger equation with a Hermitian potential that we shall assume to be central, therefore also real, with the ultimate goal of understanding the Green's function necessary for our unitarization prescription.  The following discussion only examines the analytic properties necessary for our analysis. For a more detailed and complete look at the analytic properties of radial wave functions, we refer to~\cite{Newton:1960nws, Chadan:1977pq, Newton:1982qc, Rakityansky:2022jost}.

\subsection{Radial Schrödinger equation and solutions \label{sec:Jost_SEandSolutions}}

We begin with the radial Schrödinger equation with a real central potential, $V(r) \in \RR$,
\begin{equation}\label{eq:RadialSE}
\qty[ \dv[2]{r} - \dfrac{\ell(\ell + 1)}{r^2} + k^2 ] 
u_{k,\ell}(r) =
2\mu V(r) u_{k,\ell}(r) .
\end{equation}
with $u_{k,\ell}(r) \equiv kr\psi_{k,\ell}(r)$. In the following, we investigate the solutions of \cref{eq:RadialSE} for $k\in \CC$. Physically meaningful scattering-state solutions correspond to $k\geqslant0$, while, as we shall see, bound-state solutions may arise for discrete values of purely imaginary $k$.

In particular, we are interested in studying potentials which satisfy
\begin{align}
\int_0^{\infty} \dd r\ r\ |V(r)| < \infty .
\label{eq:V_ConvergenceCondition}
\end{align}
While the Coulomb potential does not strictly meet this criterion, it shares many, if not all, of the analytic properties we will deduce. For a potential that satisfies the condition \eqref{eq:V_ConvergenceCondition}, hence growing more slowly than $1/r^2$ at $r\to 0$ and decreasing faster than $1/r^2$ at $r\to \infty$, \cref{eq:RadialSE} admits two independent solutions, scaling as $u(r) \propto r^{\ell+1}$ and $r^{-\ell}$ near the origin, which we now discuss.

\subsubsection{Regular solution}

We define the \textit{regular solution}, $\varphi_{k,\ell}(r)$, that vanishes at the origin, with its normalization determined by 
\begin{equation}
\label{eq:RegularSol_Asymptote0}
\lim_{r\to 0}
\qty
{ (2\ell + 1)!! 
\, r^{-(\ell + 1)} 
\, \varphi_{k,\ell}(r) } 
= 1.
\end{equation}
Crucially, the regular solution is a real function since the differential equation and the boundary conditions are real (meaning that for $k,r\in\RR$, it is real). Furthermore, $\varphi_{k,\ell}$ is a function of $k^2$ only, since $k$ enters into \cref{eq:RadialSE} via $k^2$ and the boundary condition \eqref{eq:RegularSol_Asymptote0} is $k$-independent~\cite{Chadan:1977pq, Newton:1982qc, Rakityansky:2022jost}. That is,
\begin{subequations}
\label{eq:RegularSol_Properties}
\label[pluralequation]{eqs:RegularSol_Properties}
\begin{align}
[\varphi_{k,\ell}(r)]^* &= \varphi_{k^*,\ell}(r),    
\label{eq:RegularSol_Real}
\\
\varphi_{-k,\ell}(r) &= \varphi_{k,\ell}(r).   
\label{eq:RegularSol_Reflection}
\end{align}
\end{subequations}

\subsubsection{Jost solution}

We define another solution to \cref{eq:RadialSE}, the \textit{Jost solution} $H_{k,\ell}(r)$, that satisfies
\begin{align}
\label{eq:JostSol_AsymptoteInf}
\lim_{r\to\infty}
\qty{
e^{-\ii\pi\ell/2}e^{-\ii kr}
H_{k, \ell}(r)
}
= 1.
\end{align}
It can be shown that the Jost solution is an analytic function of $k$ in the upper half complex plane~\cite{Newton:1982qc}. As such, for $\Im k > 0$ the boundary condition \eqref{eq:JostSol_AsymptoteInf} implies that
\begin{align}
H_{-k, \ell}(r) 
= 
(-1)^{\ell} [H_{k^*, \ell} (r)]^{*} 
\ \overset{k\in \RR}{\longrightarrow} \ 
(-1)^{\ell} H_{k, \ell}^* (r).
\label{eq:H(-k)}
\end{align}
From \cref{eq:JostSol_AsymptoteInf} we can compute the Wronskian\footnote{By Abel's identity, the Wronskian of two solutions to \cref{eq:RadialSE} is $r$-independent. Thus, it is sufficient to only examine the $r\to\infty$ limit.}
\begin{equation}
{\cal W}
\qty{
H_{k, \ell}(r), H_{-k, \ell}(r)
}
=
(-1)^{\ell + 1}2\ii k ,
\end{equation}
which demonstrates that for $k\neq 0$ the solutions $H_{\pm k,\ell}(r)$ are independent. 

\subsubsection{Jost function}

The independence of $H_{k,\ell}(r),H_{-k,\ell}(r)$ allows us to expand the regular solution as
\begin{align}
\varphi_{k,\ell}(r)
=
\dfrac{\ii}{2}
k^{-(\ell + 1)}
\qty[
\fs_{\ell}(k)
H_{-k, \ell}(r)
-
(-1)^\ell
\fs_\ell(-k)
H_{k, \ell}(r)
]
,
\label{eq:RegularSol_expansion}
\end{align}
where we took into account that $\varphi_{k,\ell}(r) = \varphi_{-k,\ell}(r)$. The function $\fs_{\ell}(k)$ is known as the \textit{Jost function}. This function can be directly defined through the Wronskian\footnote{The definition of the Jost function varies across the literature. We follow the convention of Ref.~\cite{Chadan:1977pq} where the zeros of the Jost function lie in the upper half complex plane.}
\begin{equation}\label{eq:JostFunctionDefn}
\fs_\ell(k) \equiv (-k)^\ell
{\cal W}\qty{H_{k,\ell}(r),\varphi_{k,\ell}(r)} .
\end{equation}
Considering \cref{eq:RegularSol_Properties,eq:H(-k),eq:JostFunctionDefn} the Jost function satisfies
\begin{align}
\label{eq:Jost_Analytic_Contin}
\fs_\ell(-k) 
\ \overset{\Im k \geqslant 0}{=}
[\fs_\ell(k^*)]^*
\ \overset{k\in \RR}{\longrightarrow} \ \fs_\ell^*(k).
\end{align}
The restriction in the first equality is due to the fact that the Jost solution, $H_{k,\ell}(r)$, is only analytic in the upper-half plane~\cite{Newton:1982qc}. For real momenta, we define the \emph{phase shift}, $\theta_\ell (k)$, via
\begin{align}
\fs_\ell(k) = 
|\fs_\ell(k)| \, e^{-\im \theta_\ell(k)} ,
\qquad k\in \RR.   
\label{eq:theta_def}
\end{align}
By virtue of \cref{eq:Jost_Analytic_Contin},
\begin{align}
\theta_\ell(-k) =  -\theta_\ell (k),  
\qquad k\in \RR,
\label{eq:theta_reflection}
\end{align}
up to integer multiples of 2$\pi$. From the above, we find the asymptotic behavior of the regular solutions at $r\to \infty$, 
\begin{subequations}
\label{eq:RegularSol_AsymptoteInf}
\label[pluralequation]{eqs:RegularSol_AsymptoteInf}
\begin{align}
\varphi_{k,\ell}(r)
&\overset{r\to\infty}{\longrightarrow}
\dfrac{\im}{2k^{\ell+1}} \qty[
 \fs_\ell( k) e^{-\im(kr-\pi\ell/2)}
-\fs_\ell(-k) e^{\im(kr-\pi\ell/2)}
]
\label{eq:RegularSol_AsymptoteInf_kComplex}
\\[1ex ]
&\overset{k\in\RR}{\longrightarrow}
\dfrac{|\fs_{\ell}(k)|}
{k^{\ell + 1}}
\sin[kr - \ell\pi/2 + \theta_\ell(k)].
\label{eq:RegularSol_AsymptoteInf_kReal}
\end{align}
\end{subequations}

\bigskip

We may similarly determine the asymptotic behavior of the Jost solution, $H_{k,\ell}(r)$ at $r\to0$. In this limit, we expand in the irregular and regular solutions, as follows
\begin{align}
H_{k,\ell}(r) \overset{r\to0}{\longrightarrow} 
\alpha_{k,\ell} r^{-\ell} +
\beta_{k,\ell} \, \varphi_{k,\ell} (r),
\label{eq:JostSolution_expansion}
\end{align}
where 
\begin{align}
\alpha_{k,\ell} 
= \lim_{r\to 0} 
\dfrac
{{\cal W} \qty{H_{k,\ell}(r),\varphi_{k,\ell}(r)}}
{{\cal W} \qty{r^{-\ell},\varphi_{k,\ell}(r)}}
= \dfrac{(2\ell-1)!! \, \fs_\ell(k)}{(-k)^\ell}.
\end{align}
Because this is non-zero, except at the zeros of the Jost function (cf.~\cref{sec:Jost_Poles}), we need not consider the contribution from the regular solution, which is subdominant in the $r\to0$ limit. Thus,
\begin{align}
H_{k,\ell} (r) \overset{r\to0}{\longrightarrow}
(2\ell-1)!! \, \fs_\ell(k) \ (-kr)^{-\ell} .
\label{eq:JostSol_Asymptote0}
\end{align}

\subsubsection{Orthonormality of regular and Jost solutions}

For any two solutions, $u_{k,\ell}(r)$ and $u_{k',\ell}(r)$,
Schrödinger's \cref{eq:RadialSE} implies
\begin{align}
\label{eq:Wronskians_Deriv}
\dv{r}\ {\cal W}
\qty{u_{k,\ell}(r), u_{k',\ell}(r) }
&=(k^2 - k'^2) \,
u_{k,\ell}(r) u_{k',\ell}(r) .
\end{align}
From this, and taking into account the asymptotic behavior of the regular and Jost solutions at the origin and at infinity, \cref{eq:RegularSol_Asymptote0,eq:RegularSol_AsymptoteInf,eq:JostSol_Asymptote0,eq:JostSol_AsymptoteInf}, we find
\begin{subequations}
\label{eq:Orthonorm_Regular+Jost}
\label[pluralequation]{eqs:Orthonorm_Regular+Jost}
\begin{align}
\int_0^\infty dr 
\,\varphi_{k, \ell} (r) 
\,\varphi_{k',\ell} (r)    
&= \dfrac{ 
\lim_{r\to\infty} {\cal W} \qty{
\varphi_{k, \ell} (r),
\varphi_{k',\ell} (r) }
}{k^2-k'^2}
\nn \\
&=
\dfrac{\im}{4(kk')^{\ell+1}}
\lim_{r\to\infty}
\Bigg[
\dfrac{
 \fs_\ell(k)\fs_\ell(-k') \, e^{-\im(k-k')r}
-\fs_\ell(-k)\fs_\ell(k') \, e^{+\im(k-k')r}
}{k-k'}
\nn\\
&+(-1)^\ell
\dfrac{
 \fs_\ell(-k)\fs_\ell(-k') \,e^{+\im(k+k')r}
-\fs_\ell( k)\fs_\ell (k') \,e^{-\im(k+k')r}
}{k+k'}
\Bigg],
\label{eq:Orthonorm_RegularSol}
\\[1em]
%%%%%%%%%%%%
\int_0^\infty dr 
\,H_{k, \ell}  (r) 
\,H_{k',\ell}^*(r)    
&= (-1)^\ell \ 
\dfrac{ 
\lim_{r\to\infty} {\cal W} 
\qty{ H_{k, \ell} (r), H_{-k',\ell} (r) }
}{k^2-k'^2}
= -\im \lim_{r\to\infty} \dfrac{e^{\im (k-k')r}}{k-k'} , 
\label{eq:Orthonorm_JostSol}
\\[1em]
%%%%%%%%%%%%
\int_0^\infty dr 
\,\varphi_{k, \ell}  (r) 
\,H_{k',\ell} (r)    
&= \dfrac{ 
\lim_{r\to\infty} {\cal W} 
\qty{\varphi_{k, \ell} (r), H_{k',\ell} (r)}
+(-k')^{-\ell} \fs_\ell(k')}
{k^2-k'^2}
\label{eq:Orthonorm_RegularJost}
\\
&=
-\dfrac{1}{2k^{\ell+1}}
\lim_{r\to\infty}
\qty[
(-1)^\ell\dfrac{\fs_\ell(k) e^{-\im(k-k')r}}{k-k'}
+\dfrac{\fs_\ell(-k) e^{\im [(k+k')r]}}{k+k'}
]
+\dfrac{\fs_\ell(k')}{(-k')^\ell(k^2-k'^2)}.
\nn 
\end{align}
\end{subequations}

\bigskip

\noindent For real momenta, $k,k'\in \RR$, considering the distributional limit\footnote{More precisely, the $r\to\infty$ limit on the half-line ($r\geqslant 0$), understood in the distributional sense, contains a principal-value contribution. In the combinations that enter the $\varphi$ (and hence ${\cal F}$ and ${\cal G}$) orthonormalities this PV part cancels, and for the remaining relations we retain only the $\delta$-parts since the PV terms are not used below.}
\begin{align}
\lim_{r\to+\infty} \dfrac{e^{\pm \im q r}}{q}
=\pm\im \pi \delta(q) ,
\qquad q\in \RR,
\label{eq:DeltaFunction_Limit}
\end{align}
and \cref{eq:Jost_Analytic_Contin}, we find
\begin{subequations}
\label{eq:Orthonorm_Regular+Jost_RealMomenta}
\label[pluralequation]{eqs:Orthonorm_Regular+Jost_RealMomenta}
\begin{align}
\int_0^\infty dr 
\,\varphi_{k, \ell} (r) 
\,\varphi_{k',\ell} (r)    
&=  
\dfrac{|\fs_\ell(k)|^2}{k^{2\ell+2}}
\dfrac{\pi}{2}[\delta(k-k') + \delta(k+k')] ,
\label{eq:Orthonorm_RegularSol_RealMom}
\\
\int_0^\infty dr 
\,H_{k, \ell}   (r) 
\,H_{k',\ell}^* (r)    
&= \pi \, \delta(k-k') ,
\label{eq:Orthonorm_JostSol_RealMom}
\\
\int_0^\infty dr 
\,\varphi_{k,\ell}(r) 
\,H_{k',\ell}(r)
&=-\dfrac{\im\pi}{2 (-k')^{\ell+1}} \left[
\fs_\ell(k)\,\delta(k-k')
+\fs_\ell(k')\,\delta(k+k')
\right]
+ 
\dfrac{\fs_\ell(k')}{(-k')^\ell}
\ \PV\dfrac{1}{k^2-k'^2}.
\label{eq:Orthonorm_RegularJostSol_RealMom}
\end{align}
\end{subequations}

\bigskip

For momenta at which the Jost function vanishes, which, as we shall see, occurs for bound states, \cref{eqs:Orthonorm_Regular+Jost} do not suffice to determine the normalization. The normalization of bound states is discussed in \cref{sec:Jost_Poles}.

\subsubsection{Physical, incoming-wave, outgoing-wave and irregular solutions}

We now define the solutions to \cref{eq:RadialSE},
\begin{subequations}
\label{eq:OtherSolutions_Def}
\label[pluralequation]{eqs:OtherSolutions_Def}
\begin{align}
\label{eq:PhysicalSolution}
{\cal F}_{k,\ell}(r) 
&\equiv 
\sqrt{\sym_\ell} \,
\dfrac{\ii^\ell k^{\ell + 1}}{\fs_\ell(k)}\varphi_{k,\ell}(r),
\\[1ex]
\label{eq:OutgoingWaveSolution}
{\cal H}_{k,\ell}^{(+)}(r)
&\equiv 
\sqrt{\sym_\ell} \,
(-\ii)^{\ell + 1}
\,\dfrac{\fs_\ell(-k)}{\fs_\ell(k)}
\, H_{k,\ell}(r),
\\[1ex]
\label{eq:IncomingWaveSolution}
{\cal H}_{k,\ell}^{(-)}(r)
&\equiv 
\sqrt{\sym_\ell} \,
\ii^{\ell + 1}H_{-k,\ell}(r),
\\[1ex]
\label{eq:IrregularSolution}
{\cal G}_{k,\ell}(r) 
&\equiv
\dfrac{1}{2\im} \left[ 
{\cal H}_{k,\ell}^{(+)} (r) - {\cal H}_{k,\ell}^{(-)} (r)
\right] .
\end{align}
\end{subequations}
Note that 
\begin{subequations}
\label{eq:Wavefunctions_Reflection}
\begin{align} 
{\cal H}^{(\pm)}_{k,\ell} (r) 
&={\cal F}_{k,\ell}  (r)
\pm \im {\cal G}_{k,\ell} (r),
\label{eq:H+-wrtFG}
\\ 
{\cal F}_{-k,\ell} (r) 
&=(-1)^{\ell+1} 
\dfrac{\fs_\ell(k)}{\fs_\ell(-k)}
{\cal F}_{k,\ell}(r),
\label{eq:PhysicalSol_Reflection}
\\
{\cal H}_{-k,\ell}^{(+)} (r) 
&= (-1)^{\ell+1}
\dfrac{\fs_\ell(k)}{\fs_\ell(-k)}
{\cal H}_{k,\ell}^{(-)} (r) ,
\label{eq:WavesSols_Reflection}
\\
{\cal G}_{-k,\ell} (r) 
&= (-1)^{\ell}
\dfrac{\fs_\ell(k)}{\fs_\ell(-k)}
{\cal G}_{k,\ell} (r) .
\label{eq:IrregularSol_Reflection}
\end{align}
\end{subequations}
Considering the asymptotic behaviors of $\varphi_{k,\ell}$ and $H_{k,\ell}$, given by \cref{eq:RegularSol_AsymptoteInf,eq:RegularSol_Asymptote0,eq:JostSol_AsymptoteInf,eq:JostSol_Asymptote0}, the asymptotic behaviors of the solutions \eqref{eq:OtherSolutions_Def} are the following: 
\begin{subequations}
\label{eq:WF_Asymptotes0}
\label[pluralequation]{eqs:WF_Asymptotes0}
\begin{align}
{\cal F}_{k,\ell}(r) 
~~\overset{r\to0}{\longrightarrow}~~ 
&
+\sqrt{\sym_\ell} \,
\im^\ell
\times
\dfrac{1}{\fs_\ell(k)}
\dfrac{(kr)^{\ell + 1}}{(2\ell + 1)!!},
\label{eq:WF_Asymptotes0_F}
\\
{\cal H}_{k,\ell}^{(+)} (r) 
~~\overset{r\to0}{\longrightarrow}~~ 
&
-\sqrt{\sym_\ell} \, 
\im^{\ell+1} 
\times
\fs_\ell(-k) \ (2\ell - 1)!! \, (kr)^{-\ell} ,
\\
{\cal H}_{k,\ell}^{(-)} (r) 
~~\overset{r\to0}{\longrightarrow}~~ 
&
+\sqrt{\sym_\ell} \, 
\im^{\ell+1} 
\times
\fs_\ell(-k) \ (2\ell - 1)!! \, (kr)^{-\ell} ,
\\
{\cal G}_{k,\ell}(r) 
~~\overset{r\to0}{\longrightarrow}~~ 
&
-\sqrt{\sym_\ell} \, \im^\ell
\times
\fs_\ell(-k) \ (2\ell - 1)!! \, (kr)^{-\ell} ,
\label{eq:WF_Asymptotes0_G}
\end{align}
\end{subequations}
and
\begin{subequations}
\label{eq:WF_AsymptotesInf}
\label[pluralequation]{eqs:WF_AsymptotesInf}
\begin{align}
%%%%%%%%%%%%%%%%%%%%%
{\cal F}_{k,\ell}(r) 
~~\overset{r\to\infty}{\longrightarrow}~~
&
+\dfrac{\sqrt{\sym_\ell}}{2\im} \qty[
\dfrac{\fs_\ell(-k)}{\fs_\ell(k)} e^{\im kr}
-e^{-\im(kr-\ell\pi)} ]
&&
\overset{k\in\RR}{\longrightarrow} \ 
+\dfrac{\sqrt{\sym_\ell}}{2\im}
\qty(
 e^{ \im k r} e^{2\im \theta_\ell(k)}
-e^{-\im (k r - \ell \pi)}
),
\label{eq:WF_AsymptotesInf_F}
\\
%%%%%%%%%%%%%%%%%%%%%
{\cal H}_{k,\ell}^{(+)}(r) 
~~\overset{r\to\infty}{\longrightarrow}~~
&
-\sqrt{\sym_\ell} \, \im
\,\dfrac{\fs_\ell(-k)}{\fs_\ell(k)}
\, e^{\im k r}
&&
\overset{k\in\RR}{\longrightarrow} \ 
- \sqrt{\sym_\ell} \, \ii e^{+\ii[kr + 2\theta_\ell (k)]},
\label{eq:WF_AsymptotesInf_Hplus}
\\
%%%%%%%%%%%%%%%%%%%%%
{\cal H}_{k,\ell}^{(-)}(r) 
~~\overset{r\to\infty}{\longrightarrow}~~
&
+ \sqrt{\sym_\ell} \, \ii e^{-\ii(kr - \ell\pi)},
&&
\label{eq:WF_AsymptotesInf_Hminus}
\\
%%%%%%%%%%%%%%%%%%%%%
{\cal G}_{k,\ell}(r) 
~~\overset{r\to\infty}{\longrightarrow}~~
&
-\dfrac{\sqrt{\sym_\ell}}{2}
\qty[
\,\dfrac{\fs_\ell(-k)}{\fs_\ell(k)}
\, e^{\im k r}
+ e^{-\im (kr-\ell\pi)}]
&&
\overset{k\in\RR}{\longrightarrow} \
-\dfrac{\sqrt{\sym_\ell}}{2}
\qty(
 e^{ \im k r}e^{2\im \theta_\ell(k)}
+e^{-\im (k r - \ell \pi)} 
).
\label{eq:WF_AsymptotesInf_G}
\end{align}
\end{subequations}
${\cal F}_{k,\ell}$ is the \emph{physical} scattering-state solution and belongs to the family of regular wavefunctions, owing to its normalization (see below).  
${\cal G}_{k,\ell}$ is a purely irregular solution, while 
${\cal H}_{k,\ell}^+$ and ${\cal H}_{k,\ell}^-$ represent outgoing and incoming waves, respectively.  
From \cref{eq:WF_Asymptotes0_F}, we see that the amplitude of the Jost function is related directly to the Sommerfeld factors for annihilation, via~\cite{Cassel:2009wt}
\begin{align}
\label{eq:JostFunction_Sommerfeld}
S_\ell(k)
\equiv
\abs{
\frac{(2\ell + 1)!!}{\sqrt{\sym_\ell} \, k^{\ell + 1}(\ell + 1)!}
\dv[\ell + 1]{r}\ {\cal F}_{k,\ell}(r)
}_{r = 0}^2
=
|\fs_\ell(k)|^{-2}
.
\end{align}

The wavefunctions ${\cal F}_{k,\ell}(r)$ and ${\cal G}_{k,\ell}(r)$ are real up to the same $\ell$- and $k$-dependent (but $r$-independent) phase. In particular,
\begin{subequations}
\label{eq:FG_from_FGstar}
\label[pluralequation]{eqs:FG_from_FGstar}
\begin{align}
[{\cal F}_{k,\ell} (r)]^* 
&=(-1)^{\ell} 
\, \dfrac{\fs_\ell(k^*)}{\fs_\ell(-k^*)}
\, {\cal F}_{k^*,\ell} (r),
\label{eq:F_from_Fstar}
\\
[{\cal G}_{k,\ell} (r)]^* &=(-1)^{\ell} 
\, \dfrac{\fs_\ell(k^*)}{\fs_\ell(-k^*)}
\, {\cal G}_{k^*,\ell} (r) ,
\label{eq:G_from_Gstar}
\end{align}
\end{subequations}
where we used \cref{eq:RegularSol_Properties,eq:H(-k)}.
This, in turn, implies
\begin{subequations}
\label{eq:rephasings_FF-GG-FG}
\label[pluralequation]{eqs:rephasings_FF-GG-FG}
\begin{align}
\label{eq:rephasing_FF}
{\cal F}_{k,\ell}(r) \, [{\cal F}_{k,\ell}(r')]^*
&=
[{\cal F}_{k,\ell}(r)]^* \, {\cal F}_{k,\ell}(r'),
\\[1ex]
\label{eq:rephasing_GG}
{\cal G}_{k,\ell}(r) \, [{\cal G}_{k,\ell}(r')]^*
&=
[{\cal G}_{k,\ell}(r)]^* \, {\cal G}_{k,\ell}(r'),
\\[1ex]
\label{eq:rephasing_FG}
{\cal F}_{k,\ell}(r) \, [{\cal G}_{k,\ell}(r')]^*
&=
[{\cal F}_{k,\ell}(r)]^* \, {\cal G}_{k,\ell}(r')
.
\end{align}
\end{subequations}

\bigskip

For real momenta, $k,k'\in\RR$, the solutions \eqref{eq:OtherSolutions_Def} obey the following orthonormality conditions, derived from \cref{eqs:Orthonorm_Regular+Jost_RealMomenta}, 
\begin{subequations}
\label{eq:Orthonorm_FGH+H-}
\label[pluralequation]{eqs:Orthonorm_FGH+H-}
\begin{align}
\int_0^\infty dr 
\,{\cal F}_{k, \ell} (r) 
\,{\cal F}_{k',\ell}^* (r)    
&= 
\sym_\ell \ 
\dfrac{\pi}{2}[\delta(k-k') + \delta(k+k')] ,
\label{eq:PhysicalSol_Orthonormality}
\\
\int_0^\infty dr 
\,{\cal G}_{k, \ell} (r) 
\,{\cal G}_{k',\ell}^* (r)    
&= 
\sym_\ell \ 
\dfrac{\pi}{2}[\delta(k-k') + \delta(k+k')] ,
\label{eq:IrregularSol_Orthonormality}
\\
\int_0^\infty dr 
\,{\cal H}_{k, \ell}^{(+)}   (r) 
\,[{\cal H}_{k',\ell}^{(+)} (r)]^*
&=
\int_0^\infty dr 
\,{\cal H}_{k, \ell}^{(-)}   (r) 
\,[{\cal H}_{k',\ell}^{(-)} (r)]^*    
= \sym_\ell \ 
\pi \, \delta(k-k') .
\label{eq:WavesSol_Orthonormality}
\end{align}
\end{subequations}

\subsubsection{Bound states}

Possible bound-state solutions to \cref{eq:RadialSE} may arise at (discrete) imaginary values of the momentum, $k\to \im \kappa_{n\ell}$, with $\kappa_{n\ell}\in \RR$, as we shall see in \cref{sec:Jost_Poles}. We shall denote these solutions by $u_{n\ell}(r) \to \hat{\cal B}_{n\ell}(r) \equiv {\cal B}_{n\ell}(r) / (\im \kappa_{n\ell})$. The solutions ${\cal B}_{n\ell}$ must be regular at the origin and vanish at infinity. We define them via the boundary conditions
\begin{align}
\label{eq:BoundSol_Asymptotes_0&Inf}
\lim_{r\to 0} \ [r^{-\ell-1} {\cal B}_{n\ell} (r)] 
= {\rm constant} \in \CC,    
\qquad \text{and} \qquad
\lim_{r\to \infty} {\cal B}_{n\ell} (r) = 0,    
\end{align}
such that 
\begin{align}
\int_0^\infty dr 
\, {\cal B}_{n\ell} (r) 
\, {\cal B}_{n'\ell}^* (r)
= \sym_\ell \, \delta_{nn'}. 
\label{eq:BoundSol_Orthonormality}
\end{align}
In \cref{eq:BoundSol_Orthonormality}, we have incorporated the orthogonality condition arising from \cref{eq:Wronskians_Deriv} when applied to different bound-state solutions. Similarly, from  \cref{eq:Wronskians_Deriv}, applied to a physical scattering-state and a bound-state solution, we obtain
\begin{align}
\int_0^\infty dr \, {\cal B}_{n\ell} (r) \, {\cal F}_{k,\ell}^* (r)
= 0. 
\label{eq:BoundScatteringSol_Orthonormality}
\end{align}

\subsection{Scattering-state solutions using the Frobenius method \label{sec:Jost_Frobenius}}

For some applications, it is necessary to consider higher order terms in the asymptotic scalings of the wavefunctions at $r\to 0$, with respect to those given in \cref{eqs:WF_Asymptotes0}. These corrections can be obtained using the Frobenius method~\cite{DLMF2.7.i}. (See also \cite{Parikh:2024mwa} for a similar analysis.)

We focus on potentials that grow at most as $1/r$ at $r\to 0$, and expand them in a Laurent series, 
\begin{align}
V(r) =
-\sum_{n=-1}^\infty \alpha_n \, r^n,
\label{eq:CentralV_Laurent}
\end{align}
where $\alpha_n \in \RR$ are constants of mass dimension $n+1$. 
Since $V(r)$ is at most $1/r$ as $r\to 0$, $r=0$ is a regular singular point of \cref{eq:RadialSE}. 
We therefore use a Frobenius ansatz for the solution to the Schr\"odinger \cref{eq:RadialSE}, and expand it as follows,
\begin{align}
\label{eq:WF_Expansion_u}
u_{k,\ell} (r) = 
r^{s} \sum_{n=0}^\infty A_\ell^{(n)} r^n .  
\end{align}
Inserting \cref{eq:CentralV_Laurent,eq:WF_Expansion_u} into \eqref{eq:RadialSE}, we obtain
\begin{align}
\dfrac{1}{2\mu}
\sum_{n=0}^{\infty}
\qty[(n+s)(n+s-1)-\ell(\ell+1)]
\, A_\ell^{(n)}
\, r^n
+ 
\dfrac{k^2}{2\mu} 
\sum_{n=0}^{\infty} A_\ell^{(n)} \, r^{n+2}
+
\sum_{n=0}^{\infty} 
\sum_{p=-1}^{\infty}
\alpha_p  
\, A_\ell^{(n)} 
\, r^{p+n+2} 
=0.
\end{align}
Equating the coefficient of each power of $r$ yields the indicial equation from the $r^0$ term,
\begin{align}
\qty[s(s-1)-\ell(\ell+1)] \, A_\ell^{(0)} = 0
\qquad\Rightarrow\qquad
s=\ell+1 \quad \text{or} \quad s=-\ell .
\label{eq:Frobenius_IndicialEq}
\end{align}
For $n\geqslant 1$, we obtain the recursion
\begin{align}
\dfrac{1}{2\mu}
\qty[(n+s)(n+s-1)-\ell(\ell+1)] A_\ell^{(n)}
+\alpha_{-1} \, A_\ell^{(n-1)}
+\dfrac{k^2}{2\mu} A_\ell^{(n-2)}
+\sum_{m=0}^{n-2} 
\alpha_m 
\, A_\ell^{(n-2-m)}
=0,
\label{eq:Frobenius_Recursion}
\end{align}
where it is understood that $A_\ell^{(-1)} = A_\ell^{(-2)}=0$ and the sum is absent for $n<2$. The two values of $s$ found in \cref{eq:Frobenius_IndicialEq} correspond to the regular and irregular solutions.

\paragraph{Regular solution.} Taking into account \cref{eq:WF_Asymptotes0_F}, we expand the regular solution as follows
\begin{align}
{\cal F}_{k,\ell}(r) 
=&
+\sqrt{\sym_\ell} \,
\im^\ell
\times
\dfrac{1}{\fs_\ell(k)}
\dfrac{(kr)^{\ell + 1}}{(2\ell + 1)!!}
\sum_{n=0}^\infty f_\ell^{(n)} (k) \, r^n,
\label{eq:WF_Expansion_F} 
\end{align}
where the coefficients $f_\ell^{(n)}(k)$ have mass dimensions $n$ and obey the recursion relation \eqref{eq:Frobenius_Recursion} with $s=\ell+1$. We obtain in particular, 
\begin{subequations}
\label{eq:RegularSolution_Coeff}
\label[pluralequation]{eqs:RegularSolution_Coeff}
\begin{gather}
f_\ell^{(0)}=1,
\qquad
f_\ell^{(1)}=-\dfrac{\mu\alpha_{-1}}{\ell+1},
\label{eq:RegularSolution_Coeff_0and1}
\\
f_\ell^{(n+2)} (k) = 
-\dfrac{2\mu}{(n+2)(n+2\ell+3)}
\qty(
\dfrac{k^2}{2\mu} f_\ell^{(n)}(k)
+ \alpha_{-1} f_\ell^{(n+1)}(k)
+ \sum_{m=0}^{n} \alpha_{m} \, f_\ell^{(n-m)}(k)
).
\label{eq:RegularSolution_Coeff_n+2}
\end{gather}
\end{subequations}
Note that $f_\ell^{(0)}$ and $f_\ell^{(1)}$ are momentum independent, and $f_\ell^{(n)} \in \RR$, for all $n$ and $k\in \RR$.

\paragraph{Irregular solution.} 

For $s=-\ell$, the coefficient multiplying $A_\ell^{(n)}$ in \cref{eq:Frobenius_Recursion} vanishes at $n=2\ell+1$. At this resonant order the recursion no longer determines $A_\ell^{(2\ell+1)}$ but instead imposes a compatibility (solvability) condition on the lower-order coefficients. For generic potentials this condition is not satisfied by a pure Frobenius series with $s=-\ell$, and the second independent solution must therefore include a logarithmic admixture of the regular solution. We thus take
\begin{align}
{\cal G}_{k,\ell}(r) 
=&
-\sqrt{\sym_\ell} \, \im^\ell
\times
\fs_\ell(-k) \ (2\ell - 1)!! 
\, \qty{
(kr)^{-\ell}
\sum_{n=0}^{\infty} g_\ell^{(n)} (k) \, r^n
- 2\rho_\ell(k) 
\, \ln (\mu r) 
\, (\mu r)^{\ell + 1}
\sum_{n=0}^\infty f_\ell^{(n)} (k) \, r^n
},
\label{eq:WF_Expansion_G}  
\end{align}
where we took into account \cref{eq:WF_Asymptotes0_G}, and the coefficients $g_\ell^{(n)}(k)$ and $\rho_\ell(k)$ have mass dimensions $n$ and $0$, respectively. $\rho_\ell(k)$ is fixed by the solvability condition at the resonant order, whereas $g_\ell^{(2\ell+1)} (k)$ is fixed by the boundary conditions at $r\to\infty$, given by \cref{eq:WF_AsymptotesInf_G}. Although it is not possible to determine $g_{\ell}^{(2\ell+1)}$ explicitly, the boundary condition \eqref{eq:WF_AsymptotesInf_G} implies that all coefficients in the expansion \eqref{eq:WF_Expansion_G} must be purely real, 
$\rho_{\ell} (k) \in \RR$ and 
$g_{\ell}^{(n)} (k) \in \RR$, for all $n$ and  $k\in \RR$.

Note that the scale in the logarithm of \eqref{eq:WF_Expansion_G} is conventional. If $\mu\to\tilde{\mu}$, then $\ln(\mu r)=\ln(\tilde{\mu} r)+\ln(\mu/\tilde{\mu})$, so the change induces an additive contribution proportional to the regular Frobenius series. Equivalently, it corresponds to a different choice of the free (resonant) non-logarithmic coefficient, with induced shifts in higher-order $g_\ell^{(n)}$ that leave the full solution unchanged.

\bigskip
\noindent
$\bm{\ell=0.}$
To determine $g_0^{(n)}(k)$ and $\rho_0(k)$, we substitute \cref{eq:WF_Expansion_G} and \cref{eq:CentralV_Laurent} into \cref{eq:RadialSE}, and match separately the coefficients of $r^n$ and $r^n\ln(\mu r)$. We find,
\begin{subequations}
\label{eq:IrregularSolution_Coeff_ell=0}
\label[pluralequation]{eqs:IrregularSolution_Coeff_ell=0}
\begin{align}
\label{eq:IrregularSolution_Coeff_ell=0_rho+g0+g1}
\rho_{\ell=0}(k) = \alpha_{-1},
\qquad
g_{0}^{(0)} &=1,
\qquad
g_{0}^{(1)}:~\text{fixed by \cref{eq:WF_AsymptotesInf_G}},
\end{align}
\begin{align}
&g_{0}^{(n+2)} (k) 
= -\dfrac{2\mu}{(n+2)(n+1)} \times
\nn \\
&\times
\qty(
\dfrac{k^2}{2\mu} \, g_{0}^{(n)}(k)
+ \alpha_{-1} \, g_{0}^{(n+1)}(k)
+ \sum_{m=0}^{n} \alpha_{m} \, g_{0}^{(n-m)}(k)
- \alpha_{-1} \, (2n+3)\, f_{0}^{(n+1)}(k)
), 
\quad n\geqslant 0.
\label{eq:IrregularSolution_Coeff_ell=0_g}
\end{align}
\end{subequations}
Combining the above, we find the near-origin expansion explicitly for this case,
\begin{align}
\label{eq:WF_Asymptotes0ext_G_ell=0}
{\cal G}_{k,0}(r) =
-\sqrt{\sym_0}\,\fs_0(-k) \, \times
\qty[
1
+ g_0^{(1)}(k)\, r
- 2(\mu \alpha_{-1}) \, r \, \ln(\mu r)
+ {\cal O} (r^2,r^2\ln r)
].
\end{align}
The reflection properties of the irregular solution and the Jost function, given by \cref{eq:IrregularSol_Reflection,eq:Jost_Analytic_Contin}, imply that $g_0^{(1)} (k) = g_0^{(1)} (-k)$.

\bigskip
\noindent
$\bm{\ell\geqslant 1.}$ Similarly, we find, 
\begin{subequations}
\label{eq:IrregularSolution_Coeff_ell>0}
\label[pluralequation]{eqs:IrregularSolution_Coeff_ell>0}
\begin{align}
\label{eq:IrregularSolution_Coeff_ell>0_rho}
\rho_{\ell}(k) &=
\dfrac{1}{2\ell+1}
\dfrac{1}{\mu^{\ell} \, k^{\ell}}
\qty(
\dfrac{k^2}{2\mu} \, g_{\ell}^{(2\ell-1)}(k)
+ \alpha_{-1} \, g_{\ell}^{(2\ell)}(k)
+ \sum_{m=0}^{2\ell-1} \alpha_{m} \, g_{\ell}^{(2\ell-1-m)}(k)
),
\end{align}
\begin{align}
\label{eq:IrregularSolution_Coeff_ell>0_g}
g_{\ell}^{(0)}=1,
\qquad
g_{\ell}^{(1)}=\dfrac{\mu\alpha_{-1}}{\ell},
\qquad
g_{\ell}^{(2\ell+1)}(k):~\text{fixed by \cref{eq:WF_AsymptotesInf_G}},
\end{align}
and for $n\neq 2\ell-1$,
\begin{align}
\label{eq:IrregularSolution_Coeff_ell>0_g_recursion}
&g_{\ell}^{(n+2)} (k) 
=-\dfrac{2\mu}{(n+2)(n-2\ell+1)} \times
\\[1ex]
&\times \qty(
\dfrac{k^2}{2\mu} \, g_{\ell}^{(n)}(k)
+ \alpha_{-1} \, g_{\ell}^{(n+1)}(k)
+ \sum_{m=0}^{n} \alpha_{m} \, g_{\ell}^{(n-m)}(k)
- 
\mu^\ell k^\ell \rho_{\ell}(k)
\, (2n-2\ell+3)\, f_{\ell}^{(n-2\ell+1)}(k)
),
\nn
\end{align}
\end{subequations}
where it is understood that $f_\ell^{(n)} = 0$ for $n<0$, and the last term contributes only for $n\geqslant 2\ell$, since it involves $f_\ell^{(n-2\ell+1)}$ and we set $f_\ell^{(n)}=0$ for $n<0$.
Note that $g_{\ell}^{(0)}$ and $g_{\ell}^{(1)}$ are momentum-independent. The near-origin expansion is
\begin{align}
\label{eq:WF_Asymptotes0ext_G_ell>0}
{\cal G}_{k,\ell} (r) =    
-\sqrt{\sym_\ell} \, \im^\ell
\times
\fs_\ell(-k) \ (2\ell - 1)!! 
(kr)^{-\ell}
\, \qty[
1 + r \ \dfrac{\mu \alpha_{-1}}{\ell}
+ {\cal O}(r^2,r^{2\ell+1} \ln r)
]. 
\end{align}

\subsection{Analytic properties of Jost functions \label{sec:Jost_Poles}}

We now examine the zeros of the Jost function.   We focus on solutions in the upper complex plane, 
\begin{align}\label{eq:JostZerosDefn}
\fs_\ell(k_0) = 0,
\quad
k_0\in\CC_+
.
\end{align}
From \cref{eq:JostFunctionDefn} it follows that, when $k = k_0$, the regular and Jost solutions are linearly dependent, 
\begin{align}\label{eq:RegularJostRelation}
H_{k_0,\ell}(r)
= b_0 \, \varphi_{k_0,\ell}(r) ,
\end{align}
where \cref{eq:RegularJostRelation} defines $b_0 \in \CC$. 
In view of the asymptotic behavior \eqref{eq:JostSol_AsymptoteInf} of $H_{k,\ell}$ at $r\to \infty$, the left-hand side of \cref{eq:RegularJostRelation} vanishes at infinity for $\Im k_0 > 0$. The boundary conditions of the regular solution further imply that the right-hand side of \cref{eq:RegularJostRelation} vanishes at the origin. Hence, both sides of \cref{eq:RegularJostRelation} are square integrable. 
Since $k_0^2$ is an eigenvalue of a Hermitian operator, as per \cref{eq:RadialSE}, corresponding to a square-integrable function, $k_0^2$ must be real.
This implies that $k_0$ is purely imaginary. 

We will now prove that the zeros of the Jost functions are simple zeros, i.e., that $\fs_\ell'(k_0)\neq 0$. Differentiating \cref{eq:JostFunctionDefn} with respect to $k$ and evaluating the resulting expression at $k = k_0$, yields
\begin{align}\label{eq:JostFunctionDerivative}
\fs_\ell'(k_0)
=
b_0^{-1} \, (-k_0)^\ell
{\cal W}\qty{H_{k_0,\ell}'(r),H_{k_0,\ell}(r)}
+
b_0 \, (-k_0)^\ell
{\cal W}\qty{\varphi_{k_0,\ell}(r), \varphi_{k_0,\ell}'(r)}
\end{align}
where $' \equiv \dd/\dd k$.  In order to evaluate the Wronskians, we consider \cref{eq:Wronskians_Deriv} for $\{\varphi_{k,\ell}, \varphi_{k',\ell}\}$ and for $\{H_{k,\ell}, H_{k',\ell}\}$ differentiate them  with respect to $k'$ and set $k = k' = k_0$, to obtain
\begin{subequations}
\label{eq:Wronskians_DerivDeriv}
\label[pluralequation]{eqs:Wronskians_DerivDeriv}
\begin{align}
\dv{r}\ {\cal W}
\qty{\varphi_{k_0,\ell}(r), \varphi_{k_0,\ell}'(r)}
&=-2k_0 \qty[\varphi_{k_0,\ell}(r)]^2,
\\
\dv{r}\ {\cal W}
\qty{H_{k_0,\ell}'(r), H_{k_0,\ell}(r)}
&=+2k_0 [H_{k_0,\ell}(r)]^2.
\end{align}
\end{subequations}
Integrating both sides, recalling that $\varphi_{k_0,\ell}(0) = 0$ and $\lim_{r\to\infty} H_{k_0,\ell}(r) =0$, we find
\begin{subequations}
\label{eq:Wronskians_Integral}
\label[pluralequation]{eqs:Wronskians_Integral}
\begin{align}
\label{eq:RegularWronskian}
{\cal W}
\qty{\varphi_{k_0,\ell}(r), \varphi_{k_0,\ell}'(r)}
&=-2k_0 \int_0^r \dd r'\ \qty[\varphi_{k_0,\ell}(r')]^2 ,
\\
\label{eq:JostWronskian}
{\cal W}
\qty{H_{k_0,\ell}'(r), H_{k_0,\ell}(r) }
&=-2k_0 \int_{r}^\infty \dd r' \qty[H_{k_0,\ell}(r')]^2.
\end{align}
\end{subequations}
Combining \cref{eq:RegularJostRelation,eq:JostFunctionDerivative,eq:Wronskians_Integral} gives
\begin{align}\label{eq:JostDerivative}
\fs_\ell'(k_0)
&=
2b_0 \, (-k_0)^{\ell+1}
\int_0^\infty
\dd r\ \qty[\varphi_{k_0,\ell}(r)]^2
.
\end{align}
Due to the boundary condition \eqref{eq:JostSol_AsymptoteInf}, $b_0\neq 0$. Furthermore, since the regular solution is a function of $k^2$, $\varphi_{k_0,\ell}(r)$ is real for purely imaginary $k_0$. This implies that \cref{eq:JostDerivative} cannot vanish. Since $\fs_\ell'(k_0) \neq 0$ when $\fs_\ell(k_0) = 0$, we conclude that the zero at $k_0$ is always simple.

The solutions \eqref{eq:RegularJostRelation}, being square-integrable with energies ${\cal E} = k_0^2/(2\mu) <0$, correspond to bound states. Setting $k_0 \to \im \kappa_{n\ell}$ and $b_0 \to b_{n\ell}$, with $\kappa_{n\ell} >0$, and defining
\begin{align}\label{eq:BoundStates_NormFactor}
|N_{n\ell}|^2 \equiv
\int_0^\infty \dd r\ 
\qty[\varphi_{\ii\kappa_{n\ell},\ell}(r)]^2
= \dfrac
{\fs_\ell'(\ii\kappa_{n\ell})}
{2b_{n \ell} \, (-\ii\kappa_{n\ell})^{\ell + 1}},
\end{align}
the bound-state wavefunctions normalized according to \cref{eq:BoundSol_Orthonormality} are simply
\begin{align}\label{eq:BoundStateDefn}
{\cal B}_{n\ell}(r) \equiv 
\dfrac{\sqrt{\sym_\ell} \ \varphi_{\ii\kappa_{n\ell},\ell}(r)}{N_{n\ell}} .
\end{align}
Note that the complex phase of $N_{n\ell}$ is arbitrary, and can render the wavefunctions ${\cal B}_{n\ell}(r)$ complex.

\bigskip

If the Jost function has no zeros, then there are no regular solutions that vanish at spatial infinity (cf.~\cref{eq:RegularSol_AsymptoteInf}), that is, the potential does not support bound levels.

\subsection{The Green's Function}\label{sec:Jost_Greens}

\subsubsection{Jost function representation
\label{sec:Jost_Greens_JostRep}}

Having developed the above machinery, we will now use the regular solution, the Jost solution and the Jost function to construct the Green's function corresponding to \cref{eq:RadialSE}. In particular, we want to solve 
\begin{align}\label{eq:GreensDifferentialEquation}
[{\cal S}_\ell(r) - {\cal E}_{\vb{k}}]
G_{k,\ell}(r,r')
=
\delta(r - r'),
\end{align}
where
\begin{align}
{\cal S}_\ell(r)
\equiv
-\dfrac{1}{2\mu}\dv[2]{r} + \dfrac{\ell(\ell + 1)}{2\mu r^2} + V(r)
,
\end{align}
and ${\cal E}_{\vb{k}} = \vb{k}^2/(2\mu)$. The Green's function is a solution of Schrödinger's equation for $r\neq r'$, continuous at $r=r'$ with discontinuous derivative, as follows from \cref{eq:GreensDifferentialEquation},  
\begin{align}
\label{eq:GreensFun_DiscDeriv}
\eval{
\dv{r} G_{k,\ell}(r,r') 
}_{r = r'-\epsilon}^{r = r' + \epsilon}
= -2\mu .
\end{align}
We are interested in a Green's function that is regular at $r\to 0$ and behaves as an outgoing (incoming) wave for $k>0$ ($k<0$) at $r\to \infty$. 
It follows that
\begin{align}
\label{eq:GreensFun_Asymptotes}
G_{k,\ell}(r,r') =
\begin{cases}
a_{k,\ell}(r') \, \varphi_{k,\ell}(r) , & r < r',
\\[1ex]
b_{k,\ell}(r') \, H_{k,\ell}(r) , & r > r'.
\end{cases}
\end{align}
Continuity at $r=r'$ gives
$a_{k,\ell}(r') = c_{k,\ell}\, H_{k,\ell}(r')$
and
$b_{k,\ell}(r') = c_{k,\ell}\, \varphi_{k,\ell}(r')$, while the discontinuity \eqref{eq:GreensFun_DiscDeriv} implies
$c_{k,\ell} 
= 2\mu /{ \cal W}\qty{H_{k,\ell}(r),\varphi_{k,\ell}(r)}
= 2\mu(-k)^\ell / \fs_\ell(k)$.

Collecting the above, the Green's function can be concisely written as
\begin{align}\label{eq:AnalyticGreens}
G_{k,\ell}(r,r')
&=
\dfrac{2\mu(-k)^\ell}{\fs_\ell(k)}
\, \varphi_{k,\ell}(r_<)
\, H_{k,\ell}(r_>) 
\ \overset{k\in\RR}{\longrightarrow} \ 
\dfrac{\im \, 2\mu}{\sym_\ell \, k}
\, {\cal F}_{k,\ell}^*(r_<)
\, {\cal H}_{k,\ell}^{(+)} (r_>) 
,
\end{align}
where $r_< \equiv \min\{r,r'\}$, $r_> \equiv \max\{r,r'\}$, and in the second step, we used the solutions defined in \cref{eqs:OtherSolutions_Def}. 
This reproduces the Green's function presented in \eqref{eq:GreensFunction_FH}.

\subsubsection{Spectral decomposition}

For $k\in \RR$, the $r$ dependence of the Green's function can be expanded in terms of the eigenfunctions  of the Hermitian operator ${\cal S}_\ell$ that are consistent with the desired asymptotic scalings \eqref{eq:GreensFun_Asymptotes}. Considering the small $r$ behavior, we write
\begin{align}\label{eq:GreensExpansion}
G_{k,\ell}(r,r')
=
\int_{-\infty}^\infty
\dd q \ 
\chi_{k,\ell}(q,r')
\, \varphi_{q,\ell}(r)
+
\sum_n 
\chi_{k,\ell}(n,r')
\, {\cal B}_{n\ell}(r)
,
\end{align}
where we can take $\chi_{k,\ell} (-q,r) = \chi_{k,\ell} (q,r)$, since $\varphi_{q,\ell}(r)$ is even in $q$ (cf.~\cref{eq:RegularSol_Reflection}) and we integrate over a symmetric $q$ interval. The functions $\chi_{k,\ell} (q,r')$ will determine the large $r$ behavior of \cref{eq:GreensExpansion}. 
Inserting \cref{eq:GreensExpansion} into \cref{eq:GreensDifferentialEquation} gives
\begin{align}
\int_{-\infty}^\infty
\dd q\ 
\chi_{k,\ell}(q,r')
[{\cal E}_{\vb{q}} - {\cal E}_{\vb{k}}]
\varphi_{q,\ell}(r)
+
\sum_n 
\chi_{k,\ell}(n,r')
[{\cal E}_{n\ell} - {\cal E}_{\vb{k}}]
{\cal B}_{n\ell}(r)
= \delta(r - r') .
\end{align}
Acting with 
$\int_0^\infty \dd r \, \varphi_{\tilde{q},\ell} (r)$
and
$\int_0^\infty \dd r \, {\cal B}_{\tilde{n}\ell}^* (r)$ 
to project on scattering and bound states, and using the orthonormality \cref{eq:Orthonorm_RegularSol_RealMom,eq:BoundSol_Orthonormality,eq:BoundScatteringSol_Orthonormality}, valid for real momenta, yields
\begin{subequations}
\label{eq:GreensFun_ExpansionCoeff}
\label[pluralequation]{eqs:GreensFun_ExpansionCoeff}
\begin{align}
\label{eq:GreensFun_ExpansionCoeff_Scatt}
\dfrac{1}{2} [\chi_{k,\ell}(q,r') + \chi_{k,\ell}(-q,r') ] 
=\chi_{k,\ell}(q,r')
&=
\dfrac{1}{\pi} \,
\dfrac{q^{2\ell+2}}{|\fs_\ell(q)|^2} \,
\dfrac{\varphi_{q,\ell}(r')}
{{\cal E}_{\vb{q}} - {\cal E}_{\vb{k}}}
=
\dfrac{2\mu}{\pi} \,
\dfrac{q^{2\ell+2}}{|\fs_\ell(q)|^2} \,
\dfrac{\varphi_{q,\ell}(r')}{q^2 - k^2} ,
\\[1ex]
\label{eq:GreensFun_ExpansionCoeff_Bound}
\chi_{k,\ell}(n,r') &=
\dfrac{1}{\sym_\ell}
\dfrac{{\cal B}_{n\ell}^*(r')}
{{\cal E}_{n\ell} - {\cal E}_{\vb{k}}}
=- \dfrac{2\mu}{\sym_\ell}
\dfrac{{\cal B}_{n\ell}^*(r')} {\kappa_{n\ell}^2 + k^2} .
\end{align}
\end{subequations}
Combining the above, and exchanging $\varphi_{q,\ell}$ for ${\cal F}_{q,\ell}$ according to \cref{eq:PhysicalSolution}, leads to the spectral decomposition of the Green's function presented in \cref{eq:GreensFunction_SpectralDecomp},
\begin{align}\label{eq:SpectralGreens}
G_{k,\ell}(r,r')
=
\dfrac{2\mu}{\sym_\ell}
\times
\qty[
\frac{1}{\pi}
\int_{-\infty}^\infty
\dd q\ 
\dfrac{{\cal F}_{q,\ell}(r){\cal F}_{q,\ell}^*(r')}
{q^2 - k^2 - \ii\epsilon}
-
\sum_n
\frac{
{\cal B}_{n\ell}(r)
{\cal B}_{n\ell}^*(r')
}{\kappa_{n\ell}^2 + k^2}
] , 
\qquad k > 0, 
\end{align}
where the prescription $\epsilon \to 0^+$ ensures the desired asymptotic behavior at large $r$ given in \cref{eq:GreensFun_Asymptotes}, for the physical range, $k>0$, as we show next. 

\paragraph{Caveat for long-range potentials.}
For Coulomb-like interactions, which do not formally satisfy the convergence condition \eqref{eq:V_ConvergenceCondition}, the wavefunctions carry logarithmic phases (e.g., $(\mu\alpha/q)\ln(2qr)$, cf.~\cref{sec:Jost_Coulomb}), and their
analytic continuation in $q$ depends on branch choices. Consequently, the $q\to -q$ evenness implicit in writing
$q\in(-\infty,\infty)$ and the short-range relation $\fs_\ell(-q)=\fs_\ell^*(q)$ should not be used blindly.
In practice it is safer to formulate the continuum integral using the physical region $q>0$, and only then extend by symmetry if the relevant branches are controlled. (See also Ref.~\cite{Flores:2026yay}.)

\subsubsection{Equivalence of Green's function representations \label{sec:Jost_Greens_Equivalence}}

Our principal goal is to now demonstrate the equivalence of \cref{eq:AnalyticGreens,eq:SpectralGreens}. To do so, we must evaluate the integral
\begin{align}
\label{eq:Integral_I_def}
I \equiv
\dfrac{1}{\sym_\ell \pi}
\int_{-\infty}^\infty
\dd q \ 
\dfrac{{\cal F}_{q,\ell}(r) {\cal F}_{q,\ell}^*(r')}{q^2 - k^2 - \ii \epsilon}
=
\dfrac{1}{\pi}
\int_{-\infty}^\infty \dd q \ 
\dfrac{q^{2\ell + 2}}{|\fs_\ell(q)|^2}
\dfrac{\varphi_{q,\ell}(r)\varphi_{q,\ell}(r')}{q^2 - k^2 - \ii \epsilon},
\end{align}
where in the second equality, we used \cref{eq:PhysicalSolution}. 

We begin by assuming that $r < r'$ and expanding $\varphi_{q,\ell}(r')$ in terms of Jost solutions
\begin{align}
I = 
\frac{1}{\pi}
\int_{-\infty}^\infty \dd q \
\dfrac{q^{2\ell + 2}}{|\fs_\ell(q)|^2}
\dfrac{\varphi_{q,\ell}(r)}{q^2 - k^2 - \ii \epsilon} \ 
\dfrac{\ii}{2} q^{-(\ell + 1)}
\qty[
\fs_{\ell}(q)  H_{-q, \ell}(r')
-(-1)^\ell \fs_\ell(-q)  H_{q, \ell}(r')
] .
\end{align}
To continue, we define
\begin{subequations} \label{eq:GreenFunctionContourIntegrals}
\begin{align}
I_1 &\equiv 
\dfrac{\ii}{2\pi}
\int_{-\infty}^\infty \dd q \, 
\dfrac{q^{\ell + 1}}{\fs_\ell(-q)}
\dfrac{\varphi_{q,\ell}(r) H_{-q,\ell}(r')}{q^2 - k^2 - \ii \epsilon}
=
\left[(-1)^{\ell+1}
\dfrac{\ii}{2\pi}
\int_{-\infty}^\infty  \dd q \, 
\dfrac{q^{\ell + 1}}{\fs_\ell(q)}
\dfrac{\varphi_{q,\ell}(r) H_{q,\ell}(r')}{q^2 - k^2 + \ii \epsilon}
\right]^* ,
\\[1ex]
I_2 &\equiv 
(-1)^{\ell+1}
\dfrac{\ii}{2\pi}
\int_{-\infty}^\infty
\dd q \,
\dfrac{q^{\ell + 1}}{\fs_\ell(q)}
\cdot
\dfrac{\varphi_{q,\ell}(r) H_{q,\ell}(r')}{q^2 - k^2 - \ii \epsilon} ,
\end{align}
\end{subequations}
such that $I = I_1 + I_2$. We may evaluate these integrals using Cauchy's Residue theorem. To select the appropriate contour, we must consider the asymptotic behavior at infinity of the Jost solutions, given in \cref{eq:JostSol_AsymptoteInf}.

We evaluate $I_1^*$ and $I_2$ by closing the integration contour in the \emph{upper half} of the complex $q$ plane, since $H_{q,\ell}(r')\sim \exp[+\im qr']$. 
The contribution from the large semicircle in the upper half-plane vanishes as its radius is taken to infinity, because the integrand is exponentially suppressed by 
$\exp[2\im\theta_\ell(k)]\exp[-(\Im q) \, (r'+r)]
+(-1)^{\ell+1}\exp[-(\Im q) \, (r'-r)]$.
The poles within the contour are 
$q =-k +\im \epsilon$ and $q =+k +\im \epsilon$ for $I_1^*$ and $I_2$, respectively, in addition to those due to the simple zeros of the Jost function (cf.~\cref{sec:Jost_Poles}), at $q=\im \kappa_{n\ell}$. Therefore, we find
\begin{subequations}
\begin{align}
I_1^* &=
-\dfrac{\ii^{\ell}}{ \sqrt{\sym_\ell}\,2k}
{\cal F}_{-k,\ell}(r) H_{-k,\ell}(r')
+
\sum_n
\dfrac{(-\ii\kappa_{n\ell})^{\ell + 1}}{
\fs_\ell'(\ii\kappa_{n\ell})}
\dfrac{
\varphi_{\ii\kappa_{n\ell}, \ell}^{} (r) 
H_{\ii\kappa_{n\ell},\ell}^{} (r')
}
{\kappa_{n\ell}^2 + k^2} ,
\\
I_2 &=
+\dfrac{\ii^\ell}{\sqrt{\sym_\ell} 2k}
{\cal F}_{+k,\ell}(r) H_{+k,\ell}(r')
+
\sum_n
\dfrac{(-\ii\kappa_{n\ell})^{\ell + 1}}{\fs_\ell'(\ii\kappa_{n\ell})}
\cdot
\dfrac{\varphi_{\ii\kappa_{n\ell},\ell}^{} (r) H_{\ii\kappa_{n\ell},\ell}^{}(r')}{\kappa_{n\ell}^2 + k^2} .
\end{align}
\end{subequations}
From this, it follows that $I_1=I_2$, where we used the properties of the wavefunctions described in the previous sections, including (cf.~\cref{eq:H(-k),eq:Jost_Analytic_Contin}):
$[H_{\im \kappa_{n\ell},\ell}^{} (r')]^* 
=(-1)^\ell H_{\im \kappa_{n\ell},\ell}^{} (r')$
and
$[\fs_\ell(\im x)]^* = \fs_\ell (\im x)$ for $x\in\RR$, thus by Taylor expanding in $x$,
$[\fs_\ell'(\im \kappa_{n\ell})]^* =-\fs_\ell' (\im \kappa_{n\ell})$. 
The above imply that
\begin{align}
I = 
\dfrac{\ii}{\sym_\ell \, k} \,
{\cal F}_{k,\ell}^* (r) 
{\cal H}_{k,\ell}^{(+)}(r')
+ 
\dfrac{1}{\sym_\ell}
\sum_n
\dfrac{{\cal B}_{n\ell}(r){\cal B}_{n\ell}^*(r')}{\kappa_{n\ell}^2 + k^2},
\qquad r<r',
\qquad k\in\RR,
\end{align}
where we made use of \cref{eq:OtherSolutions_Def,eq:F_from_Fstar,eq:RegularJostRelation,eq:BoundStateDefn}. 
Performing a similar calculation of $I$ for $r > r'$, we obtain the same result with $r\leftrightarrow r'$. Therefore, we conclude that
\begin{align}\label{eq:I_Compute_WO_Arc}
I 
=
\dfrac{\ii}{\sym_\ell k}
{\cal F}_{k,\ell}^*(r_<) {\cal H}_{k,\ell}^{(+)}(r_>)
+
\dfrac{1}{\sym_\ell}
\sum_n
\dfrac{
{\cal B}_{n\ell}(r) {\cal B}_{n\ell}^*(r')
}
{\kappa_{n\ell}^2 + k^2}
,
\end{align}
where we recall $r_< \equiv \min\{r,r'\}$, $r_> \equiv \max\{r,r'\}$.
After inserting this result into \cref{eq:SpectralGreens}, we recover \cref{eq:AnalyticGreens}.

%%%
%%%

\subsection{The Coulomb wave functions \label{sec:Jost_Coulomb}}

Next, we illustrate how the Jost function formalism manifests for a Coulomb potential, 
\begin{align}
V^C(r) = -\alpha/r.
\label{eq:CoulombPotential}
\end{align} 
The superscript $C$ here and below indicates that the solutions correspond to a Coulomb potential. We define $\zeta \equiv \mu\alpha/k$. For $\alpha >0$ ($\alpha <0$) the potential is attractive (repulsive).

Because the Coulomb potential is long-ranged, the Jost boundary condition
\cref{eq:JostSol_AsymptoteInf} is modified by the Coulomb logarithmic phase. For $r>0$ and $k\in\CC_+$ (assuming the principal branch of $\ln z$ and $k\neq 0$),
\begin{align}
\lim_{r\to\infty}
\Big[
e^{-\ii\pi\ell/2}\,e^{-\ii kr}\,(2kr)^{-\ii \zeta}\,
H^C_{k,\ell}(r)
\Big]=1,
\qquad \zeta\equiv \mu\alpha/k .
\end{align}
The regular and the Jost solutions are
\begin{subequations}
\label{eq:Coulomb_Solutions}
\label[pluralequation]{eqs:Coulomb_Solutions}
\begin{align}
\label{eq:Coulomb_RegularSol}
\varphi_{k,\ell}^C(r) &=
\dfrac{2^{\ell} \Gamma(\ell+1)}{\Gamma(2\ell + 2)}r^{\ell + 1}
e^{+\ii kr}
\, {}_1F_{1}
\qty(1 + \ell - \ii\zeta; 2\ell + 2;-2\ii kr) ,
\\
\label{eq:Coulomb_JostSol}
H_{k,\ell}^C(r) &= 
-\ii e^{-\pi\zeta/2} (2kr)^{\ell + 1}e^{+\ii kr}
U(1 + \ell -\ii\zeta; 2\ell + 2; -2\ii kr).
\end{align}
\end{subequations}
where ${}_1F_1$ and $U$ are the confluent hypergeometric functions of the first and second kind, respectively (see Ref.~\cite{DLMF13.2}).\footnote{
%%%%%%%%%%%%
Despite its appearance, $\varphi_{k,\ell}^C$ is a real function. This can be demonstrated with one of Kummer's transformations, i.e., \cref{eq:KummersTransform}.}
%%%%%%%%%%%%
From \cref{eq:JostFunctionDefn}, we may find the Jost function,
\begin{align}
\fs_\ell^C(k) =
\dfrac{\ell! \, e^{-\pi \zeta/2}}
{\Gamma(1 + \ell - \ii\zeta)} ,
\label{eq:JostFunction_Coulomb}
\end{align}
whose zeros correspond to 
$k \to  \ii\kappa_n = \ii\mu\alpha/n$, 
or $\im \zeta \to n$,
with $n = 1,2,\ldots$ and $n\geqslant \ell + 1$, if $\alpha >0$. There are no zeros for $\alpha <0$ in the upper half complex-$k$ plane, since a repulsive Coulomb potential does not support bound states. As expected from \cref{eq:JostFunction_Sommerfeld}, $1/|\fs_\ell^C(k)|^2$ coincides exactly with the well-known expression for the Coulomb Sommerfeld factor for annihilation~\cite{Cassel:2009wt}.

The physical solutions are found by \cref{eq:PhysicalSolution},
\begin{align}
{\cal F}_{k,\ell}^C(r)
= \sqrt{\sym_\ell} 
\,\dfrac{(2\ii)^\ell e^{\pi\zeta/2}}{\Gamma(2\ell + 2)}
\,\Gamma(1 + \ell - \ii\zeta)
\,(kr)^{\ell + 1}
\,e^{+\ii kr}
\,{}_1F_1(1 + \ell - \ii\zeta; 2\ell + 2;-2\ii kr) .
\end{align}
The Coulomb phase shift is
\begin{align}
\theta_\ell^C(k) 
\equiv -\arg [\fs_\ell^C (k)]
= \arg [\Gamma(1+\ell -\im \zeta)].    
\end{align}
At $r\to\infty$, these three solutions behave asymptotically as
\begin{align}
{\cal F}_{k,\ell}^C(r)
&\to \sqrt{\sym_\ell} 
\, \ii^\ell
\, e^{\ii\theta_\ell^C(k)}
\, \sin[kr + \zeta\ln(2kr) -\ell\pi/2 + \theta_\ell^C(k)] ,
\\[1ex]
\varphi_{k,\ell}^C(r)
&\to 
\dfrac{\ell! e^{-\pi\zeta/2}}
{k^{\ell + 1} \, |\Gamma(1 + \ell -\ii\zeta)|}
\sin[kr + \zeta\ln(2kr) -\ell\pi/2 + \theta_\ell^C(k)] ,
\\[1ex]
H_{k,\ell}^C(r)
&\to
e^{\ii[kr + \zeta\ln(2kr) +\ell\pi/2]} .
\end{align}
This behavior is consistent with our definitions of each solution up to the logarithmic phase shift which appears for the Coulomb potential.

We are interested in determining the bound-state wave functions from the regular solutions. To do so, we must calculate the normalization factor~\eqref{eq:BoundStates_NormFactor}. This requires determining the relative constant which appears when the regular and Jost solutions are evaluated at a zero of the Jost function. Using~\cref{eq:U1F1_Relation} one can show that
\begin{align}
b_{n\ell}
=
\dfrac{H_{\ii\kappa_n, \ell}^C(r)}
{\varphi_{\ii\kappa_n, \ell}^C(r)}
=(-\im)^{n+1}
\, \dfrac{2(n+\ell)!}{\ell!}
\, (-\im \kappa_n)^{\ell+1}.
\end{align}
Additionally, we must use the fact that
\begin{align}
\eval{\dv{k} \fs_\ell^C(k)}_{k = \ii\kappa_n}=
\im^n (-1)^{n-\ell} \, n \, (n-\ell-1)! \, \ell! 
\, (-\im \kappa_n)^{-1}
\end{align}
It follows then that,
\begin{align}
|N_{n\ell}|^2 
=\dfrac{\fs_\ell'(\im \kappa_n)}
{2b_{n\ell} (-\im \kappa_n)^{\ell+1}}
=\dfrac{n(\ell!)^2(n-\ell-1)!}{4(n+\ell)!}
\dfrac{1}{\kappa_n^{2\ell+3}} 
\,.
\end{align}
The normalized bound-state solution is then found from \cref{eq:BoundStateDefn,eq:Coulomb_RegularSol}
\begin{align}
{\cal B}_{n\ell}^C(r)
&= \sqrt{\sym_\ell}
\, \dfrac{\kappa_n^{1/2}}{\Gamma(2\ell + 2)}
\sqrt{\frac{\Gamma(n + \ell + 1)}{n\ \Gamma(n - \ell)}}
\, (2\kappa_n r)^{\ell + 1}
e^{-\kappa_n r}
{}_1F_{1}
\qty(1 + \ell  - n; 2\ell + 2;+2\kappa_n r) 
\nn \\
&=
\sqrt{\sym_\ell} \,
\kappa_n^{1/2}
\sqrt{\frac{\Gamma(n - \ell)}{n\ \Gamma(n + \ell + 1)}}
(2\kappa_n r)^{\ell + 1}
e^{-\kappa_n r}
L_{n - \ell - 1}^{2\ell + 1}(2\kappa_n r).
\end{align}
where we rewrote the expression in terms of associated Laguerre polynomials using~\cref{eq:Hyper1F1_Laguerre}. 
With the aid of~\cref{eq:LaguerreInt_I}, it is easy to show that ${\cal B}_{n\ell}^C(r)$ is normalized 
according to \cref{eq:BoundSol_Orthonormality}.

%%%%%%%%%%%%%%%%%%%%%%%%%%%%%%%%%%%%%%%%%%%%%%%%%%%
%%%%%%%%%%%%%%%%%%%%%%%%%%%%%%%%%%%%%%%%%%%%%%%%%%%
%%%%%%%%%%%%%%%%%%%%%%%%%%%%%%%%%%%%%%%%%%%%%%%%%%%
\clearpage
\section{Resummation  \label{sec:Resummation}}
\subsection{Schrödinger equation with a non-Hermitian non-local potential \label{sec:Resummation_SchrEq}}

Because unitarity implies a non-linear relation between scattering amplitudes, as seen directly from \cref{eq:Unitarity_Tmatrix}, it cannot be exactly satisfied by calculations truncated at a finite order in perturbation theory. Ensuring consistency of the scattering amplitudes with \cref{eq:Unitarity_Tmatrix} necessitates the appropriate resummation of all interaction kernels entering a calculation. The resummation consistently determines the self energies of the participating states, which in turn affect all scattering amplitudes. It is evident that elastic interactions contribute to the self-energy of a state. As discussed in the previous section, inelastic interactions contribute as well. 

In the following, we shall denote by 
${\cal K}^\twoPI (\vb{p'},\vb{p})$ the 2PI kernel of the state under consideration, with $\vb{p'}$ and $\vb{p}$ being, respectively, the incoming and outgoing momenta of the particles in the CM frame. Under the instantaneous approximation, the resummation of the self-energy kernel amounts to solving the Schrödinger equation. We consider Schrödinger's equation, in both momentum and position space, for a potential that is \emph{not} assumed to be local,\footnote{We draw attention to the fact that the incoming momentum in the kernel should be the one on which the state is projected, i.e. the momentum that is being integrated over.}
\begin{subequations}
\label{eq:SchrodingerEqs}
\label[pluralequation]{eqs:SchrodingerEqs}
\begin{align}
\label{eq:SchrodingerEq_momentum}
\dfrac{\vb{p}^2}{2\mu}
\tilde{\psi}(\vb{p})
-\dfrac{1}{4\mT\mu}
\int
\dfrac{\dd^3 p'}{(2\pi)^3}
{\cal K}^{\twoPI} (\vb{p'},\vb{p})
\ \tilde{\psi}(\vb{p}')
&={\cal E} \tilde{\psi}(\vb{p}) ,
\\
\label{eq:SchrodingerEq_position}
-\dfrac{\vb{\nabla}^2}{2\mu} \psi(\vb{r})
+\int \dd^3 \vb{r'} 
\, {\cal V} (\vb{r'},\vb{r}) 
\, \psi(\vb{r'})
&= {\cal E} \psi(\vb{r}),
\end{align}
\end{subequations}
where ${\cal E}\equiv E - \mT$, with $E$ being the total energy of the state, $\mT$ and $\mu$ are the total and reduced masses of the interacting particles. The Fourier transformations of the wavefunction and interaction kernel relating \cref{eq:SchrodingerEq_momentum,eq:SchrodingerEq_position} are
\begin{align}
\label{eq:FT_wavefunctions}
\psi(\vb{r})
&=
\int
\dfrac{\dd^3 p}{(2\pi)^3}
e^{\im \vb{p}\cdot \vb{r}}
\tilde{\psi}(\vb{p}), 
\\
\label{eq:FT_potential}
{\cal V}(\vb{r'},\vb{r})
&=
-
\dfrac{1}{4\mT\mu}
\int
\dfrac{\dd^3 p'}{(2\pi)^3}
\dfrac{\dd^3 p}{(2\pi)^3} 
\, e^{+\im \vb{p}\cdot \vb{r}}
\, {\cal K}^{\twoPI} (\vb{p'},\vb{p})
\, e^{-\im \vb{p'}\cdot \vb{r'}}
.
\end{align}

Note that in the non-relativistic approximation leading to Schrödinger's equation, the total energy in the CM frame in the \emph{interaction term} is set to $\sqrt{s} \to \mT$ (see e.g.~Ref.~\cite{Petraki:2015hla}). This replacement can be made in all instances where the $s$ dependence is not associated with a threshold crossing. 
We return to this point in the following, when introducing the various contributions to the non-relativistic potential.

\subsubsection*{Partial-wave expansion}

We shall assume for simplicity that ${\cal K}^{\twoPI} (\vb{p'},\vb{p})$ does not depend on any external vector, such as a spin direction or a background field. We may then expand it into partial waves in accordance with \cref{eq:Mcal2to2_PW},
\begin{subequations}
\label{eq:PW_Potential}
\label[pluralequation]{eqs:PW_Potential}
\begin{align}
\label{eq:PW_Potential_mom}
{\cal K}^{\twoPI} (\vb{p'},\vb{p})
&=
16 \pi
\sum_\ell
(2\ell + 1)
P_\ell(\vb{\hat{p}'}\cdot \vb{\hat{p}})
{\cal K}_\ell^{\twoPI} (p',p) ,
\\
\label{eq:PW_Potential_pos}
{\cal V}(\vb{r'},\vb{r})
&=
\dfrac{1}{4\pi}
\sum_\ell
(2\ell + 1)
P_\ell(\vb{\hat{r}'}\cdot \vb{\hat{r}})
{\cal V}_\ell(r',r) ,
\\
\label{eq:PW_Potential_pos-mom}
{\cal V}_\ell(r',r) &= 
-\dfrac{1}{4\mT \mu}
\ \dfrac{16}{\pi^2}
\int_0^\infty \dd p' \, p'^2 \, j_\ell (p'r') 
\int_0^\infty \dd p  \, p^2  \, j_\ell (pr) 
\ {\cal K}_\ell^{\twoPI} (p',p) .
\end{align}
\end{subequations}
where $j_\ell$ are the spherical Bessel functions of first kind.

\Cref{eqs:SchrodingerEqs} admit a 3-dimensional continuum of scattering-state solutions, $\psi_{\vb{k}} (\vb{r})$ and $\tilde{\psi}_{\vb{k}} (\vb{p})$, with the wavevector $\vb{k}=\mu \vb{v}_{\rm rel}$ denoting the (classical) momentum of the interacting particles in the CM frame, $\vb{v}_{\rm rel}$ being their relative velocity. The energy eigenvalues ${\cal E}_{\vb{k}} = \vb{k}^2/2\mu$ correspond to the kinetic energy of the system in the CM frame. Depending on the potential, \cref{eqs:SchrodingerEqs} may also admit a discrete spectrum of bound-state solutions, with ${\cal E} <0$ being the binding energy; we shall enumerate these solutions with the discrete quantum number $n$. We analyze the scattering-state and bound-state wavefunctions in partial waves as follows
\begin{subequations}
\label{eq:PW_Wavefun}
\label[pluralequation]{eqs:PW_Wavefun}
\begin{align}
%%%%%%%%%
\label{eq:PW_Wavefun_mom}
\tilde{\psi}_{\vb{k}}(\vb{p}) 
&= \sum_{\ell}(2\ell + 1)
P_\ell(\hat{\vb{k}}\cdot\hat{\vb{p}})
\tilde{\psi}_{k, \ell}(p),
&\quad
\tilde{\psi}_{n\ell m}(\vb{p}) 
&= \tilde{\psi}_{n \ell}(p)
Y_{\ell m} (\Omega_{\vb{p}}),
\\
%%%%%%%%%
\label{eq:PW_Wavefun_pos}
\psi_{\vb{k}}(\vb{r}) 
&= \sum_{\ell}(2\ell + 1)
P_\ell(\hat{\vb{k}}\cdot\hat{\vb{r}})
\psi_{k, \ell}(r),
&\quad
\psi_{n\ell m}(\vb{r}) 
&= \psi_{n \ell}(r)
Y_{\ell m} (\Omega_{\vb{r}}),
\\
%%%%%%%%%
\label{eq:PW_Wavefun_pos-mom}
\psi_{k, \ell}(r) 
&= \dfrac{\im^\ell}{2\pi^2} 
\int_0^\infty \dd p 
\, p^2 
\, \tilde{\psi}_{k,\ell}(p) 
j_\ell (pr) ,
&\quad
\psi_{n\ell}(r) 
&= \dfrac{\im^\ell}{2\pi^2} 
\int_0^\infty \dd p \, p^2 \, 
\tilde{\psi}_{n\ell}(p) 
j_\ell (pr) ,
\\
%%%%%%%%%
\label{eq:PW_Wavefun_pos-mom_inv}
\tilde{\psi}_{k, \ell}(p) 
&= 4\pi(-\im)^\ell 
\int_0^\infty \dd r 
\, r^2 
\, \psi_{k,\ell}(r) 
\, j_\ell (pr) ,
&\quad
\tilde{\psi}_{n\ell}(p) 
&= 4\pi (-\im)^\ell  
\int_0^\infty \dd r 
\, r^2 
\, \psi_{n\ell}(r) 
\, j_\ell (pr) ,
\end{align}
\end{subequations}
where $k = |\vb{k}|$ and $p = |\vb{p}|$. With the above, and setting\footnote{
%%%%
In \cref{eq:WF_u(r)_def}, the extra factor of $k$ in the definition of the rescaled scattering-state wavefunction is standard convention that ensures $u_{k,\ell} (r)$ is dimensionless and has the standard asymptotic behavior discussed in \cref{sec:Jost}.}
%%%%
%
\begin{align}
\label{eq:WF_u(r)_def}
u_{k,\ell} (r) = k \, r \, \psi_{k,\ell} (r)   
\qquad \text{and} \qquad
u_{n\ell} (r) = r\, \psi_{n\ell} (r)  ,
\end{align}
for the scattering and bound states, 
we obtain the radial Schrödinger equation
\begin{align}
\label{eq:SchroedingerEq_Radial}
\left(
-\dfrac{1}{2\mu}\dv[2]{r} 
+\dfrac{\ell(\ell+1)}{2\mu \, r^2} 
\right) u_{\ell} (r) 
+ \int_0^\infty \dd r'
\, r' \, r
\, {\cal V}_\ell(r',r)
\, u_{\ell} (r') 
= {\cal E} \, u_{\ell} (r),
\end{align}
where we omitted the principal quantum number, $k$ or $n$, for generality. Note that for the discrete spectrum, the energy eigenvalues may depend both on $n$ and $\ell$, ${\cal E} \to {\cal E}_{n\ell}$.

\subsubsection*{Central potentials}
If 
${\cal K}^{\twoPI}(\vb{p'},\vb{p}) 
={\cal K}^{\twoPI}(|\vb{p'-p}|)$, 
then we obtain a central potential,
${\cal V}(\vb{r'},\vb{r}) = V(r) \, \delta^3(\vb{r'-r})$,
with\footnote{\label{foot:LocalvsCentralPotentials}
%%%%%%%%%%%
The more relaxed assumption 
${\cal K}^{\twoPI}(\vb{p'},\vb{p}) 
={\cal K}^{\twoPI}(\vb{p'-p})$ 
leads to local but not necessarily central potentials, 
${\cal V}(\vb{r'},\vb{r}) = V(\vb{r}) \, \delta^3(\vb{r'-r})$.
In this case, the scalar kernel ${\cal K}^{\twoPI}(\vb{p'-p})$ must depend also on an external vector, which renders the partial-wave analysis of \cref{eqs:PW_Potential} inapplicable. 
}
%%%%%%%%%%%
%
\begin{align}
\label{eq:CentralPotential}
V(r) 
= -\dfrac{1}{4\mT \mu}
\int \dfrac{\dd^3\vb{q}}{(2\pi)^3} 
\, e^{-\im \vb{q}\cdot\vb{r}}
\,{\cal K}^{\twoPI} (q) 
= -\dfrac{1}{4\mT \mu}
\dfrac{1}{2\pi^2 r}
\int_0^\infty \dd q \, q \, 
{\cal K}^{\twoPI} (q) 
\, \sin (qr).   
\end{align}
Upon analyzing into partial waves according to
\cref{eq:PW_Potential_pos,eq:PW_Potential_mom}, we find
\begin{subequations}
\label{eq:CentralPotential_PW}
\label[pluralequation]{eqs:CentralPotential_PW}
\begin{align}
\label{eq:CentralPotential_PW_Position}
{\cal V}_\ell (r',r) 
&= \dfrac{1}{4\pi}
\int \dd\Omega_{\vb{r'}} \, \dd \Omega_{\vb{r}} 
\, P_{\ell}( \vb{\hat{r}'} \cdot \vb{\hat{r}})
\, {\cal V}(\vb{r'},\vb{r})
= \dfrac{\delta(r'-r)}{r^2}  V(r) ,
\\
\label{eq:CentralPotential_PW_MomentumSpace}
{\cal K}_\ell^{\twoPI} (p',p) 
&= -\mT \mu \int_0^\infty \dd r \, r^2 
\, j_\ell (p'r)\, j_\ell (p r) \, V(r).
\end{align}
\end{subequations}
The above shows that for central potentials, the partial-wave kernels are not independent, since they all derive from the same radial function $V(r)$
~\cite{Yamaguchi:1954mp, Flores:2024sfy}, and therefore cannot, in general, account for the self-energy kernels arising from independent inelastic processes~\cite{Flores:2024sfy}.

\subsubsection*{Decomposition of the potential}

As in \cref{sec:Unitarity_Kernel}, the kernel and the potential can be decomposed in Hermitian and anti-Hermitian components,
\begin{subequations}
\label{eq:KandV_decomp}
\label[pluralequation]{eqs:KandV_decomp}
\begin{align}
{\cal K}^{\twoPI} (\vb{p'},\vb{p}) &= 
{\cal K}_H^{\twoPI} (\vb{p'},\vb{p}) +
{\cal K}_A^{\twoPI} (\vb{p'},\vb{p}),
&
{\cal V} (\vb{r'},\vb{r}) &= 
{\cal V}_H (\vb{r'},\vb{r}) +
{\cal V}_A (\vb{r'},\vb{r}),
\\
{\cal K}_H^{\twoPI} (\vb{p'},\vb{p}) &\equiv
\dfrac{1}{2} \left[
{\cal K}^{\twoPI} (\vb{p'},\vb{p}) + 
{\cal K}^{\twoPI \, \dagger} (\vb{p'},\vb{p})
\right],
&
{\cal V}_H (\vb{r'},\vb{r}) &\equiv
\dfrac{1}{2} \left[
{\cal V} (\vb{r'},\vb{r}) + 
{\cal V}^\dagger (\vb{r'},\vb{r})
\right],
\\
{\cal K}_A^{\twoPI} (\vb{p'},\vb{p}) &\equiv
\dfrac{1}{2} \left[
{\cal K}^{\twoPI} (\vb{p'},\vb{p}) - 
{\cal K}^{\twoPI\, \dagger} (\vb{p'},\vb{p})
\right],
&
{\cal V}_A (\vb{r'},\vb{r}) &\equiv
\dfrac{1}{2} \left[
{\cal V} (\vb{r'},\vb{r}) - 
{\cal V}^\dagger (\vb{r'},\vb{r})
\right],
\end{align}
\end{subequations}
where
\begin{align}
{\cal K}^{\twoPI\, \dagger} (\vb{p'},\vb{p}) = 
{\cal K}^{\twoPI \, *} (\vb{p},\vb{p'})
\qquad \text{and} \qquad
{\cal V}^\dagger (\vb{r'},\vb{r}) = 
{\cal V}^* (\vb{r},\vb{r'}).
\label{eq:KandV_dagger}
\end{align}
We reiterate that, in general, the Hermitian and anti-Hermitian parts of the potentials are not real and imaginary, respectively. This would require TRI, which imposes 
${\cal K}^{\twoPI} (\vb{p},\vb{p'}) = {\cal K}^{\twoPI} (\vb{p'},\vb{p})$ 
and 
${\cal V} (\vb{r},\vb{r'}) = {\cal V} (\vb{r'},\vb{r})$. 
No such assumption is made here.

We analyze the Hermitian and anti-Hermitian components of the kernel and the potential in partial waves analogously to \cref{eqs:PW_Potential}.

\subsection{Hermitian potential \label{sec:Resummation_ReV}}

Considering a Hermitian potential only, ${\cal V}^{(0)}(r',r)$, we denote by
$u_\ell^{(0)}(r)$ the solution to the corresponding radial Schrödinger equation, 
\begin{align}
\label{eq:SchroedingerEq_Real}
{\cal S}_\ell u_{k,\ell}^{(0)}(r) 
=
{\cal E} u_{k,\ell}^{(0)}(r) 
\end{align}
where ${\cal S}_\ell$ is the integro-differential operator
\begin{align}
\label{eq:SchroedingerOperator_Real}
{\cal S}_\ell u_\ell^{(0)}(r) 
\equiv
\left(
-\frac{1}{2\mu}\dv[2]{r} 
+ \frac{\ell(\ell+1)}{2\mu \, r^2} 
\right)u_\ell^{(0)}(r) 
+
\int_0^\infty \dd r' \, r' \, r 
\ {\cal V}^{(0)}_{\ell}(r',r)
\ u_\ell^{(0)}(r') .
\end{align}
In most phenomenological applications, the Hermitian potential is dominated by contributions that are central, and is therefore also real, 
${\cal V}^{(0)}_{\ell}(r',r) \to [V(r)/r^2] \, \delta(r'-r)$, with $V(r)\in\RR$. 
However, non-local contributions to the Hermitian potential can be generated by on-shell and off-shell inelastic interactions, as suggested by the Feshbach projection discussed in \cref{sec:OpticalPotential} (cf.~\cref{eq:Vopt_H}). Having a particular form, such contributions will be included separately, as will be shown in~\cref{sec:Resummation_FullSol}. ${\cal V}^{(0)}_\ell(r',r)$ will thus not be the only part of the Hermitian potential. 

Since ${\cal V}^{(0)}_\ell (r',r)$ does not include contributions from inelastic vertices, it does not involve any threshold crossings. We may therefore make the replacement $\sqrt{s} \to \mT$. With this approximation, ${\cal S}_\ell$ does not depend on ${\cal E}$, and \cref{eq:SchroedingerEq_Real} becomes a linear eigenvalue equation. This, in turn, permits a simple spectral decomposition of the Green's function associated with \cref{eq:SchroedingerEq_Real}, as we discuss below. By contrast, a strong dependence of the Hermitian kernel on the incoming energy would lead to a non-linear eigenvalue problem, requiring a more complex spectral decomposition.

\subsubsection*{Spectrum}

For the baseline Hermitian problem, we will consider ${\cal V}^{(0)}_\ell(r,r')$ to be real and local, as designated above and analyzed in \cref{sec:Jost}. As in \cref{sec:Jost}, we denote by ${\cal F}_{k,\ell}(r)$ and ${\cal G}_{k,\ell}(r)$ the two linearly independent scattering-state solutions of \cref{eq:SchroedingerEq_Real}, corresponding to the regular and irregular families, respectively.  
We also introduce the linear combinations 
${\cal H}_{k,\ell}^{(\pm)}(r)
= {\cal F}_{k,\ell}(r) \pm \im\, {\cal G}_{k,\ell}(r)$, which represent outgoing and incoming waves.
We recall the energy eigenvalues of the scattering-state solutions, ${\cal E}_{\vb{k}} = \vb{k}^2/(2\mu)$. The potential may additionally support bound-state solutions, which we denote by ${\cal B}_{n\ell}(r)$, with ${\cal E}_{n\ell} =-\kappa_{n\ell}^2/(2\mu)$ being the energy eigenvalues. All solutions are characterized by their asymptotic behavior at $r \to 0$ and/or $r \to \infty$. The asymptotic behavior, normalization and analytic properties of these wavefunctions have been defined precisely in \cref{sec:Jost}.

\subsubsection*{Green's function}

We will be interested in the Green's function of the operator corresponding to \cref{eq:SchroedingerEq_Real}, i.e.,~the solution to the differential equation,
\begin{align}\label{eq:GreensDefn_DiffEq}
({\cal S}_\ell - {\cal E}_{\vb{k}})
G_{k,\ell}(r,r')
= \delta(r - r') ,
\end{align}
that behaves as a regular solution at $r\to0$ and an outgoing wave at $r\to\infty$. 
It is possible to explicitly construct the Green's function, and to derive its spectral decomposition using the complete set of eigenstates obtained from solving \cref{eq:SchroedingerEq_Real}. These two methods yield
\begin{subequations}
\label{eq:GreensFunction}
\label[pluralequation]{eqs:GreensFunction}
\begin{empheq}[box=\myshadebox]{align}
G_{k,\ell} (r,r') 
&=
+\dfrac{2\mu \im}{\sym_\ell k} \, 
{\cal F}_{k,\ell}^*(r_<)
{\cal H}_{k,\ell}^{(+)}(r_>) 
\label{eq:GreensFunction_FH}
\\[1ex]
&=
\dfrac{2\mu}{\sym_\ell} \left[
\dfrac{1}{\pi}
\int_{-\infty}^\infty \dd q \ \frac
{{\cal F}_{q,\ell}(r) \, {\cal F}_{q,\ell}^*(r')}
{q^2-k^2-\im \epsilon}     
- \sum_{n} \frac{{\cal B}_{n\ell} (r) {\cal B}_{n\ell}^* (r') }{\kappa_{n\ell}^2+k^2}
\right] ,
\label{eq:GreensFunction_SpectralDecomp}
\end{empheq}
\end{subequations}
where $r_< \equiv \min\{r,r'\}$, $r_> \equiv \max\{r,r'\}$,  $\kappa_{n\ell}^2 =-2\mu {\cal E}_{n\ell} >0$ and  $\epsilon \to 0^+$. \Cref{eqs:GreensFunction} are both very important for the unitarization prescription; their derivation and proof of equivalence can be found in \cref{sec:Jost_Greens}.\footnote{
%%%%%%%%%%%%%%%%%%
We recall from \cref{sec:Jost_Greens} that the integration range $q \in (-\infty,\infty)$ in \cref{eq:GreensFunction_SpectralDecomp} assumes even wavefunctions. For long-range (Coulomb-like) potentials, which do not formally satisfy the convergence condition \eqref{eq:V_ConvergenceCondition}, logarithmic phases can break this evenness and make analytic continuation in complex $q$ branch-dependent. In such cases, it is safer to restrict the integral to the physical region $q>0$, invoking symmetry and complex continuation only when the branch structure is under control.
}
%%%%%%%%%%%%%%%%%%%

\subsection{Anti-Hermitian potential \label{sec:Resummation_ImV}}

\subsubsection*{From the unitarity relation}
Inelastic processes generate anti-Hermitian contributions to the 2PI kernel that must be resummed to obtain the complete wavefunctions.  The anti-Hermitian part of the potential can be deduced from the unitarity relation underpinning the optical theorem~\cite{Flores:2024sfy}, reviewed in \cref{sec:Unitarity}. Although we do not assume TRI, we show the additional relations that would apply if it were imposed.

As in \cite{Flores:2024sfy}, we shall consider for simplicity 2-to-2 inelastic processes only. From \cref{eqs:ImK=AA_PW}, 
\begin{subequations}
\label{eq:K2PI_Im}
\label[pluralequation]{eqs:K2PI_Im}
\begin{align}
{\cal K}^{\twoPI}_{A,\ell} (p',p) 
\overset{\phantom{\TRI}}{=} \
&\im
\sum_j
\dfrac{2k^j(s)}{\sym^j\!\sqrt{s}}
\, {\cal A}_{\ell}^{\inel,j \,*} (p,k^j)
\, {\cal A}_{\ell}^{\inel,j} (p',k^j)
\label{eq:K2PI_Im_1}
\\
\overset{\TRI}{=} \
&\im 
\sum_j
\dfrac{2k^j(s)}{\sym^j\!\sqrt{s}}
\, {\cal A}_{\ell}^{\inel,j} (p,k^j) 
\, {\cal A}_{\ell}^{\inel,j \,*} (p',k^j),
\label{eq:K2PI_Im_2}
\end{align}
\end{subequations}
where the index $j$ denotes the inelastic channel and encompasses all relevant discrete indices that characterize it. $\sym^j$ stands for the symmetry factor of the state arising from the inelastic interaction.  
We recall from \cref{sec:Unitarity_Kernel} that
${\cal A}_{\ell}^{\inel,j} (p,k^j)$ is the irreducible inelastic amplitude (i.e., all initial-state 2PI factors have been amputated), and that the products of the inelastic interaction are on-shell; the magnitude of their momenta, $k^j(s)$, is thus fully determined by the total energy imparted in the system, parametrized by $s$. Any $s$ dependence other than threshold behavior can be eliminated by setting $s \to \mT^2$, consistently with the non-relativistic approximation adopted for the Hermitian potential of \cref{sec:Resummation_ReV}. Threshold behavior, however, must be kept explicit. 
We reiterate that \cref{eq:K2PI_Im_2} assumes TRI and we do not use it here. \Cref{eq:K2PI_Im,eq:PW_Potential_pos-mom} imply that the anti-Hermitian potential in position space is a sum\footnote{
%%%%%%%%%%%%
Inverse decay (2-to-1) processes can also be accommodated by the form \eqref{eq:V_Im} of the imaginary potential, as follows from \cref{eq:ImK=AA_PW_1part}. 
On the other hand, inelastic channels with final states of multiplicity $\geqslant 3$ render the imaginary potential an integro-sum of separable potentials, as seen from \cref{eq:ImK=AA_PW_MultiPart}. While this can be incorporated in the regularization procedure, it introduces technical complications that would obscure the primary goal of the present analysis, which is encoding the non-analytic behavior of inelastic amplitudes. We leave this extension for future work. 
} 
%%%%%%%%%%%%
of \emph{non-local, separable} potentials~\cite{Flores:2024sfy}
\begin{empheq}[box=\myshadebox]{align}
\label{eq:V_Im}
{\cal V}_{A,\ell}(r',r)= 
- \im \sum_j 
(\etaAell^j)^2
\ \nu_\ell^{j}(r') \, \nu_\ell^{j*}(r)
\ \overset{\TRI}{=} \
- \im \sum_j 
(\etaAell^j)^2 
\ \nu_\ell^{j*} (r') \, \nu_\ell^{j}(r) ,
\end{empheq}
where
\begin{subequations}
\label{eq:A-nu_relation}
\label[pluralequation]{eqs:A-nu_relation}
\begin{empheq}[box=\myshadebox]{align}
\etaAell^j \, \nu_\ell^j(r)
\equiv
(-\im)^\ell 
\sqrt{\dfrac{8k^j(s)}{\sym^j\pi^2 \mT^2 \mu}}
\int_0^\infty
\dd p \, p^2 \, j_\ell(pr)
\, {\cal A}_{\ell}^{\inel, j}(p,k^j),
\label{eq:nu_def}
\end{empheq}
with the inverse being
\begin{align}
{\cal A}_{\ell}^{\inel, j}(p,k^j)
&=
\im^\ell
\ \etaAell^j 
\sqrt{\dfrac{\sym^j \mT^2 \mu}{2 \, k^j(s)}}
\int_0^\infty \dd r
\ r^2
\ j_\ell(pr)
\ \nu_\ell^j(r).
\label{eq:nu_def_inv}
\end{align}
\end{subequations}
In \cref{eq:V_Im,eq:A-nu_relation}, $\etaAell^j \in \RR$ parametrize the strength of the anti-Hermitian potentials; we have factored these parameters outside the functions $\nu_\ell^j (r)$ for later convenience.\footnote{
%%%%%%%%%
Allowing for complex values of $\etaA^j$ would mean that $(\etaAell^j)^2$ in \cref{eq:V_Im} should be replaced by $|\etaAell^j|^2$. Since all of our results depend on $|\etaAell^j|^2$, we set $\etaA^j \in \RR$ and include complex phases in the $\nu_\ell^j (r)$ factors, without loss of generality. 
The factors $\etaAell^j \nu_\ell^j(r)$ are specified by \cref{eq:K2PI_Im,eq:V_Im}, and \cref{eq:PW_Potential_pos-mom} that relates them, up to an $r$-independent phase. In \cref{eq:nu_def}, the phase $(-\im)^\ell$ is chosen to simplify subsequent expressions, in particular \cref{eqs:Minel_def} below.} 
%%%%%%%%%
For inelastic channels that go on-shell only for $\sqrt{s}>\mT$, the $s$ dependence due to threshold crossing should be included in $\etaAell^j$; we leave, however, any such dependence implicit. 
Each inelastic channel $j$ thus contributes a rank-one separable kernel.

\subsubsection*{From the continuity equation and LSZ reduction}

The form \eqref{eq:V_Im} of the imaginary potential with the specification of \cref{eq:nu_def} is also supported by the continuity equation and LSZ reduction. Considering the probability current,  
$\vb{j}_{\vb{k}} (\vb{r})\equiv 
\Im [\psi_{\vb{k}}^* (\vb{r}) 
\nabla \psi_{\vb{k}} (\vb{r})] / \mu$, 
the total inelastic cross-section is the loss of flux through a closed surface $S$, normalized to the incoming flux,
\begin{subequations}
\label{eq:InelCrossSe}
\label[pluralequation]{eqs:InelCrossSe}
\begin{align}
\sigma^\inel 
&= -\dfrac{\mu}{k} \oint_S \dd\vb{S} \cdot \vb{j}_{\vb{k}} (\vb{r})
= -\dfrac{\mu}{k} \int \dd^3 \vb{r}    
\, \nabla \cdot \vb{j}_{\vb{k}} (\vb{r})
\label{eq:InelCrossSec_divOFj}
\\[0.5ex]
&= -\dfrac{\mu}{k} 
\int \dd^3 \vb{r} \, \dd^3 \vb{r'} 
\, 2 \Im \left[ 
\psi_{\vb{k}}^* (\vb{r}) 
\,{\cal V} (\vb{r'},\vb{r})
\,\psi_{\vb{k}} (\vb{r'})
\right] 
\label{eq:InelCrossSec_ImV}
\\[0.5ex]
&=+\dfrac{1}{4\mT k} 
\int 
\dfrac{\dd^3 \vb{p }}{(2\pi)^3} 
\dfrac{\dd^3 \vb{p'}}{(2\pi)^3} 
\, 2 \Im \left[ 
\tilde\psi_{\vb{k}}^* (\vb{p}) 
\,{\cal K}^{\twoPI} (\vb{p'},\vb{p})
\,\tilde\psi_{\vb{k}} (\vb{p'})
\right] 
\label{eq:InelCrossSec_ImK}
\\[0.5ex]
&=-\dfrac{\im}{4\mT k} 
\int 
\dfrac{\dd^3 \vb{p }}{(2\pi)^3} 
\dfrac{\dd^3 \vb{p'}}{(2\pi)^3} 
\,\tilde\psi_{\vb{k}}^* (\vb{p}) 
\left[
 {\cal K}^{\twoPI} (\vb{p'},\vb{p})
-{\cal K}^{\twoPI *} (\vb{p},\vb{p'})
\right]
\,\tilde\psi_{\vb{k}} (\vb{p'}) 
\label{eq:InelCrossSec_Kdiff}
\\
&=-\dfrac{\im}{4\mT k} 
\times 2
\int 
\dfrac{\dd^3 \vb{p }}{(2\pi)^3} 
\dfrac{\dd^3 \vb{p'}}{(2\pi)^3} 
\,\tilde\psi_{\vb{k}}^* (\vb{p}) 
\,{\cal K}_A^{\twoPI} (\vb{p'},\vb{p})
\,\tilde\psi_{\vb{k}} (\vb{p'}) ,
\label{eq:InelCrossSec_KA}
\end{align}
\end{subequations}
where \eqref{eq:InelCrossSec_divOFj} arises from Gauss' divergence theorem, \eqref{eq:InelCrossSec_ImV} expresses the continuity equation obtained using the Schrödinger \cref{eq:SchrodingerEqs}, and 
\eqref{eq:InelCrossSec_ImK} is found by Fourier transforming according to \cref{eq:FT_wavefunctions,eq:FT_potential}. 
To reach \eqref{eq:InelCrossSec_Kdiff}, we expanded $\Im[\cdot]$ in terms of the difference between the conjugate quantities and swapped integration variables. Finally, we used \cref{eq:KandV_decomp,eq:KandV_dagger}.

On the other hand, the inelastic amplitudes can be expressed as the convolution of the irreducible inelastic amplitude with the wavefunction of the incoming state (see e.g.~\cite{Petraki:2015hla} for derivation via LSZ reduction),
\begin{align}
{\cal M}^{\inel,j} (\vb{k},\tau^j) = 
\int \dfrac{\dd^3\vb{p}}{(2\pi)^3} 
\, \tilde{\psi}_{\vb{k}} (\vb{p})
\, {\cal A}^{\inel,j} (\vb{p},\tau^j) ,
\label{eq:AmplitudeFull=psi*A}
\end{align}
with the total inelastic cross-section given by the standard expression
\begin{align}
\sigma^\inel (\vb{k}) = 
\dfrac{1}{4\mT k}
\sum_j
\int \dd_\subsqrts\tau^j \, |{\cal M}^{\inel,j}(\vb{k},\tau_j)|^2 ,   
\label{eq:sigma_def}
\end{align}
where the phase-space element $\dd_\subsqrts\tau^j$ has been defined in \cref{eq:PhaseSpaceMeasure_def}, and we used the non-relativistic approximation for the flux factor, $\sqrt{s} \to \mT$.
Comparing \cref{eq:InelCrossSec_KA} to \cref{eq:AmplitudeFull=psi*A,eq:sigma_def}, 
we can identify 
\begin{align} 
{\cal K}_A^{\twoPI} (\vb{p'},\vb{p})
= \dfrac{\im}{2}
\sum_j
\int \dd_\subsqrts\tau^j
\, {\cal A}^{\inel,j \,*} (\vb{p},\tau^j)
\, {\cal A}^{\inel,j} (\vb{p'},\tau^j),
\label{eq:ImK_ContEq}
\end{align}
which agrees exactly with the expression \eqref{eq:UnitarityRelation_Kernel_ac} deduced from the unitarity relation.

\subsection{Full potential and solution of the Schrödinger equation \label{sec:Resummation_FullSol}}

Inelastic amplitudes that do not decrease at high momenta generate separable potentials of the form \eqref{eq:V_Im} that lead to divergences and require renormalization. In \cref{sec:Renorm}, we show that renormalizing these potentials requires Hermitian counterterms. This is expected, since inelastic interactions contribute Hermitian terms to the 2PI kernel, as in the optical potential obtained via the Feshbach projection in \cref{sec:OpticalPotential}. Under certain approximations valid for contact inelastic interactions, which are the source of the divergences, the induced Hermitian potentials have the same momentum dependence as their anti-Hermitian counterparts, but different couplings; this was shown in \cref{sec:OpticalPotential_H}. To include these terms and treat the divergences properly, we extend \cref{eq:V_Im} to complex couplings. The full potential we consider is, therefore,
\begin{empheq}[box=\myshadebox]{align}
\label{eq:V_Tot}
{\cal V}_{\ell}(r',r)=
{\cal V}^{(0)}_{\ell}(r',r)
-\sum_j (\eta_\ell^j)^2 \, 
\nu_\ell^{j}(r') \, \nu_\ell^{j*}(r),
\end{empheq}
with $[{\cal V}^{(0)}_{\ell}(r',r)]^\dagger ={\cal V}^{(0)}_{\ell}(r',r)$ as in \cref{sec:Resummation_ReV},  
and 
\begin{subequations}
\label{eq:etas_def}
\label[pluralequation]{eqs:etas_def}
\begin{align}
\eta_\ell^j \equiv \etaRell^j + \im \etaIell^j,
\label{eq:eta_def}
\end{align}
\begin{align}
\etaRell^j \equiv 
\qty(
\dfrac
{\sqrt{(\etaHell^j)^2+(\etaAell^j)^4}+\etaHell^j}
{2}
)^{1/2} 
\quad \text{and} \quad 
\etaIell^j \equiv 
\qty(
\dfrac
{\sqrt{(\etaHell^j)^2+(\etaAell^j)^4}-\etaHell^j}
{2}
)^{1/2}  
,
\label{eq:etaRI_def}
\end{align}
such that 
\begin{empheq}[box=\myshadebox]{align}
(\eta_\ell^j)^2 
=\etaHell^j+\im (\etaAell^j)^2 ,   
\label{eq:eta_squared}
\end{empheq}
\end{subequations}
with $\etaHell^j, \etaAell^j \in \RR$, where $\etaHell^j$  parametrizes the strength of the Hermitian separable potential. This permits $\etaHell^{j}\neq0$ even if $\etaAell^{j}=0$, and accommodates either $\etaHell^{}$ sign, as motivated by \cref{eq:OpticalPot_HermitianCoupling}.

Considering this potential, the radial Schrödinger equation \eqref{eq:SchroedingerEq_Radial} for scattering states becomes
\begin{align}\label{eq:SchroedingerEq}
({\cal S}_{\ell} - {\cal E}_{\bf k})\, u_{k,\ell} (r) 
=+ 
\sum_j
\left\{  
(\eta_\ell^j)^2 \ 
r \, \nu^{j*}_{\ell} (r) 
\int_0^\infty \dd r' \, 
r'\, \nu^j_{\ell}  (r')
\, u_{k,\ell} (r')
\right\}.
\end{align}
To solve it, we proceed as in Ref.~\cite{Flores:2024sfy}. 
We first define
\begin{subequations}
\label{eq:Mhat_def}
\label[pluralequation]{eqs:Mhat_def}
\begin{align}
\hat{M}_{\ell,\unreg}^j (k) &\equiv 
\sqrt{\dfrac{2\mu}{\sym_\ell k}}
\int_0^\infty \dd r\ r\ {\cal F}_{k,\ell}(r)\nu_\ell^j (r) ,
\label{eq:Mhat_unreg_def}
\\
\hat{M}_{\ell,\reg}^j (k) &\equiv 
\sqrt{\dfrac{2\mu}{\sym_\ell k}}
\int_0^\infty \dd r\ r\ u_{k,\ell}(r)\nu_\ell^j (r),
\label{eq:Mhat_reg_def}
\end{align}
\end{subequations}
where the pre-factor of the integrals is introduced for later convenience. With this, we can write an implicit solution of \cref{eq:SchroedingerEq} as follows
\begin{align}
\label{eq:Wavefunction_sol_implicit}
u_{k,\ell}(r) = {\cal F}_{k,\ell}(r) 
+ 
\sum_j
\qty[ 
(\eta_\ell^j)^2 
\int_0^\infty \dd r' r' 
G_{k,\ell} (r,r') 
\nu_\ell^{j*} (r') ]
\sqrt{\dfrac{\sym_\ell k}{2\mu}}
\hat{M}_{\ell,\reg}^{j} (k) ,
\end{align}
Defining the \emph{regularization matrix} $\NN_\ell(k)$,
\begin{empheq}[box=\myshadebox]{align}
\label{eq:Nmatrix_def}
[\NN_\ell(k)]^{ij}
\equiv
\delta^{ij} 
- 
\eta_\ell^i \eta_\ell^j \, 
\int_0^\infty \dd r  \, r
\int_0^\infty \dd r' \, r' 
\left[ \nu_\ell^i(r) 
\, G_{k,\ell}(r,r')
\, \nu_\ell^{j*}(r') \right] ,
\end{empheq}
\cref{eq:Wavefunction_sol_implicit} implies
\begin{align}
\sum_j [\NN_\ell (k)]^{ij} \eta_\ell^j \hat{M}_{\ell,\reg}^{j} (k) 
= \eta_\ell^i\hat{M}_{\ell,\unreg}^{i} (k), 
\label{eq:N*Mreg=Munreg_IndicesFormat}
\end{align}
Inverting \cref{eq:N*Mreg=Munreg_IndicesFormat}, we obtain the solution to \cref{eq:SchroedingerEq},
\begin{empheq}[box=\myshadebox]{align}
\label{eq:Wavefunction_sol}
&u_{k,\ell}(r) = {\cal F}_{k,\ell}(r) 
\nn \\
&+ 
\sum_{i,j} \eta_\ell^i \eta_\ell^j
\left(\int_0^\infty
\dd r''\, r'' \, 
G_{k,\ell}(r,r'')
\nu_\ell^{i*}(r'')
\right)
\left[\NN_\ell^{-1}(k)\right]^{ij}
\left(\int_0^\infty \dd r' \, r' \, 
{\cal F}_{k,\ell} (r')
\nu_\ell^j (r')
\right) .
\end{empheq}
\Cref{eq:Wavefunction_sol} generalizes the solution of Ref.~\cite{Flores:2024sfy}, to include both Hermitian and anti-Hermitian separable potentials, as given by \cref{eq:V_Tot}.

The sum of separable potentials in \cref{eq:V_Tot} is an operator of rank lower or equal to the number of inelastic channels (assuming $\nu_\ell^j(r) \in L^2$). By Weyl's theorem on the essential spectrum~\cite{kato1980perturbation}, a finite-rank perturbation of an operator leaves its essential spectrum unchanged. In particular, the continuous spectrum of ${\cal S}_\ell$ is preserved. Consequently, the energies of the scattering eigenstates of the full Hamiltonian span the same continuum as those of ${\cal S}_\ell$, while the eigenfunctions are modified according to \cref{eq:Wavefunction_sol}.

\subsection{Inelastic amplitudes: definitions and matrix notation \label{sec:Resummation_InelAmplitudesDef}}

To proceed, we first define the following inelastic amplitudes. 

\subsubsection*{Scattering-state inelastic scatterings}

We define the \emph{unregulated} and \emph{regulated} inelastic amplitudes as
\begin{subequations}
\label{eq:Minel_def}
\label[pluralequation]{eqs:Minel_def}
\begin{align}
{\cal M}_{\ell,\unreg}^{\inel, j}(q)
&\equiv
\dfrac{1}{2\pi^2}
\int_0^\infty \! \dd p \, p^2 
\, \tilde\psi_{q,\ell}^{(0)} (p)
\, {\cal A}_\ell^{\inel, j} (p,k^j)
=
\dfrac{\etaAell^j}{q} 
\sqrt{\dfrac{\sym^j \mT^2\mu}{2k^j(s)}}
\int_0^\infty \! \dd r \, r 
\, {\cal F}_{q,\ell}(r) 
\, \nu_\ell^j(r),
\label{eq:Minel_def_unreg}
\\[1ex]
%%%%%%%%%%%%%%
{\cal M}_{\ell,\reg}^{\inel, j}(q)
&\equiv
\dfrac{1}{2\pi^2}
\int_0^\infty \! \dd p \, p^2 
\, \tilde\psi_{q,\ell} (p)
\, {\cal A}_\ell^{\inel, j} (p,k^j)
=
\dfrac{\etaAell^j}{q} \sqrt{\dfrac{\sym^j \mT^2\mu}{2k^j(s)}}
\int_0^\infty \! \dd r \, r 
\, u_{q,\ell}(r) 
\, \nu_\ell^j(r),
\label{eq:Minel_def_reg}
\end{align}
\end{subequations}
where the former neglects the separable potentials and the latter includes them. Both include the potential of \cref{sec:Resummation_ReV}. 
The rescaled versions of \cref{eqs:Minel_def} can be obtained according to \cref{eq:Amplitudes_Rescaled_def},
\begin{align}
M_{\ell,{\rm (un)reg}}^{\inel, j} (q) \equiv 
\sqrt{\dfrac{4qk^j(s)}{\sym_\ell\sym^j \mT^2}} \, 
{\cal M}_{\ell,{\rm (un)reg}}^{\inel, j} (q) .
\label{eq:M_def}
\end{align}
Moreover, we define the `hatted' versions of the inelastic amplitudes by factoring out the absorptive coupling constants; this form will be useful for handling in a unified manner the effect of both the dispersive and absorptive separable potentials,
\begin{subequations}
\label{eq:Mhat-M_relation}
\label[pluralequation]{eqs:Mhat-M_relation}
\begin{align}
{\cal M}_{\ell,{\rm (un)reg}}^{\inel, j}(q) 
&= \etaAell^j \, 
\hat{\cal M}_{\ell,{\rm (un)reg}}^{j}(q) ,
\label{eq:Mhat-M_relation_cal}
\\
M_{\ell,{\rm (un)reg}}^{\inel,j}(q) 
&= \etaAell^j \, 
\hat{M}_{\ell,{\rm (un)reg}}^{j}(q) .
\label{eq:Mhat-M_relation_straight}
\end{align}
\end{subequations}
\Cref{eq:Mhat-M_relation_cal} constitutes the definition of $\hat{\cal M}_{\ell,{\rm (un)reg}}^{j}(q)$, while \cref{eq:Mhat-M_relation_straight} arises from \cref{eq:Mhat_def,eq:M_def,eq:Minel_def}.

Note that in \cref{eq:Minel_def,eq:M_def,eq:Mhat-M_relation,eq:N-Mreg-Munreg_MatrixFormat}, the momentum $q$ is, in general, off shell; we shall use the off-shell amplitudes in \cref{sec:Resummation_Wmatrix}. The off-shell momenta $q$ are independent of $s$.  
When on-shell, we typically denote the momentum by $k$, which is related to the total energy of the system via $\sqrt{s} = \mT + {\cal E}_{\vb{k}}$. We reiterate that the momenta of the products of the inelastic interactions, $k^j(s)$, are on-shell, hence determined by $s$. 

\subsubsection*{Bound-state decay}

The inelastic vertices generating the inelastic amplitudes \eqref{eqs:Minel_def} also induce decay of the bound-state spectrum of \cref{eq:SchrodingerEqs}. In analogy to \cref{eq:Minel_def_unreg,eq:Mhat-M_relation_cal}, we define the unregulated decay amplitudes of the $n\ell$ bound state into the $j$ channel ,  
\begin{subequations}
\label{eq:M_BSD_def}
\label[pluralequation]{eqs:M_BSD_def}
\begin{align}
{\cal M}_{n\ell,\unreg}^{\BSD, j}
&\equiv
\dfrac{1}{2\pi^2 \sqrt{2\mu}}
\int_0^\infty \dd p \, p^2 
\, \tilde\psi_{n\ell}^{(0)} (p)
\, {\cal A}_\ell^{\inel, j} (p,k^j)
=
\etaAell^j 
\, \hat{\cal M}_{n\ell,\unreg}^{\BSD,j} ,
\label{eq:M_BSDfull_def}
\\
\hat{\cal M}_{n\ell,\unreg}^{\BSD,j}
&\equiv
\sqrt{\dfrac{\sym^j \mT^2}{4k^j(s)}}
\int_0^\infty \!\! \dd r \ r
\ {\cal B}_{n\ell}(r)\nu_\ell^j(r).
\label{eq:M_BSDhat_def}
\end{align}
\end{subequations}
where we note the extra factor $1/\sqrt{2\mu}$ in \cref{eq:M_BSDfull_def} with respect to \eqref{eq:Minel_def_unreg} (see e.g.~\cite{Petraki:2015hla}). 
For completeness, we note that the corresponding unregulated bound-state partial-decay widths are 
\begin{align}
\Gamma_{n\ell,\unreg}^{\BSD,j} 
&=
32\pi (2\ell+1) \, 
\dfrac
{k^{\BSD,j}}
{\sym^j (\mT - |{\cal E}_{n\ell}|)^2}
\ |{\cal M}_{n\ell,\unreg}^{\BSD, j}|^2 
\nn 
\\
&\simeq
8\pi (2\ell+1) \, 
(\etaAell^j)^2
\qty|
\int_0^\infty \!\! \dd r \ r
\ {\cal B}_{n\ell}(r)\nu_\ell^j(r)
|^2 ,
\label{eq:Gamma_BSD_def}
\end{align}
where $\mT-|{\cal E}_{n\ell}|$ is the bound-state mass, and $k^{\BSD,j} \equiv k^j\qty(s\to (\mT-|{\cal E}_{n\ell}|)^2)$. In the second step, we assumed $|{\cal E}_{n\ell}| \ll \mT$, and neglected threshold corrections due to the binding energy. We shall not use \cref{eq:Gamma_BSD_def} in the following.

\subsubsection*{Matrix notation}

For convenience in the following, we will opt for matrix notation, wherever possible. The inelastic amplitudes, in the various versions defined above, will constitute vectors in the space spanned by inelastic channels. We also define the diagonal matrices $\ee$ and $\eeA$, with elements
\begin{align}
\ee^{ij} \equiv \eta_\ell^i \delta^{ij},
\qquad
\eeA^{ij} \equiv \etaAell^i \delta^{ij} ,
\qquad 
\eeH^{ij} \equiv \etaHell^i \delta^{ij}.
\label{eq:eta_matrices_def}
\end{align}
We implicitly restrict to channels for which $\eta_\ell^j\neq 0$.
It follows from \cref{eqs:etas_def}
\begin{align}
\eeA^\dagger = \eeA^{},
\qquad
\eeH^\dagger = \eeH^{},
\qquad
\ee^2 =\eeH^{} +\im \eeA^2. 
\label{eq:eta_matrices_identities}
\end{align}

\subsection{The regularization matrix and its inverse: Jost and spectral representations \label{sec:Resummation_Nmatrix}}

To compute the regulated inelastic and elastic cross-sections from \cref{eq:N-Mreg-Munreg_MatrixFormat,eq:ImaginaryPhaseShift}, we first examine regularization matrix, $\NN_\ell$, defined in  \cref{eq:Nmatrix_def}. Using the two equivalent expressions \eqref{eq:GreensFunction} for the Green's function, and the unregulated inelastic amplitudes \eqref{eq:Mhat_def}, we find
\begin{empheq}[box=\myshadebox]{align}
\label{eq:Nmatrix_Expansion}
\NN_\ell(k)
=
\ID
-\im 
\, \ee 
\, \hat{M}_{\ell, \unreg}^{}(k)
\, \hat{M}_{\ell, \unreg}^{\dagger}(k)
\, \ee 
+ \ee \, \WW_\ell(k) \, \ee ,
\end{empheq}
where the matrix $\WW_\ell(k)$ is defined as (cf.~\cref{eq:H+-wrtFG})
\begin{subequations}
\label{eq:Wmatrix_def}
\label[pluralequation]{eqs:Wmatrix_def}
\begin{empheq}[box=\myshadebox]{align}
\label{eq:Wmatrix_FG}
&\WW_\ell^{ij} (k)
\equiv
\frac{2\mu}{\sym_\ell k}
\int_0^\infty \dd r\ r\int_0^\infty \dd r'\ r'
{\cal F}_{k,\ell}^*(r_<) \,
{\cal G}_{k,\ell}(r_>) \, 
\nu_\ell^i(r) \, 
\nu_\ell^{j*}(r')
\\[1em]
%%%%%%%%%%%%%%%%%%%%%
\label{eq:Wmatrix_PV}
&=
\! \dfrac{4}{\sym_\ell \mT^2} \!
\sqrt{\dfrac{k^i(s) k^j(s)}{\sym^i \sym^j}}
\qty[ 
-\dfrac{\PV}{\pi}
\!\!
\int_{-\infty}^{\infty} 
\!
\dd q \, q^2 \,
\dfrac{
\hat{\cal M}^{i}_{\ell,\unreg} (q)  
\hat{\cal M}^{j *}_{\ell,\unreg} (q)}
{q^2-k^2} 
+
\! \sum_{n} \!
\dfrac{
\hat{\cal M}_{n\ell,\unreg}^{\BSD,i} 
\hat{\cal M}_{n\ell,\unreg}^{\BSD,j \, *} 
}
{(\kappa_{n\ell}^2 +k^2) /(2\mu)}
],
\end{empheq}
\end{subequations}
where \cref{eq:Wmatrix_PV} was obtained considering that the off-shell amplitudes $\hat{\cal M}_{\ell,\unreg}^{j} (q)$ do not have singularities on the real $q$ axis,\footnote{
%%%%%%%%
We draw attention to the fact that, in the prefactors appearing in \cref{eq:Wmatrix_PV}, the on-shell momenta of the products of the inelastic interactions, $k^j$, are fixed by $s$, which itself is determined by the on-shell initial-state momentum $k$. On the other hand, the amplitudes appearing in the integrand and the sum are evaluated at off-shell initial-state momenta $q$. This subtlety is particularly important when $k^j(s)$ are sensitive to $\sqrt{s}-\mT$, as is the case for bound-state formation~\cite{Flores:2026yay}. 
}
%%%%%%%%%
%
and using the decomposition
\begin{align}
\dfrac{1}{q^2-k^2-\im \epsilon} = 
+\im \pi \delta(q^2-k^2)
+\PV \ \dfrac{1}{q^2-k^2}.
\label{eq:propagator=delta+PV}
\end{align}

Taking into account that ${\cal F}_{k,\ell}$ and ${\cal G}_{k,\ell}$ are real up to the same $r$-independent phase (cf.~\cref{eq:rephasing_FG}), it is easy to show that $\WW_\ell(k)$ is Hermitian,
\begin{empheq}[box=\myshadebox]{align}
\WW_\ell^\dagger (k) = \WW_\ell(k) .
\label{eq:Wmatrix_Hermitian}
\end{empheq}
We emphasize that the Hermiticity of $\WW_\ell$ does \emph{not} presuppose TRI.

The inverse of \cref{eq:Nmatrix_Expansion} is obtained from the Sherman–Morrison formula~\cite{ShermanMorrison:1950},
\begin{align}
\label{eq:Nmatrix_Inverse}
\NN_\ell^{-1}
=
(\ID +\ee \, \WW_\ell \, \ee)^{-1}
+\im 
\frac{
(\ID +\ee \, \WW_\ell \, \ee)^{-1}
\ee 
\, \hat{M}_{\ell, \unreg}^{}
\, \hat{M}_{\ell, \unreg}^{\dagger}
\, \ee 
(\ID +\ee \, \WW_\ell \, \ee)^{-1}
}
{
1 - \im 
\, \hat{M}_{\ell, \unreg}^{\dagger}
\, \ee 
(\ID +\ee \, \WW_\ell \, \ee)^{-1}
\, \ee 
\, \hat{M}_{\ell, \unreg}^{}
}.
\end{align}
As we shall discuss in \cref{sec:Resummation_Wmatrix}, the $\WW$ matrix encodes the non-analytic and non-convergent behavior of the inelastic amplitudes in the complex momentum plane. \Cref{eq:Nmatrix_Inverse} generalizes the simpler expression found in Ref.~\cite{Flores:2024sfy} in the limit $\WW_\ell \to 0$ and $\etaHell \to 0$.

\subsection{Regulated phase shift and inelastic amplitudes \label{sec:Resummation_RegAmplitudes}}

\subsubsection*{Phase shift}

To compute the \emph{regulated} elastic cross-section, arising from the full potential \eqref{eq:V_Tot}, we need the phase shift for the solution of \cref{eq:Wavefunction_sol}, found by expanding in the $r\to\infty$ limit. Considering the asymptotic behaviors for ${\cal F}_{k,\ell}(r)$ and $G_{k,\ell}(r,r')$, given in~\cref{eq:WF_AsymptotesInf,eq:AnalyticGreens}, we obtain
\begin{align}
u_{k,\ell}(r) 
~~\overset{r\to\infty}{\longrightarrow}~~
+\dfrac{\sqrt{\sym_\ell}}{2\im}
\qty(
 e^{ \im k r} e^{2\im \Delta_\ell(k)}
-e^{-\im (k r - \ell \pi)}
),
\end{align}
with 
\begin{align}
\Delta_\ell (k) \equiv \theta_\ell (k) + \delta_\ell (k), 
\label{eq:Delta_ell}
\end{align} 
where $\theta_\ell (k)$ is the phase shift of ${\cal F}_{k,\ell}$, and
\begin{align}\label{eq:ImaginaryPhaseShift}
e^{2\ii\delta_\ell (k)} = 1
+2 \im 
\, \hat{M}_{\ell, \unreg}^{\dagger}(k)
\, \ee 
\, \NN_\ell^{-1}(k)
\, \ee 
\, \hat{M}_{\ell, \unreg}^{}(k).
\end{align}
Note that $\delta_\ell (k)$ is in general complex.

\subsubsection*{Inelastic amplitudes}

The regulated and unregulated amplitudes, in all four versions introduced earlier,  
\begin{align*}
{\cal M}_{\ell}^{\inel}, \quad 
\hat{\cal M}_{\ell}^{}, \quad
M_{\ell}^{\inel}, \quad
\hat{M}_{\ell},    
\end{align*}
are related by the regularization matrix defined in \cref{eq:Nmatrix_def}, according to \cref{eq:N*Mreg=Munreg_IndicesFormat} and its inversion. Expressed in matrix form, it reads
\begin{subequations}
\label{eq:N-Mreg-Munreg_MatrixFormat}
\label[pluralequation]{eqs:N-Mreg-Munreg_MatrixFormat}
\begin{align}
\hat{M}_{\ell,\unreg} (k) &=
\ee^{-1} \, \NN_\ell(k) \, \ee 
\, \hat{M}_{\ell,\reg} (k) ,
\label{eq:N*Mreg=Munreg_MatrixFormat}
\\
\hat{M}_{\ell,\reg} (k) &=
\ee^{-1} \, [\NN_\ell(k)]^{-1} \, \ee 
\, \hat{M}_{\ell,\unreg} (k) .
\label{eq:Munreg=Ninv*Mreg_MatrixFormat}
\end{align}
\end{subequations}

\subsubsection*{In terms of the $\WW_\ell$ matrix}

The form \eqref{eq:Nmatrix_Inverse} of $\NN_\ell^{-1}$ enables us to express the complex phase shift $\delta_\ell$ and the regulated inelastic amplitudes, \cref{eq:ImaginaryPhaseShift,eq:Munreg=Ninv*Mreg_MatrixFormat}, as follows
\begin{subequations}
\label{eq:MregAndPhaseShift_solution}
\label[pluralequation]{eqs:MregAndPhaseShift_solution}
\begin{empheq}[box=\myshadebox]{align}
e^{2\ii\delta_\ell}
&=
\frac{
1 + \im
\, \hat{M}_{\ell, \unreg}^{\dagger}
\, \ee 
(\ID +\ee \, \WW_\ell \, \ee)^{-1}
\, \ee 
\, \hat{M}_{\ell, \unreg}^{}
}{
1 - \im 
\, \hat{M}_{\ell, \unreg}^{\dagger}
\, \ee 
(\ID +\ee \, \WW_\ell \, \ee )^{-1}
\, \ee
\, \hat{M}_{\ell, \unreg}^{}
},
\label{eq:PhaseShift_solution}
%%%%%%
\\[1em]
\hat{M}_{\ell, \reg}^{}
&=
\ee^{-1}
\, \dfrac{
(\ID +\ee \, \WW_\ell \, \ee)^{-1}
\, \ee 
\, \hat{M}_{\ell, \unreg}^{}
}{
1 - \im
\, \hat{M}_{\ell, \unreg}^{\dagger}
\, \ee
(\ID +\ee \, \WW_\ell \, \ee)^{-1}
\, \ee
\, \hat{M}_{\ell, \unreg}^{}
}.
\label{eq:Mreg_solution}
\end{empheq}
\end{subequations}
From these results, we can derive expressions for the regulated elastic and inelastic cross-sections. To cast them into a more practical form, we define
\begin{subequations}
\label{eq:wParameters_def}
\label[pluralequation]{eqs:wParameters_def}
\begin{align}
w_{A,\ell}^j &\equiv
\left|\left[
\eeA
\, \ee^{-1} \, (\ID + \ee \, \WW_\ell \, \ee)^{-1} 
\, \ee \hat{M}_{\ell,\unreg}
\right]^j\right|^2 ,
\label{eq:wAj_def}
\\
w_\ell^{} &\equiv
-\dfrac{\im}{2}
\hat{M}_{\ell,\unreg}^{\dagger} 
\qty[\ee (\ID+\ee \, \WW_\ell \, \ee)^{-1}\ee - \hc]
\, \hat{M}_{\ell,\unreg}^{} ,
\label{eq:w_def}
\\
\tilde{w}_\ell^{} &\equiv
+\dfrac{1}{2}
\hat{M}_{\ell,\unreg}^{\dagger} 
\qty[\ee (\ID+\ee \, \WW_\ell \, \ee)^{-1}\ee + \hc]
\, \hat{M}_{\ell,\unreg}^{} ,
\label{eq:wtilde_def}
\end{align}
\end{subequations}
where by construction  $w_{A,\ell}^j,w_\ell,\tilde{w}_\ell \in \RR$. Moreover, $\sum_j w_{A,\ell}^j = w_\ell$ (we provide the proof in \cref{app:Math_Proof}). 
With these definitions, \cref{eqs:MregAndPhaseShift_solution} yield
\begin{subequations}
\label{eq:MregAndPhaseShift_solution_Simpler}
\label[pluralequation]{eqs:MregAndPhaseShift_solution_Simpler}
\begin{align}
e^{2\ii\delta_\ell}
&=
\dfrac
{1- (w_\ell- \im \tilde{w}_\ell)}
{1+ (w_\ell- \im \tilde{w}_\ell)}
=
\dfrac{1 
- w_\ell^2 
- \tilde{w}_\ell^2 
+ \im 2\tilde{w}_\ell}
{(1 + w_\ell)^2 + \tilde{w}_\ell^2} ,
\label{eq:PhaseShift_solution_Simpler}
\\
%%%%%%
\qty|M_{\ell, \reg}^{\inel,j}|^2
&=
\dfrac{w_{A,\ell}^j}
{(1 + w_\ell)^2 + \tilde{w}_\ell^2} .
\label{eq:Mreg_solution_Simpler}
\end{align}
\end{subequations}

\subsection{Regulated cross-sections \label{sec:Resummation_RegCrossSec}}

In the following, all quantities are calculated at on-shell momenta; for brevity, we do not denote the momentum argument explicitly.  
For convenience, we first define the cross-sections normalized to the unitarity limit,
\begin{subequations}
\label{eq:xyw_def}
\label[pluralequation]{eqs:xyw_def}
\begin{align}
x_{\ell, {\rm (un)reg}}^{} &\equiv 
\sigma_{\ell, {\rm (un)reg}}^\elas  /
\sigma_\ell^U ,
\label{eq:x_def}
\\
y_{\ell, {\rm (un)reg}}^j &\equiv 
\sigma_{\ell, {\rm (un)reg}}^{\inel,j}  /
\sigma_\ell^U ,
\label{eq:yj_def}
\\
y_{\ell, {\rm (un)reg}}^{} 
&\equiv \sum_j y_{\ell, {\rm (un)reg}}^j ,
\label{eq:y_def}
\end{align}
\end{subequations}
\Cref{eq:Mreg_solution_Simpler} gives $y_{\ell,\reg}^j$. To obtain $x_{\ell,\reg}$ from \cref{eq:PhaseShift_solution_Simpler}, we consider the relations~\cite{Lifshitz_RelativisticQM,Flores:2024sfy}\footnote{
The unregulated cross-sections violate unitarity, this is why \cref{eq:CrossSectionsFromPhaseShift_reg} with $\Delta_\ell \to \theta_\ell \in \mathbb{R}$ implies $\sigma_{\ell, \unreg}^{\inel} \to 0$ even when $M_{\ell,\unreg}^\inel$ is non-zero.} 
\begin{subequations}
\label{eq:CrossSectionsFromPhaseShift}
\label[pluralequation]{eqs:CrossSectionsFromPhaseShift}
\begin{align}
x_{\ell,\reg} + y_{\ell,\reg} &=  
\frac{1}{2} \left[ 1-{\rm Re} \left(e^{\im 2 \Delta_\ell}\right) \right] ,
&
y_{\ell,\reg} &= 
\frac{1}{4} \left( 1- e^{-4 {\rm Im} \Delta_\ell} \right) ,
\label{eq:CrossSectionsFromPhaseShift_reg}
\\
x_{\ell,\unreg} &=  
\frac{1}{2} \left[ 1-{\rm Re} \left(e^{\im 2 \theta_\ell}\right) \right] .
&
\label{eq:CrossSectionsFromPhaseShift_unreg}
\end{align}
\end{subequations}
Recalling that $\Delta_\ell = \theta_\ell + \delta_\ell$, we find
\begin{align}
x_{\ell,\reg}
= \frac{1}{2} \left[ 
1- 
{\rm Re} \left(e^{\im 2\theta_\ell}\right) 
{\rm Re} \left(e^{\im 2\delta_\ell}\right)
+ 
{\rm Im} \left(e^{\im 2\theta_\ell}\right) 
{\rm Im} \left(e^{\im 2\delta_\ell}\right)
\right] 
-y_{\ell,\reg} .
\label{eq:Regularization_solution_CrossSection_Elas}
\end{align}
The $\theta_\ell$-dependent factors are related to the unregulated elastic cross-section as follows
\begin{subequations}
\label{eq:RealPhaseShift_ReImComponents}
\label[pluralequation]{eqs:RealPhaseShift_ReImComponents}
\begin{align}
{\rm Re} \left(e^{\im 2\theta_\ell}\right) 
&= \cos(2\theta_\ell)
=1-2 x_{\ell,\unreg},
\label{eq:RealPhaseShift_ReComponent}
\\
{\rm Im} \left(e^{\im 2\theta_\ell}\right) 
&= \sin(2\theta_\ell)
= 
\pm \sqrt{1- \cos^2 (2\theta_\ell)} 
= 
\pm 2 \sqrt{x_{\ell,\unreg}(1-x_{\ell,\unreg})} , 
\label{eq:RealPhaseShift_ImComponent}
\end{align}
\end{subequations}
where the sign in \eqref{eq:RealPhaseShift_ImComponent} is fixed by
$\mathrm{sgn}(\Re M_{\ell,\unreg}^{\elas})$, and cannot be inferred from $\sigma_{\ell,\unreg}^{\elas}(k)$ alone.

Collecting the above, we arrive at
\begin{subequations}
\label{eq:Regularization_xy}
\label[pluralequation]{eqs:Regularization_xy}
\begin{empheq}[box=\myshadebox]{align}
\label{eq:Regularization_x}
x_{\ell, \reg} 
&= 
\dfrac{
\qty(\sqrt{x_{\ell, \unreg}} \pm \tilde{w}_\ell \sqrt{1-x_{\ell, \unreg}})^2 
+ (1 - x_{\ell, \unreg}) \, w_\ell^2
}
{(1 + w_\ell)^2 + \tilde{w}_\ell^2},
\\
\label{eq:yj_reg}
y_{\ell,\reg}^j
&= 
\dfrac{w_{A,\ell}^j}
{(1 + w_\ell)^2 + \tilde{w}_\ell^2},
\\[1ex]
\label{eq:Regularization_y}
y_{\ell,\reg} 
&= 
\dfrac{w_\ell}
{(1 + w_\ell)^2 + \tilde{w}_\ell^2} .
\end{empheq}
\end{subequations}
\Cref{eqs:Regularization_xy} encapsulate the result of the unitarization procedure.  $w_{A,\ell}, w_\ell, \tilde{w}_\ell$ are defined in \cref{eqs:wParameters_def}, and we note that 
$y_{\ell,\unreg}^j 
= |[\eeA\hat{M}_{\ell,\unreg}]^j|^2$
and
$y_{\ell,\unreg}^{} 
= \hat{M}_{\ell,\unreg}^\dagger \eeA^2 \hat{M}_{\ell,\unreg}^{}$.

\subsection{Illustrative limits}

\begin{description}

\item[Convergent and analytic separable potentials.] 
For $\WW_\ell\to 0$, \cref{eqs:wParameters_def} yield
\begin{align}
w_{A,\ell}^j =  y_{\ell,\unreg}^j, \qquad
w_\ell^{} =  y_{\ell,\unreg}^{}, \qquad
\tilde{w}_\ell^{} = 
\hat{M}_{\ell,\unreg}^\dagger 
\eeH^{}
\hat{M}_{\ell,\unreg}^{} .
\label{eq:W=0limit}
\end{align}
The regularized cross-sections obtained from \cref{eqs:Regularization_xy} using \eqref{eq:W=0limit} generalize the results of Ref.~\cite{Flores:2024sfy} to provide analytic treatment of Hermitian separable potentials. 

Considering further the fully absorptive case, $\eeH\to0$, sets
$\tilde{w}_\ell\to 0$, leading to~\cite{Flores:2024sfy}
\begin{align}
\label{eq:Regularization_xy_W=0+FullyAbsorptive}
x_{\ell, \reg} 
= \dfrac{
x_{\ell,\unreg}^{} 
+(1 - x_{\ell,\unreg}^{}) y_{\ell,\unreg}^2}
{(1 + y_{\ell,\unreg})^2},
\qquad
y_{\ell, \reg}^j 
= \dfrac{y_{\ell, \unreg}^j}{(1 + y_{\ell, \unreg})^2}.
\end{align}
For the physical interpretation of this remarkably simple result, we refer to \cite{Flores:2024sfy}.

\item[Fully elastic interactions.] For $\eeA \to 0$, the inelastic cross-sections vanish,  $w_{A,\ell} = w_\ell = 0$ and $y_{\ell,\reg}^j =y_{\ell,\reg}^{} =0$. The parameter $\tilde{w}_\ell$, found from \cref{eq:wtilde_def} with $\ee^2 = \eeH^{} \in \RR$, encodes the effect of the Hermitian separable potential~\cite{Yamaguchi:1954mp}. \Cref{eq:Regularization_x}, reducing to  
\begin{align}
x_{\ell,\reg} = \dfrac
{\qty(\sqrt{x_{\ell,\unreg}}
\pm \tilde{w}_\ell \sqrt{1-x_{\ell,\unreg}})^2}
{1+\tilde{w}_\ell^2} , 
\label{eq:FullyElasticLimit_xreg}
\end{align}
gives the elastic cross-section analytically, with the effect of any long-range (or non-separable) potential integrated in $x_{\ell,\unreg}$. For ${\cal V}^{(0)}(\vb{r'},\vb{r}) \to 0$, $x_{\ell,\unreg} \to 0$.
This limit applies to a pure $\phi^4$ theory, where the self-energy kernel arises from the elastic four-point vertex, yielding an exactly contact interaction, ${\cal V}(\vb{r'},\vb{r}) \propto \delta^3(\vb{r})\delta^3(\vb{r'})$, which is both central and separable.

\item[Single channel.] Considering both dispersive and absorptive couplings, $\eta_\ell^2 = \etaHell^{}+\im \etaAell^2$, we find
\begin{align}
w_{A,\ell}^{} = w_\ell^{} = 
|\hat{M}_{\ell,\unreg}|^2
\Im \qty[\qty(1/\eta_\ell^2+W_\ell)^{-1}]
,
\quad
\tilde{w}_\ell^{} = 
|\hat{M}_{\ell,\unreg}|^2
\Re \qty[\qty(1/\eta_\ell^2+W_\ell)^{-1}]
,
\label{eq:Limit_SingleChannel}
\end{align}
and $y_{\ell,\unreg} = \etaAell^2\,|\hat{M}_{\ell,\unreg}|^2$, 
with $W_\ell \in \RR$ being the sole element of the $\WW_\ell$ matrix. Considering further the fully absorptive and fully dispersive cases, the above simplifies to
\begin{subequations}
\label{eq:Limit_SingleChannel_AbsorptiveDispersive}
\label[pluralequation]{eqs:Limit_SingleChannel_AbsorptiveDispersive}
\begin{align}
\label{eq:Limit_SingleChannel_Absorptive}
&\eta_\ell^2 \to \im \etaAell^2:&
&w_{A,\ell}^{} = w_\ell^{} = 
\dfrac
{y_{\ell,\unreg}^{}}
{1 +\etaAell^4 W_\ell^2},&
&\tilde{w}_\ell^{} =
y_{\ell,\unreg}^{} \ 
\dfrac
{\etaAell^2 W_\ell}
{1 +\etaAell^4 W_\ell^2},&
\\[1ex]
\label{eq:Limit_SingleChannel_Dispersive}
&\eta_\ell^2 \to \etaHell^{}:&
&w_{A,\ell}^{} = w_\ell^{} = 0,&
&\tilde{w}_\ell^{} =
|\hat{M}_{\ell,\unreg}|^2
\dfrac{\etaHell^{}}{1+\etaHell^{} W_\ell^{}}.&
\end{align}
\end{subequations}
We note that the fully absorptive limit \eqref{eq:Limit_SingleChannel_Absorptive} is not attainable for interactions that require renormalization, as will be discussed in \cref{sec:Renorm}.

\end{description}

\subsection{Impact of analytic structure of inelastic amplitudes on unitarization \label{sec:Resummation_Wmatrix}}

To shed light on the origin and content of the $\WW_\ell$ matrix, we now consider its form \eqref{eq:Wmatrix_PV} arising from the spectral decomposition of the Green's function. We define the operation $\star$ as the Schwarz-reflection analytic continuation,
\begin{subequations}
\label{eq:SchwarzReflection}
\label[pluralequation]{eqs:SchwarzReflection}
\begin{align}
{\cal F}_{q,\ell}^\star (r) \equiv
[{\cal F}_{q^*,\ell}^{} (r)]^* 
= (-1)^\ell 
\, \dfrac{\fs_\ell(q)}{\fs_\ell(-q)}
\, {\cal F}_{q,\ell} (r) 
=
\sqrt{\sym_\ell}
\, \dfrac{(-\im)^\ell q^{\ell+1}}{\fs_\ell(-q)}
\, \varphi_{q,\ell} (r) 
, 
\quad q \in \CC.
\label{eq:SchwarzReflection_F}
\end{align}
We use ${\cal F}_{q,\ell}^\star$, and \cref{eq:Minel_def_unreg,eq:Mhat-M_relation_cal}, to define the Schwarz-reflected unregulated amplitudes
\begin{align}
\hat{\cal M}_{\ell,\unreg}^{j\star}(q)
\equiv 
\dfrac{1}{q} 
\sqrt{\dfrac{\sym^j \mT^2\mu}{2k^j(s)}}
\int_0^\infty \dd r
\, r 
\, {\cal F}_{q,\ell}^\star (r) 
\, \nu_\ell^{j*}(r)
=
\qty[\hat{\cal M}_{\ell,\unreg}^{j}(q^*) ]^* ,
\qquad q\in\CC,
\label{eq:SchwarzReflection_M}
\end{align}
\end{subequations}
which reduces to $\hat{\cal M}_{\ell,\unreg}^{j\star}(q) = 
\hat{\cal M}_{\ell,\unreg}^{j*}(q)$ for $q\in\RR$, and does not complex-conjugate the momentum $q$ when extending away from the real axis.

%%%%%%%%%%%%%%%%%%%%%%%%%%%%%%%%%%
\begin{figure}[t!]
\centering
\input{W_Contour}
\caption{
Example sketch of a contour (red line) required to compute the integral in \cref{eq:Wmatrix_PV}. The $\PV$ excludes the contributions from the poles at $q=\pm k$. 
The scaling of the unregulated inelastic amplitudes at $|q| \to \infty$, and their singularities on the upper complex $q$ plane (poles or branch cuts) determine the $\WW_\ell$ matrix and affect unitarization. 
}
\label{fig:W_contour}
\end{figure}
%%%%%%%%%%%%%%%%%%%%%%%%%%%%%%%%%%

Replacing $\hat{\cal M}_{\ell,\unreg}^* \to \hat{\cal M}_{\ell,\unreg}^\star$ in \cref{eq:Wmatrix_PV} renders the integrand analytic in the complex $q$ plane, up to isolated singularities and possible branch cuts. The integral can then be evaluated by closing the contour in the upper half-plane, as illustrated in \cref{fig:W_contour}. Applying the Cauchy residue theorem yields the following contributions:
\begin{subequations}
\label{eq:W_Contrib}
\label[pluralequation]{eqs:W_Contrib}
\begin{align}
\WW_\ell(k) 
= \WW_{\ell}^{\BS} (k) 
+ \WW_{\ell}^{\poles} (k)
+ \WW_{\ell}^{\arc} (k)
+ \WW_{\ell}^{\cuts} (k) ,
\tag{\ref{eq:W_Contrib}}
\end{align}
where
\begin{align}
[\WW_{\ell}^{\BS}(k)]^{ij}
&=
\dfrac{4}{\sym_\ell\, \mT^2}
\sqrt{\dfrac{k^i(s) \, k^j(s)}{\sym^i\sym^j}}
\times
2\mu \
\sum_{n}
\dfrac{
\hat{\cal M}_{n\ell,\unreg}^{\BSD,i} \,
\hat{\cal M}_{n\ell,\unreg}^{\BSD,j \, *} 
}
{\kappa_{n\ell}^2 +k^2}, 
\label{eq:W_Contrib_BoundStates}
\\[0.5ex]
%%%%%%%%%
[\WW_{\ell}^{\poles} (k)]^{ij} 
&=
\dfrac{4}{\sym_\ell\, \mT^2}
\sqrt{\dfrac{k^i(s) \, k^j(s)}{\sym^i\sym^j}}
\times (-2\im) 
\sum_{\rho}
~\operatorname*{Res}_{\substack{q = q_\rho \\ \Im q_\rho > 0}}
~\qty[
\dfrac{q^2}
{q^2-k^2}
\hat{\cal M}_{\ell,\unreg}^{i}(q) 
\hat{\cal M}_{\ell,\unreg}^{j\star}(q)
],
\label{eq:W_Contrib_OtherPoles}
\\[0.5ex]
%%%%%%%%%
[\WW_{\ell}^{\arc} (k)]^{ij} 
&=
\dfrac{4}{\sym_\ell\, \mT^2}
\sqrt{\dfrac{k^i(s) \, k^j(s)}{\sym^i\sym^j}}
\times
\dfrac{1}{\pi}
\, \lim_{\Lambda\to \infty}
\, \int_{{\cal C}_\Lambda}
\dd q 
\, \dfrac{q^2}{q^2-k^2} 
\hat{\cal M}_{\ell,\unreg}^{i}(q) 
\hat{\cal M}_{\ell,\unreg}^{j\star}(q)
,
\label{eq:W_Contrib_Infinity}
\\[0.5ex]
%%%%%%%%%
[\WW_{\ell}^{\cuts} (k)]^{ij} 
&=
\dfrac{4}{\sym_\ell\, \mT^2}
\sqrt{\dfrac{k^i(s) \, k^j(s)}{\sym^i\sym^j}}
\times
\dfrac{1}{\pi}
\sum_a
\! \int_{{\cal C}_{\cuts}^a} \!\! 
\dd q 
\dfrac{q^2}{q^2-k^2} 
\Disc\qty[
\hat{\cal M}_{\ell,\unreg}^{i}(q) 
\hat{\cal M}_{\ell,\unreg}^{j\star}(q)
],
\label{eq:W_Contrib_Cuts}
\end{align}
\end{subequations}
Several remarks are in order:
\begin{enumerate}[label=(\alph*)]
\item 
The $\PV$ in \cref{eq:Wmatrix_PV} excludes the poles at $q=\pm k$. The corresponding contribution to $\NN_\ell$ has already been included in \cref{eq:Nmatrix_Expansion}.
%%%%%
\item
The terms \eqref{eq:W_Contrib_OtherPoles} arise from poles of the integrand at $\Im(q)>0$.

If the Hermitian potential admits bound states with binding energies ${\cal E}_{n\ell}=-\kappa_{n\ell}^2/2\mu$, with $\kappa_{n\ell}>0$, then ${\cal F}_{q,\ell}(r)$, and hence $\hat{\cal M}_{\ell,\unreg}(q)$, have poles at $q=+\im\kappa_{n\ell}$, as discussed in \cref{sec:Jost_Poles}. By contrast, the bound-state poles of the Schwarz-reflected amplitude $\hat{\cal M}_{\ell,\unreg}^\star(q)$ lie at $q=-\im\kappa_{n\ell}$, outside the integration contour, and therefore do not contribute to \eqref{eq:W_Contrib_OtherPoles}. If the poles from $\hat{\cal M}_{\ell,\unreg}(q)\propto 1/\fs_\ell(q)$ remain simple and isolated, their contribution to \eqref{eq:W_Contrib_OtherPoles} may partially cancel \eqref{eq:W_Contrib_BoundStates}. At the same time, $\hat{\cal M}_{\ell,\unreg}^\star(q)\propto 1/\fs_\ell(-q)$ makes the zeros and branch cuts of $\fs_\ell(q)$ in the lower half-plane relevant for \eqref{eq:W_Contrib_OtherPoles}, as will be seen more explicitly in \cref{sec:Renorm_gamma=0}.

Importantly, forming $\hat{\cal M}_{\ell,\unreg}(q)$ by superposing the unregulated wavefunctions, ${\cal F}_{q,\ell}(r)$, with the irreducible inelastic amplitudes, $\nu_\ell(r)$, can modify the $q$-singularity structure exhibited by ${\cal F}_{q,\ell}(r)$. This must be taken into account when evaluating \eqref{eq:W_Contrib_OtherPoles}. A familiar example is the Fourier--Laplace transform, which arises in the unitarization of bound-state formation amplitudes~\cite{Flores:2026yay}.

%%%%%
\item 
The term \eqref{eq:W_Contrib_Infinity} arises from the arc ${\cal C}_\Lambda$ at infinity. This is nonzero if 
${\cal M}_{\ell,\unreg}^{\inel, j}(|q|\to\infty)\propto q^\gamma$ with $\gamma\geqslant -1/2$. Should this be the case, renormalization is required to render the regulated cross-sections finite. This will be demonstrated explicitly in \cref{sec:Renorm}.

\item
If the integrand exhibit branch cuts, the integration contour may need to be deformed accordingly. In \eqref{eq:W_Contrib_Cuts}, ${\cal C}_{\cuts}^a$ denotes the $a$-th branch cut in the upper half plane, traversed once from its branch point to its endpoint (or to infinity), and $\Disc[f(q)]\equiv f(q_+)-f(q_-)$ is the discontinuity across the cut, where $q_+$ and $q_-$ denote the limits to the point $q$ on the cut from its two sides. Depending on the branch cuts, the corresponding contributions may be finite or UV-sensitive.

\end{enumerate}
The above demonstrates that the $\WW_\ell$ matrix captures the non-analytic structure and the large-$|q|$ behavior of the unregulated inelastic amplitudes. On the other hand, in the Jost representation of the Green’s function, and consequently of the $\WW_\ell$ matrix, given in \cref{eq:Wmatrix_FG}, these features are encoded in the regular and irregular solutions of the Hermitian potential. This difference will define two distinct renormalization methods in \cref{sec:Renorm}.

%%%%%%%%%%%%%%%%%%%%%%%%%%%%%%%%%%%%%%%%%%%%%%%%%%%
%%%%%%%%%%%%%%%%%%%%%%%%%%%%%%%%%%%%%%%%%%%%%%%%%%%
%%%%%%%%%%%%%%%%%%%%%%%%%%%%%%%%%%%%%%%%%%%%%%%%%%%
\clearpage
\section{Renormalization
\label{sec:Renorm}}

Inelastic interactions can give rise to contact optical potentials that produce divergent contributions to the unitarization prescription, as discussed in \cref{sec:Resummation_Wmatrix}.  We now discuss how these divergences can be renormalized within our framework.  In \cref{sec:Renorm_Setup,sec:Renorm_W,sec:Renorm_gamma=0,sec:Renorm_gamma>0}, we perform the renormalization in momentum space, building on the discussion in \cref{sec:Resummation_Wmatrix}. In \cref{sec:Renorm_Position}, we outline an alternative approach based on position-space renormalization, and compare the two methods in \cref{sec:Renorm_MethodComparison}. We close by relating our results to the treatment of Ref.~\cite{Blum:2016nrz} in \cref{sec:Renorm_ComparisonBSS}.

\subsection{Setup
\label{sec:Renorm_Setup}}

For simplicity, we consider a single inelastic channel. 
If the Hermitian potential ${\cal V}^{(0)} (r',r)$ is well behaved at $r\to 0$, satisfying the convergence condition \eqref{eq:V_ConvergenceCondition}, then the inelastic amplitude at large momenta is insensitive to ${\cal V}^{(0)}$ and is determined by the irreducible inelastic amplitude ${\cal A}_\ell^\inel$. We shall assume that,
\begin{align}
{\cal A}_\ell^{\inel}(q)
= \etaAell^{} \, (q/\mu)^{\gamma},
\label{eq:SingleChannel_AmplitudeAtHighMomenta}
\end{align}
where $\etaAell \in \RR$ is a dimensionless constant, and $\gamma \in \ZZ_{\geqslant 0}$, such that ${\cal A}_\ell^{\inel}$ is holomorphic in $q\in\CC$. We do not specify $\gamma$ further, but note that for standard contact annihilations of non-relativistic particles into relativistic final states, the perturbative inelastic amplitudes carry the scaling \eqref{eq:SingleChannel_AmplitudeAtHighMomenta} with  $\gamma =\ell+2t$, $t \in \ZZ_{\geqslant 0}$. Extrapolated into the UV in order to evaluate the arc contribution \eqref{eq:W_Contrib_Infinity} to $\WW_\ell$, such growth generates divergences that will be computed and renormalized below.

At finite momenta, the potential ${\cal V}^{(0)}$ affects  ${\cal M}_{\ell,\unreg}^{\inel}(q)$ according to \cref{eq:Minel_def_unreg}. We thus set 
\begin{align}
{\cal M}_{\ell,\unreg}^{\inel}(q)    
= \etaAell^{} 
\, h_\ell^{} (q)
\, (q/\mu)^{\gamma}
\quad \text{with} \quad
\lim_{q\to\infty} |h_\ell^{} (q)| =1 ,
\label{eq:SingleChannel_AmplitudeAtGeneralMomenta}
\end{align}
where
\begin{align}
h_\ell(q) \equiv 
\dfrac{2(-\im)^\ell}{\pi}  
\dfrac{1}{q^{1+\gamma}}
\int_0^\infty \dd r \, r \, {\cal F}_{q,\ell} (r)
\int_0^\infty \dd p \, p^{2+\gamma} \, j_\ell(pr) .
\label{eq:h(q)}
\end{align}
A closed-form expression for $h_\ell(q)$ is not necessary;\footnote{
%%%%%%%%%%%%%
For $\gamma = \ell + 2t$ with $t \in \ZZ_{\geqslant 0}$, standard methods give (see e.g.~\cite{Cassel:2009wt,ElHedri:2016onc})
\begin{align*}
h_\ell(q) =
\dfrac{(-\im)^\ell}{q^{\ell+2t+1}}
(-1)^t 2^t t!
\dfrac{(2\ell+2t+1)!!}{(\ell+2t+1)!} 
\qty[
\dfrac{\dd^{\ell+2t+1} }{\dd r^{\ell+2t+1}}
{\cal F}_{q,\ell} (r)
]_{r\to0}  
=
\dfrac{\sqrt{\sym_\ell}}{q^{2t} \fs_\ell(q)}
(-1)^t 2^t t!
\dfrac{(2\ell+2t+1)!!}{(\ell+2t+1)!} 
\qty[
\dfrac{\dd^{\ell+2t+1} }{\dd r^{\ell+2t+1}}
\varphi_{q,\ell} (r)
]_{r\to0} .
\end{align*}
For $t = 0$, squaring the above reproduces the standard $\ell$-wave Sommerfeld factor, multiplied by the $\ell$-mode symmetry factor (cf.~\cref{eq:JostFunction_Sommerfeld}). For $t \geqslant 1$, it yields the Sommerfeld factors for $\ell$-mode higher-order corrections~\cite{ElHedri:2016onc}.
} 
%%%%%%%%%%%%%  
it suffices that, due to the polynomial scaling of the perturbative amplitude \eqref{eq:SingleChannel_AmplitudeAtHighMomenta}, $h_\ell(q)$ has no singularities in $q$ other than those inherited from the scattering-state wavefunction ${\cal F}_{q,\ell}(r)$, i.e.~$h_\ell(q) = \tilde{h}_\ell(q) / \fs_\ell(q)$, where $\fs_\ell(q)$ is the Jost function introduced in \cref{sec:Jost} and $\tilde{h}_\ell(q)$ is holomorphic in $q\in\CC$. 
As in \cref{eq:V_Tot}, we allow for a Hermitian and anti-Hermitian coupling, 
\begin{align}
\label{eq:eta_squared_Contact}
\eta_\ell^2=\etaHell^{} +\im \etaAell^2 ,   
\qquad \etaHell^{}, \etaAell^{} \in \RR,
\end{align}
for the separable potential. This will be a minimal setup to exhibit the renormalization procedure.

In this setup, the unregulated (inelastic) amplitudes stripped from their couplings are
\begin{align}
\hat{\cal M}_{\ell,\unreg} (q) 
= h_\ell(q) \, 
\qty(\dfrac{q}{\mu})^\gamma 
\quad \text{and} \quad
\hat{M}_{\ell,\unreg} (q) =
\sqrt{\dfrac{4 \mu k^f}{\sym_\ell \sym^f \mT^2}}
h_\ell(q) \, 
\qty(\dfrac{q}{\mu})^{\gamma + 1/2} ,
\label{eq:SingleChannel_Amplitudes}
\end{align}
where, as previously, $\sym_\ell$ and $\sym^f$ are the initial-state and final-state symmetry factors, and $k^f$ is the final-state CM momentum, with $k^f \simeq \mT/2$ for fully relativistic species. The matrices $\NN_\ell$ and $\WW_\ell$ reduce to a complex and a real number, $N_\ell$ and $W_\ell$, respectively, with (cf.~\cref{eq:Nmatrix_Expansion,eq:Wmatrix_PV}) 
\begin{align}
N_\ell (k) &= 1 
- \im \eta_\ell^2 |\hat{M}_{\ell,\unreg}^{} (k)|^2
+ \eta_\ell^2 W_\ell(k),
\label{eq:SingleChannel_N}
\end{align}
and
\begin{align}
\label{eq:SingleChannel_Wmatrix}
W_\ell (k) 
=
\dfrac{4k^f}{\sym_\ell \sym^f \mT^2}
\qty[ 
-\dfrac{\PV}{\pi}
\int_{-\infty}^{\infty} 
\dd q \, q^2 \,
\dfrac{
\hat{\cal M}^{}_{\ell,\unreg} (q)  
\hat{\cal M}^{\star}_{\ell,\unreg} (q)}
{q^2-k^2} 
+ 2\mu
\sum_{n}
\dfrac
{|\hat{\cal M}_{n\ell,\unreg}^{\BSD}|^2}
{\kappa_{n\ell}^2 +k^2}
] .
\end{align}
We compute $W_\ell(k)$ via contour integration below. 
The parameters defined in \cref{eqs:wParameters_def} are
\begin{subequations}
\label{eq:SingleChannel_wparameters}
\label[pluralequation]{eqs:SingleChannel_wparameters}
\begin{align}
w_\ell^{} (k) = w_{A,\ell}^{} (k) 
&= 
|\hat{M}_{\ell,\unreg} (k)|^2
\times
\Im\qty[\qty(1/\eta_\ell^2+W_\ell (k))^{-1}],
\\[1ex]
\tilde{w}_\ell (k) &
=
|\hat{M}_{\ell,\unreg} (k)|^2
\times
\Re\qty[\qty(1/\eta_\ell^2+W_\ell (k))^{-1}],
\end{align}
\end{subequations}
where we also note that 
\begin{align}
y_{\ell,\unreg} (k) = 
\etaAell^2 \times 
|\hat{M}_{\ell,\unreg}^{} (k)|^2
.  
\label{eq:SingleChannel_yunreg}
\end{align}
$\hat{M}_{\ell,\unreg} (k)$ has the very specific functional form given in \cref{eq:SingleChannel_Amplitudes}, and is free of any couplings. 
With \cref{eqs:SingleChannel_wparameters}, the regularized cross-sections can be deduced from \cref{eqs:Regularization_xy}.

\subsection{Contributions to the regularization matrix
\label{sec:Renorm_W}}

We evaluate $W_\ell(k)$ by contour integration, as described in \cref{sec:Resummation_Wmatrix}. As discussed in \cref{sec:Renorm_Setup}, for the irreducible inelastic amplitudes of \cref{eq:SingleChannel_AmplitudeAtHighMomenta}, the integrand of \cref{eq:SingleChannel_Wmatrix} inherits its analytic structure from the Jost functions entering the amplitudes, namely 
$
\hat{\cal M}_{\ell,\unreg} (q)
\hat{\cal M}_{\ell,\unreg}^\star (q)
\propto
1/[\fs_\ell(q)\fs_\ell(-q)]
$.
In the following we discuss, and compute where possible, the contributions to $W_\ell$ from poles, the arc at infinity and possible branch cuts in the upper half of the complex $q$ plane. The bound-state contribution, $W_\ell^{\BS}$, emanating directly from the spectral decomposition of the Green's function is given in \cref{eq:W_Contrib_BoundStates} and can be read off of \cref{eq:SingleChannel_Wmatrix}.

\subsubsection{Poles
\label{sec:Renorm_W_Poles}}

If the Hermitian potential ${\cal V}^{(0)}(r,r')$ supports bound states, then $\hat{\mathcal M}_{\ell,\unreg}(q)$ may have simple poles at $q = \im\kappa_{n\ell}$, with $\kappa_{n\ell} > 0$, corresponding to the zeros of the Jost function, $\fs_\ell(q)$. In addition, $\hat{\mathcal M}_{\ell,\unreg}^\star(q)$ may exhibit poles arising from the zeros of $\fs_\ell(-q)$  with $\Im q > 0$; such poles occur in complex-conjugate pairs~\cite{Taylor:1972pty}. We label them by $\tilde n$ and denote the corresponding momenta by $q_{\tilde n \ell}$. The contributions to $W_\ell$ are
\begin{align}
\label{eq:W_sing}
W_{\ell}^{\poles}(k) = 
-\dfrac{1}{\pi}
\,\dfrac{4k^f}{\sym^f \sym_\ell\, \mT^2}
\, 2\pi \im 
\Bigg(
&\sum_n
\underset{q\to\im \kappa_{n\ell}}
\Res
\qty[
\dfrac{q^2}{q^2-k^2} 
\hat{\cal M}_{\ell,\unreg} (q)
\hat{\cal M}_{\ell,\unreg}^\star (q)
]
\\
+ 
&\sum_{\tilde{n}}
\underset{q\to q_{\tilde{n}\ell}^{}}
\Res
\qty[
\dfrac{q^2}{q^2-k^2} 
\hat{\cal M}_{\ell,\unreg} (q)
\hat{\cal M}_{\ell,\unreg}^\star (q)
]
\Bigg).
\nn 
\end{align}
In a fashion similar to \cref{sec:Jost_Greens_Equivalence}, we bring the first term in a more transparent form.
From \cref{eq:PhysicalSolution,eq:BoundStates_NormFactor}, we find
\begin{align}
\underset{q\to\im\kappa_{n\ell}}
\Res[{\cal F}_{q,\ell}(r)] 
= 
-\dfrac{(-\ii)^{\ell}}
{2b_{n\ell} \, N_{n\ell}^*}
\, {\cal B}_{n\ell}(r),
\end{align}
such that
\begin{align}
\underset{q\to\im\kappa_{n\ell}}
\Res[{\cal F}_{q,\ell}(r)] \, 
{\cal F}_{\im\kappa_{n\ell},\ell}^\star(r') 
= 
\dfrac
{(-\im\kappa_{n\ell})^{\ell + 1}}
{2b_{n\ell}\, \fs_\ell(-\im \kappa_{n\ell})} \, 
{\cal B}_{n\ell}(r) \,
{\cal B}_{n\ell}^*(r') .
\end{align}
We reiterate that, due to the Schwarz reflection \eqref{eq:SchwarzReflection_F},  ${\cal F}_{q,\ell}^\star$ does \emph{not} have poles at $q =+\im \kappa_{n\ell}$; its bound-state poles are located at $q =-\im \kappa_{n\ell}$, hence lie outside the integration contour.
Evaluating \cref{eq:RegularSol_expansion} at $k = \ii\kappa_{n\ell}$ and recalling \cref{eq:RegularJostRelation}, it follows that,
\begin{align}
\dfrac{(-\ii\kappa_{n\ell})^{\ell + 1} }
{2b_{n\ell} \, \fs_\ell(-\ii \kappa_{n\ell})}
= \dfrac{\im}{4}.
\end{align}
Considering the amplitudes $\hat{\cal M}_{\ell,\unreg}$ and $\hat{\cal M}_{n\ell,\unreg}^{\BSD}$, defined via
\cref{eq:Minel_def_unreg,eq:Mhat-M_relation_cal,eq:M_BSD_def}, we find
\begin{align}
\underset{{q\to\im\kappa_{n\ell}}}
\Res
\qty[
\hat{\cal M}_{\ell,\unreg} (q)
\hat{\cal M}_{\ell,\unreg}^\star (q)
]
=-
\dfrac{\im\mu}{2\kappa_{n\ell}^2}\,
\big|\hat{\cal M}_{n\ell,\unreg}^{\BSD}\big|^2.
\label{eq:W_sing_ResToBSD}
\end{align}
Inserting \cref{eq:W_sing_ResToBSD} into \eqref{eq:W_sing}, and re-writing the second term in \cref{eq:W_sing}, we obtain
\begin{align}
\label{eq:W_poles}
W_{\ell}^{\poles}(k)
&=
-
\dfrac{1}{2}
\times \qty(
2\mu \,
\dfrac{4k^f}{\sym^f\sym_\ell\,\mT^2}
\sum_{n}
\dfrac{\big|
\hat{\cal M}_{n\ell,\unreg}^{\BSD}
\big|^2}{\kappa_{n\ell}^2+k^2}
)
\\
&-(2\im)
\dfrac{4k^f}{\sym^f\sym_\ell\,\mT^2}
\sum_{\tilde{n}}
\dfrac{q_{\tilde{n}\ell}^2}{q_{\tilde{n}\ell}^2-k^2}
\hat{\cal M}_{\ell,\unreg} (q_{\tilde{n}\ell}^{})
\Res_{q=q_{\tilde{n}\ell}^{}}
\qty[\hat{\cal M}_{\ell,\unreg}^\star (q)].
\nn
\end{align}
The first term is manifestly real. The second term is likewise real, since the poles $q_{\tilde n\ell}$ occur in complex-conjugate pairs and the corresponding residue contributions are conjugate to one another. We note a \emph{partial} cancellation between the first term in \cref{eq:W_poles} and $W_{\ell}^{\BS}$ in \eqref{eq:W_Contrib_BoundStates}. In the absence of bound states, both of them vanish, whereas the second term in \eqref{eq:W_poles} may remain non-zero. \Cref{eq:W_Contrib_BoundStates,eq:W_poles} exhibit an important model-independent feature: at sufficiently low momenta, $k \ll \min (\kappa_{n\ell}^{},|q_{\tilde{n}\ell}^{}|)$, they become insensitive to $k$.

\subsubsection{Divergence
\label{sec:Renorm_W_Divergence}}

The contribution \eqref{eq:W_Contrib_Infinity} to $W_\ell$ from the arc is
\begin{align}
W_{\ell}^{\arc} (k) =
\dfrac{4k^f}{\pi\sym^f \sym_\ell \, \mT^2}
\int_{C_\arc} 
\dd q \ q^2 \ 
\dfrac{ 
\hat{\cal M}_{\ell,\unreg}^{}(q) 
\hat{\cal M}_{\ell,\unreg}^{\star}(q)
}
{q^2-k^2} 
= 
\dfrac{4}{\pi \, \sym^f \sym_\ell}
\dfrac{k^f}{\mT^2 \mu^{2\gamma}}
\int_{C_\arc} 
\dd q \ 
\dfrac{ q^{2+2\gamma} }{q^2-k^2} ,
\label{eq:Warc_def}
\end{align}
where we used the large-$|q|$ behavior~\eqref{eq:SingleChannel_AmplitudeAtHighMomenta} of the amplitudes, noting that since $\gamma\in\mathbb{Z}_{\ge0}$, the UV factor $(q/\mu)^\gamma$ is an entire function of $q$ and therefore admits an unambiguous analytic continuation to $q\in\mathbb{C}$. 
We parameterize the upper semicircle as $q=\Lambda e^{\im\phi}$, $\phi\in[0,\pi]$, so that
$dq = \im\Lambda e^{\im\phi}\,d\phi$, and expand with respect to $k/\Lambda$, to obtain
\begin{align}
\int_{C_\arc} \! \dd q \
\dfrac{q^{2+2\gamma}}{q^2-k^2}
= \im \Lambda^{1+2\gamma} 
\sum_{j=0}^\infty
\left(\dfrac{k^2}{\Lambda^2}\right)^j
\int_0^\pi \dd\phi 
\, e^{\im (1+2\gamma-2j)\phi}
= \Lambda^{1+2\gamma} 
\sum_{j=0}^\infty
\left(\dfrac{k^2}{\Lambda^2}\right)^j
\dfrac{e^{\im (1+2\gamma-2j)\pi}-1}{1+2\gamma-2j} .
\label{eq:ArcIntegral}
\end{align}
Only the terms with non-negative powers of $\Lambda$ survive in the limit $\Lambda\to\infty$, which, for $\gamma \in \ZZ_{\geqslant 0}$, implies $j\leqslant \gamma$. Hence, the divergent contribution from the arc is
\begin{align}
W_{\ell}^{\arc} (k)
\ \overset{\Lambda \to \infty}{=} \
-\dfrac{8}
{\pi \, \sym^f \sym_\ell}
\dfrac{k^f}{\mT^2 \mu^{2\gamma}} 
\sum_{j=0}^{\gamma}
\dfrac{\Lambda^{1+2\gamma-2j}}{1+2\gamma-2j}
\, k^{2j}.
\label{eq:W_arc}
\end{align}
This is a real polynomial in $k^2$ of degree $\gamma$ with coefficients that diverge as powers of~$\Lambda$. In the following, we consider the cases $\gamma=0$ and $\gamma >0$ separately. 

Before proceeding, it is useful to assess corrections from the full momentum dependence of the unregulated amplitudes \eqref{eq:SingleChannel_Amplitudes}.
The integrand of \cref{eq:Warc_def} should contain the factor $h_\ell(q)\,h_\ell^\star(q)$, which we have approximated by unity along the arc, $\Lambda\to \infty$. General properties of the unregulated wavefunctions, discussed in \cref{sec:Jost}, imply that $h_\ell(q) \,h_\ell^\star(q)$ is even in $q$. If at large $|q|$, this factor admits an expansion in powers of $q^{-2}$, then the corresponding contributions are either sub-leading in the cutoff or vanish as $\Lambda\to\infty$. Non-analytic terms $\propto \ln q^2$ can instead produce logarithmically enhanced divergences, of the form $\Lambda^{p}\ln\Lambda$. Nevertheless, since $k$ enters $W_{\ell}^{\arc}(k)$ only through $(q^2-k^2)^{-1}$ in \cref{eq:Warc_def}, such UV corrections can only modify the divergent coefficients multiplying the same polynomial in $k^2$; they do not generate new functional dependence on $k$, hence do not change the $k$-running in this scheme.

\subsubsection{Branch cuts
\label{sec:Renorm_W_Cuts}}

If branch cuts are present in the upper half plane, the integration contour must be appropriately deformed around them, resulting in contributions, $W_\ell^{\cuts}$, of the form \eqref{eq:W_Contrib_Cuts}. If the discontinuity across a cut is integrable and the branch point lies at $|q_{a\ell}|>0$, where $a$ enumerates the branch cut, then the corresponding contribution to $W_\ell^\cuts(k)$ is finite. Its momentum dependence is determined by the $1/(q^2-k^2)$ factor in \cref{eq:W_Contrib_Cuts}, and implies the expansion
\begin{align}
W_{\ell}^{\cuts,a}(k)
=
W_{\ell}^{\cuts,a}(0)
+{\cal O} (k^2/|q_{a\ell}|^2).
\label{eq:W_cut_LowMomentum}
\end{align}
For long-range interactions with logarithmic phases, such as Coulomb-like potentials, $q_{a\ell} \to 0$; in such cases, it is safer to work directly with the physical spectral representation on $q>0$ rather than the symmetrized form appearing in \cref{eq:SingleChannel_Wmatrix}, and compute the integral along the real axis (see e.g.~\cite{Flores:2026yay}).

Branch cuts may also contribute to the UV-sensitive part of $W_\ell(k)$ if they extend to $|q|\to\infty$. In a similar fashion to the contour integral computed in \cref{sec:Renorm_W_Divergence}, their contribution is again a polynomial in $k^2$ at most of degree $\gamma$, possibly with logarithmically enhanced coefficients, and can be included into the same renormalization procedure as the arc contribution described below.

\subsection{Renormalizable theories
\label{sec:Renorm_gamma=0}}

We first focus on $\gamma=0$, which is a case of particular physical significance. The partial-wave unitarity bounds discussed in \cref{sec:Unitarity} imply that a UV-complete theory cannot sustain an amplitude that grows at large $|q|$, thereby imposing $\gamma \leqslant 0$. Hence, $\gamma=0$ is the only asymptotic scaling of a contact interaction compatible with unitarity. 

As seen in \cref{eq:W_arc}, $\gamma=0$ makes  $W_{\ell}^{\arc}$ momentum-independent, 
\begin{align}
W_{\ell}^{\arc} = 
-\qty(\dfrac{8k^f}{\pi \, \sym^f \sym_\ell\, \mT})
\dfrac{\Lambda}{\mT} .
\label{eq:W_arc_gamma=0}
\end{align}
UV-divergent contributions from possible branch cuts, discussed in \cref{sec:Renorm_W_Cuts}, can be absorbed in \cref{eq:W_arc_gamma=0}. In this case, all UV sensitivity can be absorbed into $\etaHell$ and $\etaAell$, as will be shown below. This is in contrast to the $\gamma>0$ case, that will be discussed in \cref{sec:Renorm_gamma>0}. The scaling $\gamma=0$ corresponds therefore to the unique fully renormalizable contact interaction.

\subsubsection*{Position-space potential}

Before proceeding, we note that for $\ell=0$ and a momentum-independent ${\cal A}_{\ell=0}^\inel$, as hypothesized here, using the distributional limit 
$\int_0^\infty \dd p \, p^2 j_{0} (pr) = 
(\pi/2) \, \delta(r)/r^2$, \cref{eq:nu_def}, 
yields
\begin{align}
\nu_{\ell=0} (r) = 
\sqrt{\dfrac{2k^f}{\sym^f\mT^2\mu}}
\dfrac{\delta (r)}{r^2} 
,
\label{eq:ContactPotential_nu}
\end{align}
with the full potential \eqref{eq:V_Tot} in this case being
\begin{align}
\label{eq:ContactPotential_Total}
{\cal V}_{\ell=0}^{} (r,r') =
{\cal V}_{\ell=0}^{(0)} (r,r')
-\dfrac{2\eta_0^2\, k^f}{\sym^f \mT^2 \mu}
\dfrac{\delta(r)}{r^2}
\dfrac{\delta(r')}{r'^2} .
\end{align}
This potential was considered in the unitarization analysis of Ref.~\cite{Blum:2016nrz}. In \cref{sec:Renorm_Position}, we perform renormalization in position space, in a similar fashion to Ref.~\cite{Blum:2016nrz}, using \cref{eq:ContactPotential_nu}. We then compare our results in \cref{sec:Renorm_ComparisonBSS}.

\subsubsection*{Renormalization conditions}

Collecting all contributions to $W_\ell(k)$, we split it in divergent and finite parts, as follows
\begin{subequations}
\label{eq:Wmatrix_DivFinDecomp_defs}
\label[pluralequation]{eqs:Wmatrix_DivFinDecomp_defs}
\begin{align}
\label{eq:Wmatrix_DivFinDecomp}
W_\ell(k) 
~=~
W_{\ell}^{\arc} + 
W_{\ell}^{\poles} (k) +
W_{\ell}^{\cuts} (k) + 
W_{\ell}^{\BS} (k)
~=~ 
W_{\ell}^{\Div}(k_\diamond) 
+ W_{\ell}^{\fin} (k;k_\diamond) , 
\end{align}
with
\begin{align}
\label{eq:Wmatrix_Div}
W_{\ell}^{\Div} (k_\diamond) 
&\equiv 
W_\ell(k_\diamond) 
,
\\
\label{eq:Wmatrix_Finite}
W_{\ell}^{\fin} (k;k_\diamond) 
&\equiv 
W_\ell(k) - W_\ell(k_\diamond),   
\end{align}
\end{subequations}
where $W_{\ell}^{\BS} (k)$, $W_{\ell}^{\poles} (k)$, $W_{\ell}^{\arc}$ and $W_{\ell}^{\cuts}$ are given by \cref{eq:W_Contrib_BoundStates,eq:W_poles,eq:W_arc_gamma=0,eq:W_Contrib_Cuts}, and $k=k_{\diamond}$ is the renormalization scale. By construction, $W_{\ell}^{\fin}(k_{\diamond};k_\diamond) = 0$, such that the renormalization conditions involve only $W_{\ell}^{\Div}$. Note that in the present case,  $W_{\ell}^{\Div}$ is independent of the momentum $k$, but depends on the renormalization scale $k_\diamond$.

We allow the bare couplings $\eta_\ell^2 = \etaHell^{}+\im \etaAell^2$ to depend on the cutoff $\Lambda$, such that the regularization parameters \eqref{eq:SingleChannel_wparameters}, $w_\ell(k)$ and $\tilde{w}_\ell(k)$, and therefore the regulated cross-sections, do not. For this purpose, we define the renormalization conditions
\begin{align}
1/[\etaren_\ell(k_\diamond)]^2 \equiv 
1/\eta_\ell^2 + W_{\ell}^{\Div} (k_\diamond) ,   
\label{eq:RenormConditions}
\end{align}
with the renormalized coupling, $\etaren_\ell$, decomposed as
\begin{align}
[\etaren_\ell(k_\diamond)]^2 
= \etaHellren(k_\diamond) 
+ \im [\etaAellren(k_\diamond)]^2.  
\label{eq:RenormCouplings}   
\end{align}
For later convenience, we also define the dispersive-to-absorptive coupling ratio,
\begin{align}
\beta_\ell(k_\diamond) \equiv 
\etaHellren(k_\diamond) / 
[\etaAellren(k_\diamond)]^2 .
\label{eq:beta_def}
\end{align} 

\subsubsection*{Renormalized cross-sections}

Considering \cref{eq:SingleChannel_yunreg}, we define the  renormalized unregulated inelastic cross-section divided by the unitarity cross-section, $y_{\ell,\unreg}^\ren (k;k_\diamond)$, as well as the function $z_\ell(k;k_\diamond)$ that encapsulates the effect of the finite momentum-dependent part of $W_\ell(k)$, as follows
\begin{subequations}
\label{eq:Renorm_yz}
\label[pluralequation]{eqs:Renorm_yz}
\begin{align}
y_{\ell,\unreg}^\ren (k;k_\diamond)  
&\equiv 
[\etaAellren(k_\diamond)]^2 
\times |\hat{M}_{\ell,\unreg} (k)|^2 ,  
\label{eq:Renorm_y_unreg}
\\
z_\ell (k;k_\diamond)  
&\equiv 
[\etaAellren(k_\diamond)]^2 
\times 
W_{\ell}^{\fin} (k;k_\diamond) .  
\label{eq:Renorm_z}
\end{align}
\end{subequations}
By construction, $z_\ell(k_\diamond;k_\diamond)=0$. Substituting the full expression for $W_\ell(k)$ from \cref{eq:Wmatrix_DivFinDecomp} into \cref{eqs:SingleChannel_wparameters} and imposing the renormalization conditions \eqref{eq:RenormConditions} eliminates the divergent terms, 
\begin{subequations}
\label{eq:Renorm_w-wt_pre}
\label[pluralequation]{eqs:Renorm_w-wt_pre}
\begin{align}
\label{eq:Renorm_w_pre}
w_\ell(k;k_\diamond)
&=
|\hat{M}_{\ell,\unreg} (k)|^2
\ \Im \qty[
\qty(\dfrac{1}{[\etaren_\ell(k_\diamond)]^2}
+ W_{\ell}^{\fin}(k;k_\diamond)
)^{-1} ],
\\[1ex]
\label{eq:Renorm_wt_pre}
\tilde{w}_\ell(k;k_\diamond)
&=
|\hat{M}_{\ell,\unreg} (k)|^2
\ \Re \qty[
\qty(\dfrac{1}{[\etaren_\ell(k_\diamond)]^2} 
+ W_{\ell}^{\fin}(k;k_\diamond))^{-1} 
].
\end{align}
\end{subequations}
Using the definitions of \cref{eq:RenormCouplings,eq:beta_def,eq:Renorm_yz}, this becomes
\begin{subequations}
\label{eq:Renorm_w-wt}
\label[pluralequation]{eqs:Renorm_w-wt}
\begin{align}
\label{eq:Renorm_w}
w_\ell(k;k_\diamond)
&=
\dfrac{y_{\ell,\unreg}^\ren (k;k_\diamond)}
{\qty[1+\beta_\ell(k_\diamond) 
\, z_\ell(k;k_\diamond)]^2 + z_\ell(k;k_\diamond)^2}
,
\\[1ex]
\label{eq:Renorm_wt}
\tilde{w}_\ell(k;k_\diamond)
&=
w_\ell(k;k_\diamond) \
\Big[
\beta_\ell(k_\diamond) [1+\beta_\ell(k_\diamond)\, z_\ell(k;k_\diamond)]
+z_\ell(k;k_\diamond)
\Big].
\end{align}
\end{subequations}
Inserting the expressions \eqref{eq:Renorm_w-wt} into
\cref{eqs:Regularization_xy}, we obtain $x_{\ell,\reg}$ and $y_{\ell,\reg}$,
\begin{subequations}
\label{eq:SingleChannel_RegRen_xy}
\label[pluralequation]{eqs:SingleChannel_RegRen_xy}
\begin{empheq}[box=\myshadebox]{align}
x_{\ell,\reg} (k)
&=
\dfrac{1}{R_\ell (k;k_\diamond)} 
\Bigg\{
\qty(
\sqrt{x_{\ell,\unreg} (k)} 
\ z_\ell(k;k_\diamond)
\pm 
\sqrt{1-x_{\ell,\unreg} (k)} 
\ y_{\ell,\unreg}^\ren (k;k_\diamond)
)^2
\nn
\\[1ex]
\label{eq:SingleChannel_RegRen_x}
&+
\qty(
\! \sqrt{x_{\ell,\unreg} (k)} 
\, \Big[1
+\beta_\ell (k_\diamond) \, z_\ell(k;k_\diamond)\Big]
\pm 
\!\!\sqrt{1-x_{\ell,\unreg} (k)} 
\, \beta_\ell (k_\diamond) 
\, y_{\ell,\unreg}^\ren (k;k_\diamond)
)^2
\Bigg\},
%%%%%%%%%%
\\[1.5em]
y_{\ell,\reg}^{} (k)
&=\dfrac
{y_{\ell,\unreg}^\ren(k;k_\diamond) } 
{R_{\ell} (k;k_\diamond)},
\label{eq:SingleChannel_RegRen_y}
%%%%%%%%%
\\[1.5em]
R_\ell (k;k_\diamond) 
&\equiv 
\qty[1+y_{\ell,\unreg}^\ren(k;k_\diamond)]^2 
+ z_\ell^2(k;k_\diamond)
\nn 
\\[1ex]
&
+ \beta_\ell^2 (k_\diamond) 
\ [y_{\ell,\unreg}^\ren(k;k_\diamond)]^2
+ 
\qty[1+\beta_\ell(k_\diamond) \, z_\ell(k;k_\diamond)]^2
-1.
\end{empheq}
\end{subequations}
The parameter $\beta_\ell$ appearing in \cref{eqs:SingleChannel_RegRen_xy} and defined in \eqref{eq:beta_def}, quantifies the relative importance of the dispersive coupling $\etaHellren$ with respect to the absorptive coupling $(\etaAellren)^{2}$. Both originate from the same short-distance physics, as seen in the optical-potential picture of \cref{sec:OpticalPotential}. The value of $\beta_\ell$ can be fixed by matching to an observable, either the elastic or inelastic cross-section at a chosen scale, and is thus determined from experimental input. Importantly, it also requires theoretical input, namely one of the unregulated cross-sections. Concretely, \cref{eq:SingleChannel_RegRen_y}, evaluated at $k=k_\diamond$, which sets $z_\ell(k_\diamond;k_\diamond)=0$, can be easily transformed to obtain
\begin{align}
\beta_\ell^2 (k_\diamond) = 
\dfrac{1}
{[y_{\ell,\unreg}^{\ren}(k_\diamond;k_\diamond)]^2}
\qty(
\dfrac
{y_{\ell,\unreg}^{\ren}(k_\diamond;k_\diamond)}
{y_{\ell,\reg}(k_\diamond)}
-\qty[1+
y_{\ell,\unreg}^{\ren}(k_\diamond;k_\diamond)]^2
).
\label{eq:beta}
\end{align}
In a multichannel system, determining the full set of dispersive couplings would require either measurements of the corresponding inelastic cross-sections, or, equivalently, elastic cross-section measurements at multiple momenta.

\subsubsection*{Running}

The regularized and renormalized cross-sections must not depend on the renormalization scale, $k_\diamond$, which results in the running of the renormalized couplings. We determine this by requiring
\begin{align}
\dfrac{\dd w_\ell (k;k_\diamond)}{\dd k_\diamond} = 
\dfrac{\dd \tilde{w}_\ell (k;k_\diamond)}{\dd k_\diamond} =  0,
\qquad \forall k \geqslant 0,
\label{eq:Renorm_gamma=0_RGconditions} 
\end{align}
which yields the renormalization group equations,
\begin{align}
\label{eq:RGeqs}
\dfrac{\dd}{\dd k_\diamond} 
\qty[
\dfrac{1}{[\etaren_\ell(k_\diamond)]^2} 
+W_{\ell}^{\fin}(k;k_\diamond)
] = 0 ,
\end{align}
that can be integrated to give
\begin{align}
\label{eq:RGeqs_sol}
\dfrac{1}{[\etaren_\ell (k_\diamond)]^2} 
= 
\dfrac{1}{[\etaren_\ell(\tilde{k}_{\diamond})]^2}
+ W_{\ell}^{\fin} (k;\tilde{k}_{\diamond}) 
- W_{\ell}^{\fin} (k;k_\diamond) . 
\end{align}
Note that $W_{\ell}^{\fin} (k;\tilde{k}_{\diamond}) 
- W_{\ell}^{\fin} (k;k_\diamond)$ is $k$-independent. If $W_{\ell}^{\fin}$ is momentum-independent, the couplings do not run and $z_\ell(k;k_\diamond)$ vanishes, which greatly simplifies \cref{eqs:SingleChannel_RegRen_xy}.

\subsubsection*{Discussion}

Several interesting points emerge from the above.
\begin{enumerate}[label=(\roman*)]
\item 
It is evident from \cref{eq:RenormConditions,eq:Renorm_w-wt} that, assuming only an absorptive coupling, $\eta_\ell^2 \to \im \etaAell^2$ and $\etaHell^{} \to0$, does not suffice to absorb the divergence and keep $w_\ell(k;k_\diamond)$, $\tilde{w}_\ell(k;k_\diamond)$ finite. This attests to the fact that \emph{on-shell contact inelastic interactions generate both absorptive and conservative potentials}, a notion that becomes explicit in the Feshbach formalism for the optical potential that we reviewed in \cref{sec:OpticalPotential}. 

On the other hand, it is possible to set $\etaAell \to 0$, while keeping $\etaHell \neq 0$, which allows for $\tilde{w}_\ell(k;k_\diamond)\neq0$ while $w_\ell(k;k_\diamond)\to0$. Physically, this situation can arise from inelastic vertices where the interaction products are kinematically forbidden from going on-shell, or in a theory with purely elastic interactions, such as $\phi^4$.

\item 
The momentum dependence of the finite contributions of \cref{eq:W_Contrib_BoundStates,eq:W_poles}, and implicitly of \cref{eq:W_cut_LowMomentum}, has an interesting implication. If the renormalization scale is chosen below the scales set by these singularities, and we restrict our attention to momenta in the same regime, namely
$k,k_\diamond \ll 
\min(\kappa_{n\ell}^{},|q_{\tilde{n}\ell}^{}|,|q_{a\ell}^{}|)$, 
then $z_\ell(k;k_\diamond)\approx 0$ and the renormalized couplings do not run. This considerably simplifies the unitarization formulas \eqref{eq:Regularization_xy}, rendering the low-momentum behavior of the regularized cross-sections transparent, since it depends only on $x_{\ell,\unreg}(k)$ and $y_{\ell,\unreg}^\ren(k)$. This scaling changes, of course, once larger momenta are considered.

\item 
The unitarized inelastic cross-section in \cref{eq:SingleChannel_RegRen_y} can approach or saturate the unitarity bound in the limit of small dispersive couplings, $\beta_\ell(k_\diamond) \to 0$, and $z_\ell(k;k_\diamond) \to 0$. Even in such cases, the inelastic cross-section can track the unitarity limit over a continuous range of velocities only if the unregulated inelastic cross-section exhibits the same velocity dependence as the unitarity bound (such that $y_{\ell,\unreg}$ is constant across the velocity range of interest). This happens when the Hermitian self-energy kernel is Coulombic~\cite{vonHarling:2014kha,Baldes:2017gzw,Flores:2024sfy}.
\end{enumerate}

\subsection{Effective theories 
\label{sec:Renorm_gamma>0}}

Next, we consider $\gamma>0$. Equation \cref{eq:W_arc} shows that the arc term $W_\ell^{\arc}(k)$ acquires cutoff
divergences multiplying successive powers of $k^2$, forming in particular a polynomial of degree $\gamma$.
Canceling these divergences requires contact counterterms with up to $2\gamma$ derivatives, and hence a growing
set of renormalization conditions as $\gamma$ increases. 
This UV growth also violates partial-wave unitarity at large momentum. In this framework, the loss of unitarity
tracks the loss of renormalizability: a contact interaction with $\gamma>0$ cannot represent the true UV
behavior and must be replaced above some scale by dynamics that softens the high-momentum growth. When such
scaling appears in UV-complete theories, it is an artifact of the approximations employed.

\subsubsection*{Renormalization conditions at several momenta}

To renormalize the polynomial UV sensitivity in $W_\ell^{\arc}(k)$, we may impose a set of conditions on both $w_\ell$ and $\tilde w_\ell$ at several reference momenta $k=k_j$, with $j\in [0,\gamma]$, in direct analogy with matching successive terms of the effective-range expansion in contact effective field theories~\cite{Kaplan:2005es,Bedaque:2002mn}. Concretely, we fix 
$[\etaren_\ell(k_j)]^2$, which provides $2(1+\gamma)$ renormalization conditions. The resulting renormalized parameters depend on the chosen set of momenta $\{k_j\}$, and the corresponding renormalization-group flow can be viewed as a multi-parameter (or multi-scale) generalization of the $\gamma=0$ running.

\subsubsection*{Low-momentum truncation} 

If one truncates $W_\ell^{\arc}(k)$ to leading order in $k^2$, i.e.,\ keeps only the $j=0$ term in \cref{eq:W_arc}, then $W_\ell^{\arc}$ becomes effectively momentum-independent. In such an approximation, the renormalization reduces to the $\gamma=0$ analysis, with only two parameters required to absorb the UV sensitivity. This corresponds to treating the interaction as an effective field theory valid below the scale where the higher-$k^2$ divergences become relevant.

\subsection{Renormalization in position space
\label{sec:Renorm_Position}}

It is possible to formulate the renormalization procedure in position space. We consider the renormalizable case, $\gamma = 0$, and focus on $\ell=0$ which corresponds to the $\delta$-function position-space potential \eqref{eq:ContactPotential_nu}. Following Ref.~\cite{Blum:2016nrz}, we regulate the divergence of the potential as follows 
\begin{align}
\nu_0(r) = 
\sqrt{\dfrac{2k^f}{\sym^f\mT^2\mu}}
\dfrac{\delta(r-\varepsilon)}{r^2} ,   
\label{eq:ContactPotential_nu_epsilonReg}
\end{align}
with $\varepsilon \to 0^+$. To calculate $W_0(k)$ in terms of the cutoff parameter $\varepsilon$, we recall its form \eqref{eq:Wmatrix_FG} that employs the Jost representation \eqref{eq:GreensFunction_FH} of the Green's function, and consider the asymptotic behaviors at $r\to0$ of the regular and irregular solutions for the Hermitian potential ${\cal V}^{(0)} (\vb{r},\vb{r'})$, including sub-leading corrections obtained via the Frobenius method in \cref{sec:Jost_Frobenius}. We find
\begin{align}
W_0 (k) 
&=
\dfrac{4k^f}{\sym_0 \sym^f \mT^2 k}
\ \dfrac{1}{\varepsilon^2}
\ {\cal F}_{k,0}^*(\varepsilon)
\ {\cal G}_{k,0}^{}  (\varepsilon)
\nn 
\\
&=-\dfrac{4k^f}{\sym^f \mT^2} 
\, \dfrac{1}{\varepsilon} 
\qty[1
+ \varepsilon \, f_0^{(1)}
+ {\cal O}(\varepsilon^2)]
\qty[1
+ \varepsilon \, g_0^{(1)} 
+ \varepsilon \ln (\mu\varepsilon) \, a 
+ {\cal O}(\varepsilon^2,\varepsilon^2\ln\varepsilon)
]
\nn \\
&=-\dfrac{4k^f}{\sym^f \mT^2} 
\, \qty[
\dfrac{1}{\varepsilon} 
+ a \, \ln (\mu\varepsilon)
+ \qty(f_0^{(1)}+g_{0}^{(1)} (k))
+ {\cal O}(\varepsilon,\varepsilon\ln\varepsilon)
].
\label{eq:W_Position}
\end{align}
The parameters appearing above have been defined in \cref{sec:Jost_Frobenius}: 
$f_0^{(1)}$ and $a\in \RR$ are independent of the momentum $k$ and determined solely by $\mu$ and the parameters of the Hermitian potential ${\cal V}^{(0)}$. However, $g_{0}^{(1)} (k)$ may in general depend on $k$. $W_0(k)$ can thus be split in a momentum-independent divergent part and a momentum-dependent finite part, as in \cref{eq:Wmatrix_DivFinDecomp}, with
\begin{subequations}
\label{eq:Wmatrix_DivFinDecomp_Position}
\label[pluralequation]{eqs:Wmatrix_DivFinDecomp_Position}
\begin{align}
\label{eq:Wmatrix_DivFinDecomp_Position_Wepsilon}
W_{0,\Div} (k_\diamond)
&\equiv
-\dfrac{4k^f}{\sym^f\mT^2}
\qty[
\dfrac{1}{\varepsilon}
+a\ln(\mu\varepsilon)
+f_0^{(1)} + g_0^{(1)} (k_\diamond)
],
\\
\label{eq:Wmatrix_DivFinDecomp_Position_Wfinite}
W_{0,\fin}(k;k_\diamond)
&\equiv
-\dfrac{4k^f}{\sym^f\mT^2}
\qty[g_0^{(1)}(k)-g_0^{(1)}(k_\diamond)] ,
\end{align}
\end{subequations}
with the renormalization procedure carrying through as in \cref{sec:Renorm_gamma=0}.

\subsection{On the choice of renormalization method \label{sec:Renorm_MethodComparison}}

The above demonstrates the consistency of the two renormalization methods --- based on the spectral decomposition and on the Jost representation of the Green's function --- for $s$-wave renormalizable contact interactions. In both approaches, the UV-divergent part is momentum-independent, while the momentum-dependent part is finite, and the divergence is absorbed into the complex coupling of the contact interaction. The full equivalence could be verified by evaluating the finite part of the regularization matrix in specific models using both methods, but this lies beyond the scope of the present work.

Based on the features of the two methods illustrated in this example, we find the momentum-space formulation technically advantageous and more readily applicable to a broader class of models, for the following reasons:
\begin{itemize}
\item
The momentum-space formulation eliminates the need to construct intermediate position-space optical potentials, according to \cref{eq:nu_def}, which can be ill-defined and require regularization, or simply difficult to compute analytically.

It also bypasses the explicit projection of the Green’s function, which, in the position-space formulation, entails evaluating overlap integrals between the optical potential and the regular and irregular wavefunctions of the associated Hermitian problem, according to \cref{eq:Wmatrix_FG}. While these integrals are straightforward in the case of a $\delta$-function optical model, they become significantly more involved for general optical potentials.

\item
In momentum space, the $k$-dependence is manifest, entering through the resolvent factor $(q^{2}-k^{2})^{-1}$, with $q$ ranging over different domains in the finite and UV-sensitive contributions. In position space, by contrast, the $k$-dependence is only implicit, being encoded in the short-distance behavior of the wavefunctions; beyond leading order, even its functional form is model-dependent. 

\end{itemize}

On the other hand, evaluating the Green's function via its spectral representation by complex-analysis methods requires care. For long-range interactions (notably Coulomb-type potentials), scattering wavefunctions develop additional non-analytic structure (e.g., multi-valued logarithmic phases), and the contour manipulations for the integral \eqref{eq:SingleChannel_Wmatrix} may require modification. In such cases, it is often preferable to work directly with the physical spectral representation on $q>0$~\cite{Flores:2026yay}.

We emphasize the importance of the renormalization group \cref{eq:RGeqs,eq:RGeqs_sol}, in both methods, to ensure independence of the physical cross-sections from the renormalization scale.

\subsection{Comparison with previous results
\label{sec:Renorm_ComparisonBSS}}

In Ref.~\cite{Blum:2016nrz}, Blum, Sato, and Slatyer (BSS) considered distinguishable particles interacting via a long-range real central potential,\footnote{Here, the term ``long-range" encompasses also finite-range potentials, but stands in contrast to contact potentials.}
\begin{align}
\label{eq:BSS_Potential_Total}
{\cal V}^{\BSS} (\vb{r},\vb{r'}) 
= V(r) \, \delta^3(\vb{r}-\vb{r'})
+\lambda \, \delta^3(\vb{r}) \, \delta^3(\vb{r}-\vb{r'}) ,
\end{align}
where $\lambda \in \CC$. 
Setting 
$\delta^3 (\vb{r}) = \delta(r)/(4\pi r^2)$ 
and  
$\delta^3 (\vb{r} - \vb{r'}) = [\delta(r-r')/r^2] \, \delta^2(\Omega_{\vb{r}}-\Omega_{\vb{r'}})$, 
and projecting into partial waves according to \cref{eq:CentralPotential_PW_Position}, we find\footnote{
%%%%%%%%%%%%%
As pointed out below \cref{eqs:CentralPotential_PW}, and seen explicitly in \cref{eq:BSS_Potential_Total_PW}, for central potentials, all partial waves are correlated. However, a $\delta$-function potential cannot affect higher $\ell$ modes, since the regular wavefunction vanishes at $r\to0$ (cf.~\cref{eq:WF_Asymptotes0_F}), nor does it emanate from any standard higher-$\ell$ irreducible amplitude.} 
%%%%%%%%%%%%%
%
\begin{align}
\label{eq:BSS_Potential_Total_PW}
{\cal V}_{\ell}^{\BSS} (r,r') 
= V(r) \, \dfrac{\delta(r-r')}{r^2}    
+ \dfrac{\lambda}{4\pi} 
\, \dfrac{\delta(r)}{r^2} 
\, \dfrac{\delta(r')}{r'^2} .
\end{align}
Focusing on $\ell=0$, and considering the potential \eqref{eq:ContactPotential_Total} used in our analysis, the above maps in our notation to 
$\lambda = -8\pi\eta_0^2 k^f/(\sym^f \mT^2 \mu)$, 
with $\Im(\lambda)$ parametrizing the strength of a single channel $s$-wave annihilation. In this subsection, we compare our results for this specific case, with those of Ref.~\cite{Blum:2016nrz}. To ease the notation,  we omit the label $\ell=0$.

Rewriting \cite[Eqs.~(27) and (28)]{Blum:2016nrz} as closely as possible in our notation,
\begin{subequations}
\label{eq:BSS}
\label[pluralequation]{eqs:BSS}
\begin{align}
%%%%%%%%%%%%%%%%%%%%%%%%%%%%%%%%%%%%
\label{eq:BSS_x}
x_{\reg}^{\BSS} (k) &=
\dfrac{1}{4}\left|
e^{\im 2\theta(k)}
\ \dfrac
{1-
\dfrac{k}{k_{\diamond}}
[S(k)+ \im \, T(k;k_\diamond)] 
\, 
\qty(\bar{y}_\diamond-\im \bar{\xi}_{\diamond}\sqrt{\bar{x}_{\diamond}-\bar{y}_{\diamond}^2})
}
{1+
\dfrac{k}{k_{\diamond}}
[S(k) - \im \, T(k;k_\diamond)] 
\, 
\qty(\bar{y}_\diamond-\im \bar{\xi}_{\diamond}\sqrt{\bar{x}_{\diamond}-\bar{y}_{\diamond}^2})
}
-1
\right|^2 ,
\\[1ex]
%%%%%%%%%%%%%%%%%%%%%%%%%%%%%%%%%%%%
y_{\reg}^{\BSS} (k)  
&=
\dfrac
{y_{\unreg}^{\ren}(k;k_\diamond)}
{\left|
1+
\dfrac{k}{k_{\diamond}}
[S(k) - \im \, T(k;k_\diamond)] 
\, 
\qty(\bar{y}_\diamond-\im \bar{\xi}_{\diamond}\sqrt{\bar{x}_{\diamond}-\bar{y}_{\diamond}^2})
\right|^2} ,
\label{eq:BSS_y}
\end{align}
\end{subequations}
where $\theta(k) \in [0,\pi]$ is the $s$-wave phase-shift due to the real long-range potential $V(r)$. Considering that $x_{\unreg}(k) = \sin^2\theta(k)$, it follows that 
\begin{align}
e^{\im 2\theta(k)} = 1-2x_{\unreg}(k) 
+\im 2\xi (k) \sqrt{x_{\unreg}(k) [1-x_{\unreg}(k)]}  ,
\end{align} 
with $\xi(k) \equiv {\rm sgn} [\sin 2\theta(k)] = {\rm sgn} [\Re M_\unreg^\elas(k)]$. 
As previously, $k=k_\diamond$ denotes a reference (or renormalization) momentum. Denoting by 
$\bar{\sigma}^{\elas} (k)$ and $\bar{\sigma}^{\ann}(k)$ 
the elastic and annihilation cross-sections without resummation of either the long-range or contact interactions (i.e., in most cases, the tree-level cross-sections due to the contact interactions), 
we have defined 
\begin{align}
\bar{x}_{\diamond} \equiv 
\dfrac
{\bar\sigma^{\elas} (k_\diamond)}{\sigma^U(k_{\diamond})}
\quad \text{and} \quad
\bar{y}_\diamond \equiv 
\dfrac
{\bar\sigma^{\ann} (k_{\diamond})}
{\sigma^U(k_{\diamond})} ,
\label{eq:xy_hat_RenormScale}
\end{align} 
where for an $s$-wave contact interaction at leading order, 
$\bar\sigma^{\elas} (k)$ 
and 
$k \, \bar\sigma^{\ann} (k)$
are independent of $k$. Correspondingly, $\bar{\xi}_\diamond$ denotes the sign of the real part of the elastic amplitude due to the contact interaction only, at $k=k_\diamond$. With this, we can write the unregulated annihilation cross-section at a momentum $k$, divided by the unitarity limit, as follows
\begin{align}
y_{\unreg}^{\ren} (k;k_\diamond) 
= \dfrac{k\,\sigma_{\unreg}^{\ann}(k)}{k \, \sigma^U(k)}
= \dfrac
{[k \, \bar\sigma^{\ann}(k)] \, S(k)}
{k \, \sigma^U(k)} 
= \bar{y}_\diamond 
\ \dfrac{k_{\diamond} \,  \sigma^U(k_\diamond)}{k \, \sigma^U(k)}
\ S(k) 
= \bar{y}_{\diamond} 
\ \dfrac{k}{k_{\diamond}}
\ S(k) .    
\end{align}
The functions $S(k)$ and $T(k;k_\diamond)$ are determined by the real long-range potential $V(r)$. $S(k)$ is the $s$-wave Sommerfeld factor corresponding to that potential, while $T(k;k_\diamond) \equiv [g_{0}^{(1)} (k)-g_{0}^{(1)} (k_\diamond)]/k$ encodes the momentum-dependent finite contributions to the Green's function~\cite{Blum:2016nrz}, where $g_{0}^{(1)} (k)$ is as in \cref{sec:Renorm_Position,sec:Jost_Frobenius}. Collecting the above, and comparing with our results of \cref{sec:Renorm_gamma=0,sec:Renorm_Position}, we find that \cref{eqs:BSS} reduce to \cref{eqs:SingleChannel_RegRen_xy}, if
\begin{align}
\label{eq:BSS_Correspondence}
[(k/k_\diamond) \,
\bar{y}_\diamond ] \,
T(k;k_\diamond) \to 
-z(k;k_\diamond)
\qquad\text{and}\qquad
\beta^{\BSS} (k_\diamond) 
\equiv
\bar{\xi}_\diamond 
\sqrt{\bar{x}_\diamond/\bar{y}_\diamond^2 - 1} 
\to \beta(k_\diamond).
\end{align}

Having already discussed the finite contributions to the Green's function in \cref{sec:Renorm_MethodComparison}, we now turn to the determination of $\beta$. In our formalism, $\beta(k_\diamond)$ is fixed by one observable, the elastic or inelastic cross-section, and theoretical input in the form of the unregulated cross-sections derived from the underlying Lagrangian. Indeed, \cref{eq:beta} expresses $\beta(k_\diamond)$ explicitly in terms of $y_{\reg}(k_\diamond)$ and $y_{\unreg}^\ren(k_\diamond;k_\diamond)$. Equivalently, combining \cref{eq:SingleChannel_RegRen_x,eq:SingleChannel_RegRen_y}, $\beta(k_\diamond)$ can be expressed in terms of $x_{\reg} (k_\diamond)$ and $x_{\unreg}(k_\diamond), y_{\unreg}^\ren (k_\diamond;k_\diamond)$. 

To match $\beta(k_\diamond)$ with $\beta^{\BSS}(k_\diamond)$, we must invoke the following approximations. If there exists a scale (which we can set to be $k_\diamond$), where the Hermitian potential ${\cal V}^{(0)}(\vb{r},\vb{r'})$ has negligible effect, so that $x_{\unreg} (k_\diamond) \approx 0$, then, recalling that $z(k_\diamond;k_\diamond)=0$,  \cref{eqs:SingleChannel_RegRen_xy} imply
\begin{align}
\beta^2(k_\diamond) \approx 
\dfrac{x_{\reg}(k_\diamond)}{y_{\reg}(k_\diamond) \, y_{\unreg}^\ren(k_\diamond;k_\diamond)}-1 ,  
\label{eq:beta_NoV0}
\end{align}
where $x_{\reg}(k_\diamond)$ is generated solely by the contact interaction \eqref{eq:SingleChannel_AmplitudeAtHighMomenta}, for $\gamma=0$. 
Under the further assumption that, at the same scale, it suffices to consider the contact interaction \eqref{eq:SingleChannel_AmplitudeAtHighMomenta} at lowest order, with any corrections emanating from its resummation via the potential \eqref{eq:ContactPotential_nu} being insignificant, both for the elastic and inelastic cross-sections, we may set 
$x_{\reg}(k_\diamond) \approx \bar{x}_{\diamond}$ and 
$y_{\reg}(k_\diamond) \approx y_{\unreg}^\ren(k_\diamond;k_\diamond) \approx \bar{y}_{\diamond}$, 
where $\bar{x}_{\diamond}$ and  $\bar{y}_{\diamond}$ are defined in \cref{eq:xy_hat_RenormScale}. With this, \cref{eq:beta_NoV0} reduces to the second condition \eqref{eq:BSS_Correspondence}, $\beta(k_\diamond) \approx \beta^{\BSS}(k_\diamond)$. These are the underlying assumptions in Ref.~\cite{Blum:2016nrz}. 
Although some of these approximations can be adequate (depending on the Hermitian potential and provided that the renormalization scale is taken sufficiently high, yet still within the non-relativistic regime), the proper determination of $\beta(k_\diamond)$ from  \cref{eq:beta} implies that $y_{\ell,\reg}(k_\diamond)$ cannot equal $y_{\ell,\unreg}^\ren(k_\diamond;k_\diamond)$ for any choice of $k_\diamond$, except in the trivial case of vanishing inelasticity.

%%%%%%%%%%%%%%%%%%%%%%%%%%%%%%%%%%%%%%%%%%%%%%%%%%%
%%%%%%%%%%%%%%%%%%%%%%%%%%%%%%%%%%%%%%%%%%%%%%%%%%%
%%%%%%%%%%%%%%%%%%%%%%%%%%%%%%%%%%%%%%%%%%%%%%%%%%%
\clearpage
\section{Discussion \label{sec:Discussion}}

\subsection{Methodology}

The unitarization method developed in Ref.~\cite{Flores:2024sfy} ensures that scattering amplitudes satisfy the coupled unitarity constraints governing elastic and inelastic processes. By properly resumming the inelastic contributions to the self-energy of the incoming state, in addition to those from purely elastic interactions, this approach establishes a precise relation between the regulated and unregulated amplitudes, computed with and without inclusion of the inelastic effects, respectively.

The consistency of the method relies on the form of the anti-Hermitian potential, expressed in terms of the irreducible inelastic vertices. As argued in Ref.~\cite{Flores:2024sfy}, this specific form emanates from the unitarity relation underpinning the optical theorem. In the present work, we showed that it can also be derived from the continuity equation and from the Feshbach formalism. The uniqueness and completeness of the anti-Hermitian potential underpin a correspondingly \emph{unique and complete unitarization scheme}.  Moreover, the anti-Hermitian potential -- non-local but separable -- gives rise to a Schrödinger equation that admits an analytic solution and yields a compact relation between the regulated and unregulated amplitudes, for both elastic and inelastic channels. Under suitable analyticity and convergence conditions, the regularization scheme depends solely on the unregulated inelastic amplitudes in a remarkably simple manner~\cite{Flores:2024sfy}.

In this work, we have employed the analytical solution previously obtained in Ref.~\cite{Flores:2024sfy} to systematically encode the effects of possible non-analytic or non-convergent behavior of the unregulated inelastic amplitudes, as well as the existence of bound-state solutions. This behavior is encapsulated in the $\WW_\ell$ matrix defined in \cref{eq:Wmatrix_FG}, which enters the unitarization prescription through \cref{eqs:Regularization_xy}. 
The $\WW_\ell$ matrix can be evaluated in three equivalent ways:
\begin{enumerate}[label=(\roman*)]
\item 
By (numerically) computing the integral involving  the unregulated inelastic amplitudes, given in \cref{eq:Wmatrix_PV}, along the real momentum axis. (See Ref.~\cite{Flores:2026yay} for an example.)

\item
Through contour integration of the expression  \eqref{eq:Wmatrix_PV} on the complex momentum plane, as in \cref{eqs:W_Contrib}, which makes manifest the contributions arising from the non-analyticities of the inelastic amplitudes or from their divergent behavior at large momentum, as described in \cref{sec:Renorm_W}. (See Ref.~\cite{Flores:2026yay} for an example).
 
\item 
Directly from the $\WW_\ell$ definition \eqref{eq:Wmatrix_FG}, involving position-space overlap integrals of the irreducible inelastic vertices with the regular and irregular wavefunctions of the Hermitian potential, ${\cal V}^{(0)}$, as in \cref{sec:Renorm_Position}.
\end{enumerate}

For amplitudes that do not converge at large momentum, we demonstrated two renormalization procedures within the unitarization scheme. An important outcome is that absorptive (anti-Hermitian) separable interactions require Hermitian counterterms of the same structure for renormalization to succeed. This generalizes the solution of Ref.~\cite{Flores:2024sfy}, which can now be applied to Hermitian separable potentials, even in the absence of absorptive counterparts.

\subsection{Phenomenology}

Having developed this framework, it is now possible to apply the unitarization method to a broad range of cases. Let us return to the examples mentioned in the introduction, which have provided much of the phenomenological motivation for this and other related works~\cite{Blum:2016nrz,Braaten:2017dwq,Parikh:2024mwa,Watanabe:2025kgw}.

\begin{description}

\item[Bound-state-formation] amplitudes are ultra-soft and therefore fall rapidly at large momenta, giving no contribution from the large-momentum arc in \cref{eq:W_Contrib_Infinity}. They may nevertheless contain poles or branch cuts in the complex momentum plane, which yield finite contributions to $\WW_\ell$. These may be computed either by evaluating the integral in \eqref{eq:Wmatrix_PV} along the real axis or by contour integration, as in \eqref{eq:W_Contrib}. A dedicated study is presented in~\cite{Flores:2026yay}.

Regularizing bound-state formation is phenomenologically crucial when capture occurs via emission of a scalar or vector boson that carries a conserved charge: without regularization, such processes violate unitarity at sufficiently low velocities even for arbitrarily small couplings~\cite{Oncala:2019yvj,Oncala:2021swy,Oncala:2021tkz,Harz:2018csl,Harz:2019rro,Binder:2023ckj,Beneke:2024nxh}. Even when this pathology is absent, e.g. for capture via emission of an Abelian gauge boson or a neutral scalar, unregulated cross-sections can exceed the unitarity bound at large but still perturbative couplings~\cite{vonHarling:2014kha,Petraki:2015hla,Petraki:2016cnz}. The unitarization procedure developed here applies in both cases, and requires minimal input: the unregulated bound-state formation cross-sections. (The unregulated elastic cross-sections are also required to obtain the regulated ones, but do not enter the regularization of the inelastic cross-sections.)

\item[Contact-type inelastic interactions] (such as annihilation of non-relativistic particles into relativistic species), when calculated perturbatively, violate unitarity at large enough couplings that may still be well within the perturbative regime. The problem is ameliorated if the interacting particles couple via a long-range force (e.g. a Coulomb potential) that is integrated in the computation (see e.g.~\cite{vonHarling:2014kha}), but persists nonetheless. 

In these cases, resumming the irreducible inelastic interactions introduces divergences that can be absorbed into the $\WW_\ell$ matrix and renormalized following the procedure described in \cref{sec:Renorm}. After renormalization, the resulting cross-sections satisfy unitarity. The final expressions take a form analogous to \cref{eqs:SingleChannel_RegRen_xy}, potentially extended to a multi-channel framework, with the necessary inputs consisting of the unregulated inelastic cross-sections together with the dispersive couplings induced by the inelastic vertices.

\item[Sommerfeld (parametric) resonances] arise when particles interact through a light but massive mediator and a bound level lies near threshold. For $s$-wave annihilation processes, these resonances drive the cross-sections to $\sigma_{\ell=0}^{\rm ann} \propto v_{\rm rel}^{-3}$ over the resonance-enhanced regime at low velocities, thereby violating unitarity even at small couplings. 

Equivalently, the on/off-shell unregulated partial-wave amplitudes scale as ${\cal M}_{\ell=0,\,\unreg} (q) \sim 1/q$ at small momenta. (For higher partial waves, the centrifugal barrier removes the infrared divergence at very low momenta.) Upon regularization, this infrared divergence is canceled by phase-space suppression in the integral of  \cref{eq:Wmatrix_PV} and causes no problem. 

The ultraviolet issue due to the contact part of the interaction remains, and, as mentioned above, can be treated according to \cref{sec:Renorm}, with the result of unitarization and renormalization shown in \cref{eqs:SingleChannel_RegRen_xy} for a single channel. 
\end{description}

These examples have a wide range of phenomenological applications, including dark-matter production in the early universe and indirect-detection signals. We leave detailed phenomenological studies to future work, though we note Ref.~\cite{Petraki:2025zvv}, where the simplest version of the prescription~\cite{Flores:2024sfy} has already been applied to dark-matter freeze-out. We anticipate that the unitarization framework developed in Ref.~\cite{Flores:2024sfy}, together with the extensions presented here, will prove useful across a range of physics beyond the Standard Model, and potentially also within the Standard Model, where long-standing interest in unitarization has motivated a variety of approaches, particularly in hadronic physics (see~\cite{Oller:2020guq} for a recent review).

\section*{Acknowledgments} 
We thank Alex Kusenko and Hitoshi Murayama for useful discussions. 
This work was supported by the European Union’s Horizon 2020 research and innovation programme under grant agreement No 101002846, ERC CoG CosmoChart. MMF was also supported through a FRIPRO grant of the Norwegian Research Council (project ID 353561 ‘DarkTurns’).

\clearpage

\appendix
\section*{Appendices}

\section{Mathematical formulas \label{app:Math}}

\subsection{Proof of identity  \label{app:Math_Proof}}

We aim to prove that
\begin{align}
\sum_j w_{A,\ell}^j = w_\ell , 
\label{eq:sumOfwAj=w}
\end{align}
with $w_\ell$ and $w_{A,\ell}$ defined in \cref{eq:w_def,eq:wAj_def}. We first define
$R\equiv (\ID+\ee \, \WW_\ell \ee)^{-1}$. 
Then, starting from the definition \eqref{eq:wAj_def} of $w_{A,\ell}^j$, 
\begin{align}
\sum_j w_{A,\ell}^j 
%%%%%%%%%%%%%%%%%%
&=\hat{M}_{\ell,\unreg}^\dagger\qty[
\ee^\dagger
\, R^\dagger
\, (\ee^\dagger)^{-1}
\, (\eeA^\dagger
\, \eeA^{})
\, \ee^{-1}
\, R
\, \ee
] \, \hat{M}_{\ell,\unreg}
\nn \\
%%%%%%%%%%%%%%%%%%
&=- \dfrac{\im}{2}
\hat{M}_{\ell,\unreg}^\dagger\qty[
\ee^\dagger
\, R^\dagger
\, (\ee^\dagger)^{-1}
\,\, \ee^2 \
\, \ee^{-1}
\, R
\, \ee
-\hc
] \, \hat{M}_{\ell,\unreg}
\nn \\
%%%%%%%%%%%%%%%%%%
&=- \dfrac{\im}{2}
\hat{M}_{\ell,\unreg}^\dagger\qty[
\ee^\dagger
\, R^\dagger
\, (\ee^\dagger)^{-1}
\, \ee
\, R
\, \ee
-\hc
] \, \hat{M}_{\ell,\unreg} ,
\label{eq:sumOfwAj=w_derivation}
\end{align}
where in the second line we used 
$\eeA^\dagger \eeA^{} 
= \eeA^2 
=-(\im/2) [\ee^2 - (\ee^\dagger)^2]$ (cf.~\cref{eq:eta_matrices_identities}). 
To proceed, we set 
$X \equiv \ee \, R \, \ee$ 
and 
$Y \equiv X^\dagger 
= \ee^\dagger 
\, R^\dagger 
\, \ee^\dagger$. 
Then, 
\begin{align}
Y^{-1}-X^{-1}
&=(\ee^\dagger)^{-1}
\, (R^\dagger)^{-1}
\, (\ee^\dagger)^{-1}
-
\, \ee^{-1}
\, R^{-1}
\, \ee^{-1}
\nn \\
&=
\qty[(\ee^\dagger)^{-2} +\WW_\ell ] 
- \qty[ \ee^{-2} + \WW_\ell]
=
(\ee^\dagger)^{-2} - \ee^{-2}.
\nn
\end{align}
Then, using the identity
$X-Y=Y\,(Y^{-1}-X^{-1})\,X$, 
we find
\begin{align}
\ee \, R \, \ee - \hc
&= X-Y 
\nn \\
&=
Y 
\, \qty[(\ee^\dagger)^{-2}-\ee^{-2}] 
\, X
\nn \\
&=
\ee^\dagger 
\, R^\dagger
\, (\ee^\dagger)^{-1} 
\, \ee 
\, R 
\, \ee 
-\hc ,
\end{align}
The right-hand side of the above is the square brackets in the last lines of \cref{eq:sumOfwAj=w_derivation}. Thus
\begin{align}
\sum_j w_{A,\ell}^j = -\dfrac{\im}{2}
\hat{M}_{\ell,\unreg}^\dagger\qty[
\ee \, R \, \ee - \hc
] \, \hat{M}_{\ell,\unreg}
\ \overset{\rm \cref{eq:w_def}}{=} \
w_\ell.
\end{align}

\subsection{Standard identities \label{app:Math_Identities}}

The relevant Kummer's transformation for our work is~\cite{DLMF13.2.39}
\begin{align}\label{eq:KummersTransform}
{}_1F_1(a,b,z)
=
e^{z}
{}_1F_1(b - a,b,-z)
.
\end{align}
For $m\in \ZZ_{\geqslant 0}$, the confluent hypergeometric functions of the first and second kind are related~\cite{DLMF13.2.7},
\begin{align}\label{eq:U1F1_Relation}
U(-m,b,z) =
(-1)^m 
\frac{\Gamma(b + m)}{\Gamma(b)}
{}_1F_1(-m,b,z)
.
\end{align}
For the Coulomb potential, it is standard to express the bound-state wave function in terms of the Laguerre polynomials using~\cite{DLMF13.6.19},
\begin{align}\label{eq:Hyper1F1_Laguerre}
{}_1F_1(-m,\alpha + 1, z) = \frac{\Gamma(\alpha + 1)\Gamma(m + 1)}{\Gamma(\alpha + m + 1)}L_{m}^{\alpha}(z),
\end{align}
where the Laguerre polynomials are normalized to satisfy~\cite{Arfken:2011mat},
\begin{align}\label{eq:LaguerreInt_I}
\int_0^\infty x^{\alpha + 1} e^{-x}\ [L_n^\alpha(x)]^2\ \dd x
=
\frac{\Gamma(n + \alpha + 1)}{\Gamma(n + 1)}
(2n + \alpha + 1)
.
\end{align}

\newpage

\section{Notation \label{app:Notation}}

\begin{longtable}{|
p{0.10\textwidth}| 
p{0.74\textwidth}| 
p{0.16\textwidth}|}
\hline 
\textbf{Symbol} 
& \textbf{Description} 
& \textbf{Defining Equation} 
\\
\hline \hline
$u_{k,\ell}^{(0)}(r)$ 
& Reduced wave function for Hermitian potential ${\cal V}^{(0)}$  
& \eqref{eq:SchroedingerEq_Real}
\\[1ex]
$u_{k,\ell}(r)$ 
& Reduced wave function for total potential
& \eqref{eq:SchroedingerEq}
\\[1ex]
$\varphi_{k,\ell}(r)$ 
& The regular solution & \eqref{eq:RegularSol_Asymptote0}
\\[1ex]
$H_{k,\ell}(r)$ 
& The Jost solution 
& \eqref{eq:JostSol_AsymptoteInf}
\\[1ex]
${\cal F}_{k,\ell}(r)$ 
& The physical scattering-state solution
& \eqref{eq:PhysicalSolution}
\\[1ex]
${\cal G}_{k,\ell}(r)$ 
& The irregular scattering-state solution
&\eqref{eq:IrregularSolution}
\\[1ex]
${\cal H}_{k,\ell}^{(\pm)}(r)$ 
& Outgoing/incoming spherical wave functions
& \eqref{eq:OutgoingWaveSolution}, \eqref{eq:IncomingWaveSolution}
\\[1ex]
${\cal B}_{n\ell}(r)$ 
& Bound-state solution 
& \eqref{eq:BoundStateDefn}
\\[1ex]
$G_{k,\ell}(r,r')$ 
& Radial Green's function 
& \eqref{eq:GreensDefn_DiffEq}, \eqref{eq:GreensFunction}
\\[1ex]
$\fs_\ell(k)$ 
& The Jost function 
& \eqref{eq:JostFunctionDefn}
\\[1ex] 
$\theta_\ell(k)$ 
& Phase shift due to Hermitian potential ${\cal V}^{(0)}$ 
& \eqref{eqs:WF_AsymptotesInf}
\\[1ex]
$\Delta_\ell(k)$ 
& Total phase shift 
& Above \eqref{eq:ImaginaryPhaseShift}
%%%%%%%%%%%%%%%%%%%%%%%%%%%%%%%%%
\\[1ex]
\hline
$\NN_\ell(k)$ 
& Matrix relating regulated and unregulated inelastic amplitudes
& \eqref{eq:Nmatrix_def}
\\[1ex]
$\WW_\ell(k)$ 
& Matrix encoding non-analytic and non-convergent behavior of unregulated inelastic amplitudes, contributing to $\NN_\ell(k)$
& \eqref{eq:Wmatrix_FG}
%%%%%%%%%%%%%%%%%%%%%%%%%%%%%%%%%
\\[1.5em]
\hline
${\cal M}_{\ell, {\rm (un)reg}}^{\inel, j}$ & 
$\ell$-wave (un)regulated inelastic amplitude, channel $j$
&\eqref{eqs:Minel_def}
\\[1ex]
$M_{\ell, {\rm (un)reg}}^{\inel, j}$ 
& $\ell$-wave (un)regulated inelastic amplitude, channel $j$, momentum-rescaled 
& \eqref{eq:M_def}
\\[1ex]
$\hat{\cal M}_{\ell, {\rm (un)reg}}^{j}$ 
& $\ell$-wave (un)regulated inelastic amplitude, channel $j$, inelastic coupling strength factored out
& \eqref{eq:Mhat-M_relation_cal}
\\[1ex]
$\hat{M}_{\ell, {\rm (un)reg}}^{j}$ 
& $\ell$-wave (un)regulated inelastic amplitude, channel $j$, inelastic coupling strength factored out, momentum-rescaled 
& \eqref{eqs:Mhat_def}, \eqref{eq:Mhat-M_relation_straight}
\\[1ex]
$
{\cal M}_{n\ell,{\unreg}}^{\BSD, j}$ 
& Unregulated decay amplitude of the $n\ell$ bound state into the $j$ channel
& \eqref{eq:M_BSDfull_def}
\\[1ex]
$
\hat{\cal M}_{n\ell,{\unreg}}^{\BSD, j}$ 
& Unregulated decay amplitude of the $n\ell$ bound state into the $j$ channel, inelastic coupling strength factored out
& \eqref{eq:M_BSDhat_def}
%%%%%%%%%%%%%%%%%%%%%%%%%%%%%%%%%
\\[1.5em]
\hline
$\sigma_\ell^U(k)$ 
& The unitarity cross-section 
& \eqref{eq:sigmaU_def}
\\[1ex]
$x_{\ell, {\rm (un)reg}}$ 
& (Un)regulated elastic cross-section normalized to $\sigma_\ell^U$
& \eqref{eq:x_def} 
\\[1ex]
$y_{\ell, {\rm (un)reg}}^j$ 
& (Un)regulated inelastic cross-section for channel $j$, normalized to $\sigma_\ell^U$
& \eqref{eq:yj_def} 
\\[1ex]
$y_{\ell, {\rm (un)reg}}$ 
& (Un)regulated total inelastic cross-section, normalized to $\sigma_\ell^U$
& \eqref{eq:y_def}
\\[1ex]
$w_\ell^j$ 
& Generalization of $y_{\ell, \unreg}^j$ entering the regularization prescription
& \eqref{eq:wAj_def}
\\[1ex]
$w_\ell$ 
& Generalization of $y_{\ell, \unreg}$ entering the regularization prescription
& \eqref{eq:w_def}
\\[1ex]
$\tilde{w}_\ell$ 
& Parameter encoding strength of Hermitian separable potentials, entering the regularization prescription
& \eqref{eq:wtilde_def}
\\[1ex]
\hline
\end{longtable}

%%%%%%%%%%%%%%%%%%%%%%%%%%%%%%%%%%%%%%%%%%%%%%%%%%%%%%%%
\clearpage
%\addcontentsline{toc}{section}{References}
\bibliography{Bibliography}

\end{document}

%% file: TIKZ_prelim.tex
\usepackage{tikz}
\usepackage{tikz-3dplot,circuitikz,pgfplots,tikz-cd,tikz-feynman} 
\usetikzlibrary{arrows,shapes}
\usetikzlibrary{trees,patterns}
\usetikzlibrary{matrix,arrows} 				% For commutative diagram
\usetikzlibrary{positioning}				  % For "above of=" commands
\usetikzlibrary{calc,through}				  % For coordinates
\usetikzlibrary{decorations.pathreplacing}  % For curly braces
\usepackage{pgffor}							% For repeating patterns

\usetikzlibrary{decorations.pathmorphing}	% For Feynman Diagrams
\usetikzlibrary{decorations.markings}
\tikzset{
>=stealth', %%  (Un)comment for (more) less conventional arrows
vector/.style={decorate, decoration={snake}, draw},
provector/.style={decorate, decoration={snake,amplitude=2.5pt}, draw},
antivector/.style={decorate, decoration={snake,amplitude=-2.5pt}, draw},
bigvector/.style={decorate, decoration={snake,amplitude=4pt}, draw},
fermion/.style={draw=black, postaction={decorate}, 
	decoration={markings,mark=at position .55 with {\arrow[draw=black]{>}}}},
fermionbar/.style={draw=black, postaction={decorate},
    decoration={markings,mark=at position .55 with {\arrow[draw=black]{<}}}},
fermionnoarrow/.style={draw=black},
doublefermion/.style={draw=black,double, postaction={decorate},
	decoration={markings,mark=at position .57 with {\arrow[draw=black]{>}}}},
doublefermionbar/.style={draw=black,double, postaction={decorate},
	decoration={markings,mark=at position .57 with {\arrow[draw=black]{<}}}},
doublefermionnoarrow/.style={draw=black,double},
gluon/.style={decorate, draw=black,
    decoration={coil,amplitude=4pt, segment length=5pt}},
scalar/.style={dashed,draw=black, postaction={decorate},
	decoration={markings,mark=at position .55 with {\arrow[draw=black]{>}}}},
scalarbar/.style={dashed,draw=black, postaction={decorate},
    decoration={markings,mark=at position .55 with {\arrow[draw=black]{<}}}},
scalarnoarrow/.style={dashed,draw=black},
momentum/.style={draw=black, postaction={decorate},
    decoration={markings,mark=at position 1 with {\arrow[draw=black]{>}}}},
antimomentum/.style={draw=black, postaction={decorate},
    decoration={markings,mark=at position 0.1 with {\arrow[draw=black]{<}}}}
}

\tikzstyle{block} = [draw, rectangle, minimum height=3em, minimum width=6em]
    

%% file: W_Contour.tex
\begin{tikzpicture}[scale=0.9]

    % Large dashed semicircle in the upper half-plane
    \draw[line width=2pt, red, opacity=0.5] (3,0) arc[start angle=0, end angle=180, radius=3];
    \draw[line width=2pt, red, opacity=0.5, <-] (0,3) -- (0.01,3); % Arrow for orientation

    % Deformed horizontal line: from -3 to -k, detour around -k, to +k, detour, to +3
    \draw[line width=2pt, red, opacity=0.5]
      (-3,0) -- (-2.3,0)
      arc[start angle=180, end angle=0, radius=0.3]; % bump around -k
        
    \draw[line width=2pt, red, opacity=0.5]
      (-1.7,0) -- (1.7,0)
      arc[start angle=180, end angle=0, radius=0.3]; % bump around +k

    \draw[line width=2pt, red, opacity=0.5] (2.3,0) -- (3,0); % finish line

    \draw[line width=2pt, red, opacity=0.5, ->] (-1,0) -- ++(0.001,0);
    \draw[line width=2pt, red, opacity=0.5, ->] (+1,0) -- ++(0.001,0);
    \draw[line width=2pt, red, opacity=0.5, ->] (-2.5,0) -- ++(0.001,0);
    \draw[line width=2pt, red, opacity=0.5, ->] (+2.75,0) -- ++(0.001,0);

    % Axes
    \draw[->] (-4,0) -- (4,0) node[right] {$\Re(q)$};
    \draw[->] (0,-0.5) -- (0,3.5) node[above] {$\Im(q)$};
    
    % Label the contour
    \node at (3,1.6) {${\cal C}_\Lambda$};
    
    % Origin
    \filldraw (0,0) circle (2pt);
    \node at (0.3,-0.3) {$0$};
    
    % x-like marks for +k and -k
    \node at (2,0) {\large $\times$};
    \node at (2,-0.3) {$+k$};
    
    \node at (-2,0) {\large $\times$};
    \node at (-2,-0.3) {$-k$};
    
    % Random poles inside the semicircle with labels
    \node at (1.5,1.2) {\large $\times$};
    \node at (1.8,1.2) {$q_1$};
    
    \node at (0.8,2.3) {\large $\times$};
    \node at (1.1,2.3) {$q_2$};

    \node at (-1,1.5) {\large $\times$};
    \node at (-0.7,1.5) {$q_3$};
    
    \node at (-1.8,0.7) {\large $\times$};
    \node at (-1.5,0.7) {$q_4$};

\end{tikzpicture}

%% file: Bibliography.bib
@article{Petraki:2025zvv,
    author = "Petraki, Kalliopi and Socha, Anna and Vasilaki, Christiana",
    title = "{The role of unitarisation on dark-matter freeze-out via metastable bound states}",
    eprint = "2505.20443",
    archivePrefix = "arXiv",
    primaryClass = "hep-ph",
    doi = "10.1088/1475-7516/2025/09/026",
    journal = "JCAP",
    volume = "09",
    pages = "026",
    year = "2025"
}

@article{Aydemir:2012nz,
    author = "Aydemir, Ufuk and Anber, Mohamed M. and Donoghue, John F.",
    title = "{Self-healing of unitarity in effective field theories and the onset of new physics}",
    eprint = "1203.5153",
    archivePrefix = "arXiv",
    primaryClass = "hep-ph",
    doi = "10.1103/PhysRevD.86.014025",
    journal = "Phys. Rev. D",
    volume = "86",
    pages = "014025",
    year = "2012"
}

@Article{Baldes:2017gzw,
  Title                    = {{Asymmetric thermal-relic dark matter:
 Sommerfeld-enhanced freeze-out, annihilation signals and
 unitarity bounds}},
  Author                   = {Baldes, Iason and Petraki, Kalliopi},
  Journal                  = {JCAP},
  Year                     = {2017},
  Number                   = {09},
  Pages                    = {028},
  Volume                   = {1709},

  Archiveprefix            = {arXiv},
  Doi                      = {10.1088/1475-7516/2017/09/028},
  Eprint                   = {1703.00478},
  Primaryclass             = {hep-ph},
  Reportnumber             = {DESY-17-034, NIKHEF-2017-009},
  Slaccitation             = {%%CITATION = ARXIV:1703.00478;%%}
}

@book{Lifshitz_RelativisticQM,
  title     = {Relativistic Quantum Theory, Part 1},
  author    = {Berestetskii, V. B. and Lifshitz, E. M. and Pitaevskii, L. P.},
  series    = {Course of Theoretical Physics},
  volume    = {4},
  publisher = {Pergamon Press},
  year      = {1971},
  note      = {Translated from the Russian by J. B. Sykes and J. S. Bell}
}

@article{Binder:2023ckj,
    author = "Binder, Tobias and Garny, Mathias and Heisig, Jan and Lederer, Stefan and Urban, Kai",
    title = "{Excited bound states and their role in dark matter production}",
    eprint = "2308.01336",
    archivePrefix = "arXiv",
    primaryClass = "hep-ph",
    reportNumber = "TUM-HEP 1469/23, TTK-23-21",
    doi = "10.1103/PhysRevD.108.095030",
    journal = "Phys. Rev. D",
    volume = "108",
    number = "9",
    pages = "095030",
    year = "2023"
}

@Article{Blum:2016nrz,
  Title                    = {{Self-consistent Calculation of the Sommerfeld Enhancement}},
  Author                   = {Blum, Kfir and Sato, Ryosuke and Slatyer, Tracy R.},
  Journal                  = {JCAP},
  Year                     = {2016},
  Number                   = {06},
  Pages                    = {021},
  Volume                   = {1606},

  Archiveprefix            = {arXiv},
  Doi                      = {10.1088/1475-7516/2016/06/021},
  Eprint                   = {1603.01383},
  Primaryclass             = {hep-ph},
  Slaccitation             = {%%CITATION = ARXIV:1603.01383;%%}
}

@article{Flores:2020drq,
    author = "Flores, Marcos M. and Kusenko, Alexander",
    title = "{Primordial Black Holes from Long-Range Scalar Forces and Scalar Radiative Cooling}",
    eprint = "2008.12456",
    archivePrefix = "arXiv",
    primaryClass = "astro-ph.CO",
    reportNumber = "IPMU20-0092",
    doi = "10.1103/PhysRevLett.126.041101",
    journal = "Phys. Rev. Lett.",
    volume = "126",
    number = "4",
    pages = "041101",
    year = "2021"
}

@Article{Braaten:2017dwq,
  Title                    = {{Zero-range effective field theory for resonant wino dark
 matter. Part III. Annihilation effects}},
  Author                   = {Braaten, Eric and Johnson, Evan and Zhang, Hong},
  Journal                  = {JHEP},
  Year                     = {2018},
  Pages                    = {062},
  Volume                   = {05},

  Archiveprefix            = {arXiv},
  Doi                      = {10.1007/JHEP05(2018)062},
  Eprint                   = {1712.07142},
  Primaryclass             = {hep-ph},
  Slaccitation             = {%%CITATION = ARXIV:1712.07142;%%}
}

@article{Cassel:2009wt,
    author = "Cassel, S.",
    title = "{Sommerfeld factor for arbitrary partial wave processes}",
    eprint = "0903.5307",
    archivePrefix = "arXiv",
    primaryClass = "hep-ph",
    reportNumber = "OUTP-0910P",
    doi = "10.1088/0954-3899/37/10/105009",
    journal = "J. Phys. G",
    volume = "37",
    pages = "105009",
    year = "2010"
}

@Article{ElHedri:2016onc,
  Title                    = {{A Sommerfeld Toolbox for Colored Dark Sectors}},
  Author                   = {El Hedri, Sonia and Kaminska, Anna and de Vries, Maikel},
  Journal                  = {Eur. Phys. J.},
  Year                     = {2017},
  Number                   = {9},
  Pages                    = {622},
  Volume                   = {C77},

  Archiveprefix            = {arXiv},
  Doi                      = {10.1140/epjc/s10052-017-5168-z},
  Eprint                   = {1612.02825},
  Primaryclass             = {hep-ph},
  Reportnumber             = {MITP-16-135},
  Slaccitation             = {%%CITATION = ARXIV:1612.02825;%%}
}

@Article{vonHarling:2014kha,
  Title                    = {{Bound-state formation for thermal relic dark matter and unitarity}},
  Author                   = {von Harling, Benedict and Petraki, Kalliopi},
  Journal                  = {JCAP},
  Year                     = {2014},
  Pages                    = {033},
  Volume                   = {12},

  Archiveprefix            = {arXiv},
  Doi                      = {10.1088/1475-7516/2014/12/033},
  Eprint                   = {1407.7874},
  Primaryclass             = {hep-ph},
  Reportnumber             = {NIKHEF-2014-018},
  Slaccitation             = {%%CITATION = ARXIV:1407.7874;%%}
}

@article{Harz:2019rro,
      author         = "Harz, Julia and Petraki, Kalliopi",
      title          = "{Higgs-mediated bound states in dark-matter models}",
      journal        = "JHEP",
      volume         = "04",
      year           = "2019",
      pages          = "130",
      doi            = "10.1007/JHEP04(2019)130",
      eprint         = "1901.10030",
      archivePrefix  = "arXiv",
      primaryClass   = "hep-ph",
      reportNumber   = "TUM-HEP-1186-19; Nikhef-2019-004",
      SLACcitation   = "%%CITATION = ARXIV:1901.10030;%%"
}

@Article{Harz:2018csl,
  Title                    = {{Radiative bound-state formation in unbroken perturbative
 non-Abelian theories and implications for dark matter}},
  Author                   = {Harz, Julia and Petraki, Kalliopi},
  Journal                  = {JHEP},
  Year                     = {2018},
  Pages                    = {096},
  Volume                   = {07},

  Archiveprefix            = {arXiv},
  Doi                      = {10.1007/JHEP07(2018)096},
  Eprint                   = {1805.01200},
  Primaryclass             = {hep-ph},
  Reportnumber             = {Nikhef-2018-023},
  Slaccitation             = {%%CITATION = ARXIV:1805.01200;%%}
}

@Article{Hisano:2003ec,
  Title                    = {{Explosive dark matter annihilation}},
  Author                   = {Junji Hisano and Shigeki Matsumoto and Mihoko M. Nojiri},
  Journal                  = {Phys.Rev.Lett.},
  Year                     = {2004},
  Pages                    = {031303},
  Volume                   = {92},

  Archiveprefix            = {arXiv},
  Doi                      = {10.1103/PhysRevLett.92.031303},
  Eprint                   = {hep-ph/0307216},
  Primaryclass             = {hep-ph},
  Reportnumber             = {ICRR-REPORT-500-2003-4, YITP-03-42},
  Slaccitation             = {%\%CITATION = HEP-PH/0307216;\%\%}
}

@article{Kamada:2022zwb,
    author = "Kamada, Ayuki and Kobayashi, Shin and Kuwahara, Takumi",
    title = "{Perturbative unitarity of strongly interacting massive particle models}",
    eprint = "2210.01393",
    archivePrefix = "arXiv",
    primaryClass = "hep-ph",
    doi = "10.1007/JHEP02(2023)217",
    journal = "JHEP",
    volume = "02",
    pages = "217",
    year = "2023"
}

@article{Ko:2019wxq,
    author = "Ko, Pyungwon and Matsui, Toshinori and Tang, Yi-Lei",
    title = "{Dark matter bound state formation in fermionic Z$_{2}$ DM model with light dark photon and dark Higgs boson}",
    eprint = "1910.04311",
    archivePrefix = "arXiv",
    primaryClass = "hep-ph",
    doi = "10.1007/JHEP10(2020)082",
    journal = "JHEP",
    volume = "10",
    pages = "082",
    year = "2020"
}

@article{Oncala:2021tkz,
    author = "Oncala, Ruben and Petraki, Kalliopi",
    title = "{Bound states of WIMP dark matter in Higgs-portal models. Part I. Cross-sections and transition rates}",
    eprint = "2101.08666",
    archivePrefix = "arXiv",
    primaryClass = "hep-ph",
    reportNumber = "Nikhef-2021-003",
    doi = "10.1007/JHEP06(2021)124",
    journal = "JHEP",
    volume = "06",
    pages = "124",
    year = "2021"
}

@article{Oncala:2021swy,
    author = "Oncala, Ruben and Petraki, Kalliopi",
    title = "{Bound states of WIMP dark matter in Higgs-portal models. Part II. Thermal decoupling}",
    eprint = "2101.08667",
    archivePrefix = "arXiv",
    primaryClass = "hep-ph",
    reportNumber = "Nikhef-2021-004",
    doi = "10.1007/JHEP08(2021)069",
    journal = "JHEP",
    volume = "08",
    pages = "069",
    year = "2021"
}

@article{Oncala:2019yvj,
    author = "Oncala, Ruben and Petraki, Kalliopi",
    archivePrefix = "arXiv",
    doi = "10.1007/JHEP02(2020)036",
    eprint = "1911.02605",
    journal = "JHEP",
    pages = "036",
    primaryClass = "hep-ph",
    reportNumber = "Nikhef-2019-50",
    title = "{Dark matter bound state formation via emission of a charged scalar}",
    volume = "02",
    year = "2020"
}

@Article{Petraki:2016cnz,
  Title                    = {{Radiative bound-state-formation cross-sections for dark
 matter interacting via a Yukawa potential}},
  Author                   = {Petraki, Kalliopi and Postma, Marieke and de Vries,
 Jordy},
  Journal                  = {JHEP},
  Year                     = {2017},
  Pages                    = {077},
  Volume                   = {04},

  Archiveprefix            = {arXiv},
  Doi                      = {10.1007/JHEP04(2017)077},
  Eprint                   = {1611.01394},
  Primaryclass             = {hep-ph},
  Slaccitation             = {%%CITATION = ARXIV:1611.01394;%%}
}

@Article{Petraki:2015hla,
  Title                    = {{Dark-matter bound states from Feynman diagrams}},
  Author                   = {Petraki, Kalliopi and Postma, Marieke and Wiechers, Michael},
  Journal                  = {JHEP},
  Year                     = {2015},
  Pages                    = {128},
  Volume                   = {1506},

  Archiveprefix            = {arXiv},
  Doi                      = {10.1007/JHEP06(2015)128},
  Eprint                   = {1505.00109},
  Primaryclass             = {hep-ph},
  Reportnumber             = {NIKHEF-2015-013},
  Slaccitation             = {%%CITATION = ARXIV:1505.00109;%%}
}

@Article{Tulin:2017ara,
  Title                    = {{Dark Matter Self-interactions and Small Scale
 Structure}},
  Author                   = {Tulin, Sean and Yu, Hai-Bo},
  Journal                  = {Phys. Rept.},
  Year                     = {2018},
  Pages                    = {1-57},
  Volume                   = {730},

  Archiveprefix            = {arXiv},
  Doi                      = {10.1016/j.physrep.2017.11.004},
  Eprint                   = {1705.02358},
  Primaryclass             = {hep-ph},
  Slaccitation             = {%%CITATION = ARXIV:1705.02358;%%}
}

@article{Yamaguchi:1954mp,
    author = "Yamaguchi, Yoshio",
    title = "{Two nucleon problem when the potential is nonlocal but separable. 1.}",
    doi = "10.1103/PhysRev.95.1628",
    journal = "Phys. Rev.",
    volume = "95",
    pages = "1628--1634",
    year = "1954"
}

@article{Flores:2024sfy,
    author = "Flores, Marcos M. and Petraki, Kalliopi",
    title = "{Unitarity in the non-relativistic regime and implications for dark matter}",
    eprint = "2405.02222",
    archivePrefix = "arXiv",
    primaryClass = "hep-ph",
    doi = "10.1016/j.physletb.2024.139022",
    journal = "Phys. Lett. B",
    volume = "858",
    pages = "139022",
    year = "2024"
}

@article{Parikh:2024mwa,
    author = "Parikh, Aditya and Sato, Ryosuke and Slatyer, Tracy R.",
    title = "{Regulating Sommerfeld resonances for multi-state systems and higher partial waves}",
    eprint = "2410.18168",
    archivePrefix = "arXiv",
    primaryClass = "hep-ph",
    reportNumber = "MIT-CTP/5790, OU-HET-1243",
    doi = "10.1007/JHEP12(2025)025",
    journal = "JHEP",
    volume = "12",
    pages = "025",
    year = "2025"
}

@article{Beneke:2024nxh,
    author = "Beneke, Martin and Binder, Tobias and de Ros, Lorenzo and Garny, Mathias and Lederer, Stefan",
    title = "{Perturbative unitarity violation in radiative capture transitions to dark matter bound states}",
    eprint = "2411.08737",
    archivePrefix = "arXiv",
    primaryClass = "hep-ph",
    reportNumber = "TUM-HEP-1534/24",
    doi = "10.1007/JHEP02(2025)189",
    journal = "JHEP",
    volume = "02",
    pages = "189",
    year = "2025"
}

@article{Feshbach:1958nx,
    author = "Feshbach, Herman",
    title = "{Unified theory of nuclear reactions}",
    doi = "10.1016/0003-4916(58)90007-1",
    journal = "Annals Phys.",
    volume = "5",
    pages = "357--390",
    year = "1958"
}

@article{Feshbach:1962nra,
  author       = {Feshbach, Herman},
  title        = {A Unified Theory of Nuclear Reactions II},
  journal      = {Ann. Phys.},
  volume       = {19},
  pages        = {287--313},
  year         = {1962},
  eprint       = {3901},
  archivePrefix= {inspire},
  doi          = {10.1016/0003-4916(62)90221-X}
}

@book{Rakityansky:2022jost,
  title={Jost functions in quantum mechanics},
  author={Rakityansky, Sergei A},
  year={2022},
  publisher={Springer}
}

@book{Chadan:1977pq,
    author = "Chadan, Khosrow and Sabatier, Pierre C.",
    title = "{Inverse Problems in Quantum Scattering Theory}",
    doi = "10.1007/978-3-662-12125-2",
    isbn = "978-3-662-12125-2",
    publisher = "Springer",
    series = "Theoretical and Mathematical Physics",
    year = "1977"
}

@book{Newton:1982qc,
    author = "Roger G. Newton",
    title = "{Scattering Theory of Waves and Particles}",
    edition = "2nd",
    year = "1982",
    publisher = "Springer",
    address = "Berlin, Heidelberg",
    isbn = "978-3-642-88130-5",
    doi = "10.1007/978-3-642-88128-2",
    url = "https://link.springer.com/book/10.1007/978-3-642-88128-2"
}

@article{Newton:1960nws,
    author = "Roger G. Newton",
    title = "{Analytic Properties of Radial Wave Functions}",
    journal = "Journal of Mathematical Physics",
    volume = "1",
    number = "4",
    pages = "319--347",
    month = apr,
    year = "1960",
    publisher = "American Institute of Physics",
    doi = "10.1063/1.1703665",
    url = "https://aip.scitation.org/doi/10.1063/1.1703665"
}

@book{Arfken:2011mat,
  title={Mathematical methods for physicists: a comprehensive guide},
  author={Arfken, George B and Weber, Hans J and Harris, Frank E},
  year={2011},
  publisher={Academic press}
}

@article{Watanabe:2025kgw,
    author = "Watanabe, Yuki",
    title = "{Unitarization of the Sommerfeld enhancement through the renormalization group}",
    eprint = "2508.09511",
    archivePrefix = "arXiv",
    primaryClass = "hep-ph",
    month = "8",
    year = "2025"
}

@misc{DLMF13.2.39,
  author       = {Olver, Frank W. J. and Lozier, Daniel W. and Boisvert, Ronald F. and Clark, Charles W.},
  title        = {{NIST Digital Library of Mathematical Functions, \S13.2.39}},
  year         = {2024},
  url          = {https://dlmf.nist.gov/13.2#E39},
  note         = {National Institute of Standards and Technology. Release 1.1.10, May 2024},
}

@misc{DLMF13.2.7,
  author       = {Olver, Frank W. J. and Lozier, Daniel W. and Boisvert, Ronald F. and Clark, Charles W.},
  title        = {{NIST Digital Library of Mathematical Functions, \S13.2.7}},
  year         = {2024},
  url          = {https://dlmf.nist.gov/13.2#E7},
  note         = {National Institute of Standards and Technology. Release 1.1.10, May 2024},
}

@misc{DLMF13.2,
  author       = {Olver, Frank W. J. and Lozier, Daniel W. and Boisvert, Ronald F. and Clark, Charles W.},
  title        = {{NIST Digital Library of Mathematical Functions, \S13.2}},
  year         = {2024},
  url          = {https://dlmf.nist.gov/13.2},
  note         = {National Institute of Standards and Technology. Release 1.1.10, May 2024},
}

@misc{DLMF13.6.19,
  author       = {Olver, Frank W. J. and Lozier, Daniel W. and Boisvert, Ronald F. and Clark, Charles W.},
  title        = {{NIST Digital Library of Mathematical Functions, \S13.6.19}},
  year         = {2024},
  url          = {https://dlmf.nist.gov/13.6#E19},
  note         = {National Institute of Standards and Technology. Release 1.1.10, May 2024},
}

@misc{DLMF2.7.i,
  author       = {Olver, Frank W. J. and Lozier, Daniel W. and Boisvert, Ronald F. and Clark, Charles W.},
  title        = {{NIST Digital Library of Mathematical Functions, \S2.7.i}},
  url          = {https://dlmf.nist.gov/2.7#i},
  note         = {National Institute of Standards and Technology.},
}

@book{Taylor:1972pty,
    author = "Taylor, John R.",
    title = "{Scattering Theory: The Quantum Theory of Nonrelativistic Collisions}",
    publisher = "John Wiley {\&} Sons, Inc.",
    address = "New York",
    year = "1972"
}

@article{Oller:2020guq,
    author = "Oller, J. A.",
    title = "{Unitarization Technics in Hadron Physics with Historical Remarks}",
    eprint = "2005.14417",
    archivePrefix = "arXiv",
    primaryClass = "hep-ph",
    doi = "10.3390/sym12071114",
    journal = "Symmetry",
    volume = "12",
    number = "7",
    pages = "1114",
    year = "2020"
}

@book{kato1980perturbation,
  author    = {Kato, Tosio},
  title     = {Perturbation Theory for Linear Operators},
  edition   = {2},
  year      = {1980},
  series    = {Die Grundlehren der mathematischen Wissenschaften},
  volume    = {132},
  publisher = {Springer-Verlag},
  address   = {Berlin, Heidelberg, New York},
  isbn      = {3-540-07558-5, 0-387-07558-5}
}

@article{ShermanMorrison:1950,
author = {Jack Sherman and Winifred J. Morrison},
title = {{Adjustment of an Inverse Matrix Corresponding to a Change in One Element of a Given Matrix}},
volume = {21},
journal = {The Annals of Mathematical Statistics},
number = {1},
publisher = {Institute of Mathematical Statistics},
pages = {124 -- 127},
year = {1950},
doi = {10.1214/aoms/1177729893},
URL = {https://doi.org/10.1214/aoms/1177729893}
}

@article{Bedaque:2002mn,
    author = "Bedaque, Paulo F. and van Kolck, Ubirajara",
    title = "{Effective field theory for few nucleon systems}",
    eprint = "nucl-th/0203055",
    archivePrefix = "arXiv",
    doi = "10.1146/annurev.nucl.52.050102.090637",
    journal = "Ann. Rev. Nucl. Part. Sci.",
    volume = "52",
    pages = "339--396",
    year = "2002"
}

@inproceedings{Kaplan:2005es,
    author = "Kaplan, David B.",
    title = "{Five lectures on effective field theory}",
    eprint = "nucl-th/0510023",
    archivePrefix = "arXiv",
    month = "10",
    year = "2005"
}

@article{Flores:2026yay,
    author = "Flores, Marcos M. and Petraki, Kalliopi",
    title = "{Unitarity violation and restoration in radiative bound-state formation}",
    eprint = "2602.20243",
    archivePrefix = "arXiv",
    primaryClass = "hep-ph",
    month = "2",
    year = "2026"
}
